\documentclass[preprint2,psfig]{aastex}

\def\arcs{\rlap{.}$^{\prime\prime}$}

\begin{document}


\title{The Ringed Spiral Galaxy NGC 4622. I. Photometry, Kinematics,
\\ and the Case for Two Strong Leading Outer Spiral Arms\altaffilmark{1}}


\author{Ronald J. Buta\altaffilmark{2} and Gene G. Byrd} 
\affil{Department of Physics \& Astronomy Department, University of Alabama,
    Tuscaloosa, AL 35487, USA}

\and

\author{Tarsh Freeman}
\affil{Bevill State Community College, Fayette, AL, USA}


\altaffiltext{1}{Based on observations with the NASA/ESA {\it Hubble Space
Telescope}, obtained at the Space Telescope Science Institute,
which is operated by the Association of Universities for Research
in Astronomy (AURA), Inc., under NASA Contract NAS 5-26555.}
\altaffiltext{2}{Visiting Astronomer, Cerro Tololo Inter-American Observatory.
CTIO is operated by AURA, Inc.\ under contract to the National Science
Foundation.}


\begin{abstract}
The intriguing nearly face-on southern ringed spiral galaxy NGC 4622, the first 
galaxy definitively shown to have leading spiral structure, is revisited
in this paper with new images from the {\it Hubble Space Telescope's} WFPC2,
together with ground-based optical and near-IR imaging, and a Fabry-Perot
H$\alpha$ velocity field. The data provide new information on the disk/bulge/halo
mix, rotation curve, star formation in the galaxy, and the sense of winding
of its prominent spiral arms. Previously, we suggested that the weaker,
inner single arm most likely has the leading sense, based on a numerical
simulation. Now, taking advantage of HST resolution and using de Vaucouleurs'
standard extinction and reddening 
technique to determine the near side of the galaxy's slightly tilted disk, 
we come to the more surprising conclusion that {\it the two strong outer
arms have the leading sense.} We suggest that this highly unusual configuration
may be the result of a past minor merger or mild tidal encounter.
Possible evidence for a minor merger is 
found in a short, central dust lane, although this is purely circumstantial
and an unrelated interaction with a different companion could also 
be relevant. The leading arms may be allowed to persist because NGC 4622
is dark halo-dominated (i.e., not ``maximum disk'' in the inner regions)
and displays a significantly rising rotation curve. 
The new HST observations also reveal a rich globular cluster system in the 
galaxy.  The mean color of these clusters is $(V-I)_o$ = 1.04 and the specific
frequency is 3.4$\pm$0.6. The luminosity function of these clusters
confirms the membership of NGC 4622 in the Centaurus Cluster.

\end{abstract}


\keywords{galaxies: spiral;  galaxies: photometry; galaxies: kinematics 
and dynamics; galaxies: structure}


\section{Introduction}

The sense of winding of spiral arms in a disk galaxy is an
important property that must be explained by gravitational
theories of spiral structure. Although both leading and
trailing waves are thought to be part of the dynamics
and propagation of spiral structure, the swing amplification mechanism
of Toomre (1981) demonstrates the robustness of trailing
waves over leading waves. Observationally, de Vaucouleurs (1958)
showed that in all spirals where it was possible to
determine the sense of winding of the arms directly, the
arms are trailing. De Vaucouleurs used Doppler shifts to determine which
half of the major axis of a galaxy is receding from us relative
to its center, and
then used an asymmetry in the observed dust distribution to determine
which side of the minor axis is the near side. The dust asymmetry
he used is not intrinsic but is caused by the fact that in an inclined 
galaxy, dust in the near side is silhouetted against the background starlight 
of the bulge and disk. In galaxies with significant nearly spherical
bulge components, this effect can be seen even if the inclination
is less than 45$^{\circ}$.

NGC 4622 is a nearly face-on southern spiral galaxy with an intriguing 
morphology. From a ground-based photograph, Byrd et al. (1989) pointed out 
that in addition to a pair of strong, lopsided outer arms winding outward clockwise,  
NGC 4622 has a weaker, single inner arm winding outward counterclockwise. 
Byrd et al. noted that one set of arms must be leading, a very rare 
configuration. Buta, Crocker, and Byrd (1992, hereafter BCB) showed using multi-band
surface photometry that the single inner arm is a stellar dynamical 
feature, not a result of an unusual dust distribution. NGC 4622 
thus became the most convincing case of a galaxy having leading spiral 
structure. However, these observations could not determine which
set of arms is leading. Based on theory and numerical simulations, 
Byrd, Freeman, and Howard (1993) suggested that the inner arm leads. 

We present new {\it Hubble Space Telescope} WFPC2 images of NGC 4622 
that challenge this conclusion. These images, together with a 
ground based Fabry-Perot H$\alpha$ velocity field, give the 
surprising and completely unexpected result that the two outer arms in 
NGC 4622 must be leading. The images also reveal (circumstantial)
evidence for a past merger
between NGC 4622 and a smaller galaxy. These results challenge current theories of
spiral structure in galaxies, and suggest that strong two-armed leading spirals
can result from an interaction.

The new observations are summarized in Section 2. In Section 3, we discuss the
morphology and group membership of NGC 4622, and consider its distance
which is important for the analysis in this paper. In Section 4 we
discuss the results of surface photometry and bulge, disk, and
Fourier decomposition.
Section 5 covers the velocity field and rotation of the galaxy,
while Section 6 discusses the sense of winding of the spiral arms.
Section 7 presents models that show how we can detect a near-side/far-side
reddening and extinction asymmetry to inclinations as low as 15$^{\circ}$,
at HST resolution. Sections 8 and 9 analyze the globular clusters
and associations in the galaxy. A discussion is presented in Section
10, and conclusions in Section 11.

\section{Observations}

\subsection{{\it Hubble Space Telescope} Optical Imaging}

The HST observations of NGC 4622 were secured on 25 May 2001 with
WFPC2. The center of the galaxy was placed within the WF3 and
positioned such that the whole galaxy lies within the WFPC2 field.
For most of the galaxy, the pixel size on these images is 
0\rlap{.}$^{\prime\prime}$1,
or 19.5 pc at the adopted distance of 40.2 Mpc (see Section 3).
The WF2/WF3 and WF3/WF4 boundaries cover 160$^{\prime\prime}$, or
31 kpc at the adopted distance.
Four broadband filters were used to
observe the galaxy: F336W, F439W, F555W, and F814W. These
four filters approximate the Johnson $U$, $B$, $V$, and 
Cousins $I$ photometric systems, respectively. 

The total exposure times were 2000s each for F336W
and F439W, and 1000s each for F555W and F814W. The observations were made in
a CRSPLIT mode to facilitate removal of cosmic rays. After receiving
the standard pipeline pre-processed images, the separate images were
corrected for bad pixels and columns using IRAF\footnote{
IRAF is distributed by the National Optical Astronomy Observatories,
which is operated by AURA, Inc., under contract with the National Science 
Foundation.} STSDAS routine WFIXUP, and
then the images for each filter were combined using STSDAS routine
CRREJ. The latter routine was very effective at removing most of the
cosmic rays. The gain used was 7 electrons per ADU and the 
readnoise was 5.2 electrons. 

\subsection{Ground-based Optical and Near-Infrared Imaging}

Optical images of NGC 4622 in $B$- and $I$-band filters were
obtained on 30 March 1992 UT with the 1.5-m telescope of the
Cerro Tololo Inter-American Observatory. A TEK2 1024$\times$1024
CCD was used with a gain of 1.6$e^-$ ADU$^{-1}$ and a readnoise
of 4$\pm$1 $e^-$. The scale of these images is 0\rlap{.}$^{\prime\prime}$435
pix$^{-1}$, or 84.8 pc at the adopted distance. Exposure times were 600s in $B$ and 300s in $I$.
The observations were bias-corrected, flat-fielded, and
cosmic ray cleaned using standard IRAF routines.

Near-Infrared Johnson $H$-band observations of NGC 4622 were obtained on 
9 February 1996 UT with the CTIO Infrared Imager (CIRIM) attached to the
1.5-m telescope. CIRIM was used with a gain of 9$e^-$ ADU$^{-1}$
and a readnoise of 37$e^-$. The procedures outlined by Joyce (1992)
were followed on both the acquisition and reduction of these
observations. The scale of these observations is 1\rlap{.}$^{\prime\prime}$137
pix$^{-1}$ (221.6 pc), and a total exposure time of 900s was achieved.

\subsection{Ground-Based H$\alpha$ Fabry-Perot Interferometry}

Observations at H$\alpha$ with the Rutgers Fabry-Perot Interferometer
(RFP) on the CTIO 4-m telescope were obtained by Buta and G. B. Purcell
on 1992 March 29 (UT).
Eleven frames separated by about 1$\AA$ were needed, covering a velocity
range of 500 km s$^{-1}$. The reduction of these images used standard
IRAF tasks and IRAF-based routines, and the procedure is described
by Purcell (1998) and Buta \& Purcell (1998). A preliminary reduction
and analysis of the velocity field was presented by Scott (1996).
We have used revised routines from Purcell (1998) to re-reduce
the velocity field and improve its quality. 

To match the coordinate
systems of the WFPC2, TEK2, CIRIM, and RFP images, we used an image from the
Digitized Sky Survey and STSDAS task XYEQ to measure accurate
coordinates of field stars around NGC 4622. This analysis provided
accurate checks of the scales and the relative orientations
of the images. The analysis also led to the identification of
a bright supernova (SN 2001jx) that was present at the time 
of the WFPC2 observations. Information on astrometry and photometry 
of the supernova is provided in Appendix B.

\section{Morphology, Group Membership, and Distance}

Figure~\ref{colorimage} shows a color image based on the four WFPC2 images of NGC 4622.
These images include only the part of the WFPC2 field occupied
by the galaxy, and exclude some of the surrounding field.
The color image captures all of the salient features of the galaxy.
Recent star formation is confined mainly to the two outer arms
and the southwest portion of the inner ring (see also Figure 1 of
BCB). The weaker, single
inner arm can be traced for more than 540$^{\circ}$, and only
has recent star formation near its juncture with the east outer
arm. The faint, yellow star-like objects scattered across
the central area are globular clusters (see Section 8). These
stand out in color fairly well from faint red foreground stars and
background galaxies. The color image also shows thin dust lanes
at the juncture between the inner single arm and the west outer
arm. This is the first time clear dust lanes have been identified
in NGC 4622.

The central region of NGC 4622 shows a previously unknown
feature in the new WFPC2 images: a central dust lane. This
is shown for the $V$-band and in the $V-I$ color index
in Figure~\ref{center}. The feature is well-defined and
almost splits the nucleus into equal halves. Detailed
HST studies of the central regions of spiral galaxies
in both optical (Carollo et al. 1998) and near-IR
(Carollo et al. 2002) filters reveal a wide variety
of nuclear morphological features, but nucleus-splitting
dust lanes in nearly face-on early-type spiral galaxies 
appear to be rare. The presence of such a feature in
NGC 4622 suggests that the galaxy has suffered from a minor merger 
that may have destroyed a small companion. The central dust
lane does not extend much outside the nucleus and is
sharper than most of the other dust features seen in the
WFPC2 images. The feature is most reminiscent of the striking
``X-shaped" dust lane crossing the nucleus of M51 (Grillmair et al. 1997).
At least one ``arm" of this
feature has been suggested to be an edge-on accretion disk feeding
the central AGN in M51. However, the feature in M51 is clearly more
complex than that in NGC 4622, and Grillmair et al. argue that the dust lane
obscuring the center of M51 is too asymmetric to be part
of a disk. More properties of the central dust lane in NGC 4622
are given in Section 4.3.

NGC 4622 has been assigned to Lyon Galaxy Group (LGG) number 305
(Garcia 1993). From the database available at the time, 16 galaxies
could be assigned to this group, whose mean luminosity-weighted 
heliocentric radial velocity is 4351 km s$^{-1}$. However, two other
groups are in the same general area: LGG 298 which has 54 members
at a mean radial velocity of 2900 km s$^{-1}$, and LGG 301 which
has 8 members at a mean radial velocity of 2294 km s$^{-1}$,
indicating a possible superposition of unrelated groups.
However, this is not the case.
The nature of this puzzling situation was actually clarified 
earlier by Lucey, Currie, and Dickens (1986), who found a bimodal
velocity distribution in Centaurus. These authors suggested
that although the two velocity components are separated by
about 1500 km s$^{-1}$ [with one component at a mean radial
velocity of 3000 km s$^{-1}$ (``Cen30") and the other component at
4500 km s$^{-1}$ (``Cen45")], analysis of the color-magnitude relation
for the E galaxies in each group suggested a similar distance.
NGC 4622 is close to NGC 4616, an E galaxy with a similar
redshift (see Figure 3 of BCB). Both are part of Lucey et al's subgroup 
called ``WBCen45", meaning they are in an elongated feature called
the "western branch." Lucey et al. suggested that there
was a possibility that members of WBCen45 were part of
a background group whose mean redshift is 4770 km s$^{-1}$,
based on the smaller sizes of the spirals compared to those
in WBCen30. However, this was not clearcut, and most of Cen45, 
including the main elliptical NGC 4709,
was considered to be part of the same cluster as Cen30 (corresponding
to LGG 298) but undergoing an infall, rather than being a separate 
group in the background.

Tully and Pierce (2000) have recently determined the distance to Cen30
using the Tully-Fisher relation. From 13 galaxies and a revised calibration
of the relation, they obtained a distance modulus of 33.02$\pm$0.17,
corresponding to a distance of 40.2$\pm$3.1 Mpc. The rms scatter about their 
template relation was 0.60 mag, the largest of the 12 clusters studied.
A recent study of distances from the surface brightness fluctuation
(SBF) method (Tonry et al. 2001) suggests that WBCen45 may not be
simply background galaxies. Two members of LGG 305 (NGC 4616 and NGC 4709), 
and a third likely member (ESO 323$-$34), are in the SBF database 
(Table 1 of Tonry et al. 2001) and have an unweighted mean distance 
modulus of 32.80 $\pm$ 0.16 (s.d.), corresponding to a distance of
36.3 $\pm$ 2.7 Mpc. The mean radial velocity (4787 km s$^{-1}$)
of these galaxies relative to the cosmic microwave background radiation 
is the same as that for NGC 4622 (4779 km s$^{-1}$), and their mean
distance modulus agrees within uncertainties with that for Cen30.

NGC 4616 and NGC 4603D, the two large galaxies closest to NGC 4622 in Figure 3
of BCB, have apparent $B_T$ magnitudes of 14.4 and 14.1, respectively,
compared to 13.4 for NGC 4622. If both NGC 4616 and NGC 4603D
are bonafide members of the Centaurus Cluster, then NGC 4622 is very
likely also a member and not part of the background group discussed by
Lucey, Currie, and Dickens (1986). We therefore adopt 40.2 Mpc as the
best estimate of the distance to NGC 4622. This is supported by our
analysis of the luminosity function of the globular clusters in Section 8.

\section{Surface Photometry}

Surface photometry provides a way of connecting the ground-based
and space-based observations. First we analyze the ground-based
optical and near-IR images to re-evaluate the photometric
orientation parameters of the galaxy and to assess bulge and disk
properties. Sky background levels on the optical and near-IR images
were estimated using IRAF routine IMSURFIT, by fitting a plane to
background intensities in a surrounding border, after removal of
foreground stars and other field objects. In both sets of images,
the field of view was large enough to insure accurate sky subtraction.
For the WFPC2 images, we used IRAF routine FITSKY to estimate the
average background in sections of the WF2 and WF4 fields farthest
from the center of the galaxy. Although the galaxy clearly fills the
entire WFPC2 field, this approach is adequate because for profile
modeling purposes, we use the groundbased images exclusively at the
lowest surface brightness levels, and the WFPC2 images at the highest
surface brightness levels.

\subsection{Isophotal Ellipse Fits}

The CTIO $B$ and $I$-band images are the deepest images we have
obtained of NGC 4622. The $B$-band image is displayed in Figure~\ref{deepB}
to show faint outer isophotes. These become rather round at
large radii. To re-evaluate the orientation parameters, we
first cleaned the images of foreground stars and field galaxies
using a combination of point spread function (psf) fitting
and image editing. IRAF routines ALLSTAR and IMEDIT were
used for this purpose. The cleaned images were then
block-averaged in 8$\times$8 pixel boxes and ellipses were
fitted to isophotes in steps of 0.1 mag arcsec$^{-2}$.
Figure~\ref{qpa} shows how the axis ratio and position
angle of the fitted ellipses vary in the outer parts of the
galaxy. For these fits, 2$\sigma$ rejection was used for
isophotes fainter than 26.0 mag arcsec$^{-2}$. The
position angles of the outer isophotes become fairly
constant beyond $a$ = 55$^{\prime\prime}$. Restricting
to 55$^{\prime\prime}$ $\leq$ $a$ $\leq$ 85$^{\prime\prime}$,
the mean axis ratio is 0.903$\pm$0.030 and the mean
position angle is $-$13\rlap{.}$^{\circ}$1 $\pm$ 8\rlap{.}$^{\circ}$4
(or 166\rlap{.}$^{\circ}$9 $\pm$ 8\rlap{.}$^{\circ}$4).
For an oblate spheroid with an intrinsic axis ratio of 0.2
(see Schommer et al. 1993; Tully and Pierce 2000), the mean axis ratio 
implies an inclination of 26$^{\circ}$ $\pm$ 4$^{\circ}$. However, 
the mean photometric major axis position angle is offset 35$^{\circ}$
from the line of nodes for circular rotation (Section 5). An intrinsic
distortion in the outer parts of the disk may be responsible for this offset.

\subsection{Decomposition of Major and Minor Axis Luminosity Profiles}

Analysis of the velocity field in Section 5 shows that the kinematic
line of nodes of NGC 4622 is in position angle +22$^{\circ}$. Figure~\ref{decomp}
shows mean (folded) luminosity profiles along this position angle and
the implied kinematic minor axis position angle. Plots of the profiles
versus $r^{1\over 4}$ show that the bulge can be approximated by
an $r^{1\over 4}$ law over the range 2\arcs 5 $\leq r \leq$ 11\arcs 7.
The outer parts of the profiles display an exponential decline
in surface brightness outside the bright spiral arms. Thus, we
have carried out a standard bulge/disk decomposition by fitting
a combination of the following equations to limited ranges of 
the profiles:

$$\mu^I = a^I + b^I r^{1\over 4} \eqno{1a}$$
$$\mu^{II} = a^{II} + b^{II} r \eqno{1b}$$

\noindent
where the superscript I refers to the bulge and the superscript II refers
to the disk. Although a more generalized law for the bulge, such as
a S\'ersic (1968) law with variable bulge exponent (Mollenhoff \& Heidt
2001) might provide a better description of the bulge profile (see,
e.g., Section 4.3), we have used the de Vaucouleurs $r^{1\over 4}$ law
as an approximation in order to take advantage of the Young (1976)
asymptotic volume density for the $r^{1 \over 4}$ law,
for the purposes of modeling the near side/far side reddening asymmetry
(see Section 7 and Appendix A.) 

For $B$ and $I$, only points in the ranges 
2\arcs 5 to 16\arcs 0 and 53\arcs 0 to 60\arcs 0
were used for the major axis fits, and 2\arcs 5 to 16\arcs 0
and 53\arcs 0 to 65\arcs 0 for the minor axis fits. These ranges
avoid the humps due to the bright spiral arms. Inside 2\arcs 5,
the light distribution departs from the $r^{1\over 4}$ law (see next
section).
We used composite profiles for these fits, using the HST data
from 2\arcs 5 to 6\arcs 0, and the ground-based profiles for all
larger radii. Table 1 summarizes the results of the fits for
the parameters $a^I$, $b^I$, $a^{II}$, and $b^{II}$. The
combined solutions are plotted as solid curves in
Figure~\ref{decomp} while the separate solutions are
plotted on the low resolution ground-based major axis
profiles in Figure~\ref{lowresprofs}.  Since
the isophotes of the bulge are round, the scatter in the
parameter $b^I$ reflects mainly fitting uncertainties. The models
do not reveal a significant difference in bulge effective 
radius between the $B$ and $I$ filters, therefore we adopt an
average slope $<b^I>$ = 4.7585$\pm$0.0652, corresponding
to an effective radius $r_e^I$ = 9\arcs 38 $\pm$ 0\arcs 56 or
1.83 $\pm$ 0.11 kpc.

For the $H$-band decomposition, we could not derive a reliable bulge
model directly, given the low resolution. Instead, we forced the
slope $b^I$ to be 4.7585 and solved for the other three parameters
in equations 1a and 1b. Table 1 summarizes these parameters for
both the major and minor axes, and Figure~\ref{lowresprofs} shows
the solutions for the major axis. For the fits, we used points
in the range 3\arcs 0 - 16\arcs 0 and 53\arcs 0 - 85\arcs 0 for the major axis
and 3\arcs 0 - 16\arcs 0 and 45\arcs 0 - 65\arcs 0 for the minor axis.

The parameters in Table 1 for the slope $<b^{II}>$ are also
not significantly different between $B$ and $I$, along each
axis. The averages yield a major axis effective radius
$a_e^{II}$ = 25\arcs 97 and a minor axis effective radius 
of $b_e^{II}$ = 23\arcs 30. The axis ratio, $b_e/a_e$ = 0.897,
is consistent with an inclination of 26$^{\circ}$. The mean
$B$-band central surface brightness $<a^{II}>$ = 22.29$\pm$0.09
corresponds [for Galactic extinction $A_B$ = 0.598 mag (Schlegel
et al. 1998) and an
inclination of 26$^{\circ}$] to a corrected value $B(0)_c$ = 
21.81 mag arcsec$^{-2}$, within the original range of this 
parameter noted by Freeman (1970).

The parameters from this decomposition for $B$ and $I$ agree well
with those derived by BCB. From purely ground-based profiles
based on TI CCD images, the average bulge effective radius
from $B$ and $I$ filters was 8\arcs 8$\pm$0\arcs 4 while the
average disk major axis effective radius was 
28\arcs 5 $\pm$ 0\arcs 9. 

The decompositions indicate that the bulge is a significant
contribution to the total luminosity of NGC 4622. To determine
the relative flux contribution of the bulge, we computed total
magnitudes in $B$, $I$, and $H$ by extrapolating
the bulge/disk decomposition profiles. This led to
$B_T$ = 13.44 and $I_T$ = 11.09, to be compared with 13.37
and 11.13, respectively, derived by BCB. The total $H$-band
magnitude was found to be $H_T$ = 9.16. Using the mean
values of $a^I$ in all three filters, an effective radius of
9\arcs 38, and the revised total magnitudes derived
here, we find that the bulge of NGC 4622 contributes
39\%, 51\%, and 58\% of the $B$, $I$, and $H$-band luminosities,
respectively. The $B$-band value is consistent
with a Hubble type of Sa according to Table 3B of
Simien and de Vaucouleurs (1986). The other photometric
parameters are consistent with this same Hubble type
based on Figures 4 and 6 of Simien and de Vaucouleurs
(1986). Thus, in spite of the apparent asymmetries,
NGC 4622 has normal parameters for its type.

For the adopted distance of 40.2 Mpc, NGC 4622 has a total
absolute magnitude of $M_B^{b,k,i}$ = $-$20.3. This would make it
an intermediate luminosity Sa spiral.

\subsection{The Central Light Distribution and Dust Lane}

The $r^{1\over 4}$ law does not describe the central
light distribution of NGC 4622. There is excess light
over the $r^{1\over 4}$ law defined by the outer
parts of the bulge in this region. This excess was
evident even in the groundbased study of BCB, where 
the residual $I$-band light distribution showed a 
small ring of excess light around the center. A better
representation of the light distribution in this
region is provided by a "Nuker" law, which describes
the central regions of many ellipticals (Lauer et al.
1995; Faber et al. 1997). The 
Nuker law includes the parameters $\alpha$, $\beta$,
and $\gamma$ that define its shape and two other
parameters, $\mu_b$ and $r_b$, that describe the
surface brightness and radius of the "break" point
on the profile. For ellipticals, if the parameter
$\gamma$ $<$ 0.3, the galaxy is a "core type"
while if $\gamma$ $>$ 0.3, the galaxy is a
"power law" type.

The central light profile in NGC 4622 was derived
by fitting free ellipses to the central two-dimensional
light distribution, using the STSDAS routine ELLIPSE (Jedrezjewski
1987). These profiles are shown for $B$, $V$, $I$, $B-V$,
and $V-I$ in Figure~\ref{nukers}. The central dust lane impacts 
these profiles and causes the small dips in the $B$ and
$V$ profiles and the small peaks in $B-V$ and $V-I$.
The $I$-band profile is least affected so we fit only
this profile for the Nuker law parameters. Excluding
points having $log r $ $>$ 2.575 ($r >$ 376 pc),
we derive $\alpha$ = 2.188 $\pm$ 0.086, $\beta$ = 1.208 $\pm$ 0.010,
$\gamma$ = 0.059 $\pm$ 0.019, $\mu_b(I)$ = 15.016 $\pm$ 0.010,
and $r_b$ = 67.88 $\pm$ 0.97 pc. Thus, the central profile
is a "core" type. In ellipticals, core types are characteristic
of the more massive and luminous cases. Kormendy (1987) reviews
many of the known properties of cores in E galaxies and bulges.

Figure~\ref{nukers} also shows the ellipticity $\epsilon$
and position angle $\phi$ of the fitted ellipses near the
center. Beyond $r$ = 125 pc ($log(r)$ = 2.1), the ellipticity
is below 0.05.

The properties of the central dust lane were derived by fitting ellipses
to isochromes in the central 1$^{\prime\prime}$ radius. The IRAF
routine ELLIPSE was also used for this purpose. Figure~\ref{qpanuc}
shows the axis ratio and position angle of the fitted ellipses
as a function of radius. The isochromes have a nearly constant
axis ratio of $<q>$ = 0.38 $\pm$ 0.01 from 0\arcs 15 $\leq$ $a$ $\leq$
0\arcs 70. Out to its apparent edge, the diameter of the feature is 1\arcs 4 =
273 pc while its width along its minor axis is 0\arcs 53 or 104 pc.
Figure~\ref{qpanuc} shows that over the same radius range, the position
angle increases from 60$^{\circ}$ to about 65$^{\circ}$, a possible
signature that the dust lane represents an edge-on, slightly warped
central dust disk. Although ionized gas was not detected in this region
in the groundbased H$\alpha$ Fabry-Perot interferometry, high resolution
H$\alpha$ observations might verify this possibility.

\subsection{Fourier Decomposition} 

Fourier decomposition provides another useful way of dissecting
the structure of NGC 4622. It was used by BCB to analyze
their groundbased images, and we repeat their analysis
here using the WFPC2 data. We investigate the Fourier
structure of the stellar background in the $I$-band WFPC2
image, as well as the variation of relative Fourier
amplitudes and phases with radius in $B$, $V$, and $I$.

Figure~\ref{Fourier} shows four images that duplicate
those shown by BCB but which are at higher resolution.
The figure shows the $m$=0, 1, and 2 Fourier components,
in addition to an image that sums the $m$ = 0-6 components,
based on a routine that computes circular averages centered on the nucleus.
The $m$=1 image shows a complete inner spiral arm that clearly winds
at least 540$^{\circ}$, and which is distinct from all the
other structure in the map. The map also shows that there
is some asymmetry within the bright bulge region, especially
south of the center. The $m$=2 map reveals the two bright
outer arms and little structure in the region dominated
by the single inner arm. The $m$=0 image is dominated
by a ring that represents the average feature seen.

The $m$=0-6 image provides a very good approximation
to the total stellar background, and we show in Figure~\ref{onlyclusters}
the difference between the full mosaic $I$-band image and
this summed Fourier image. The figure highlights the smallest-scale
structures, and so provides an intriguing
map of the distribution of star clusters, both young and old,
in NGC 4622. The map also reveals lanes of dust on the west and
south sides of the inner ring, but is surprisingly less sensitive
to the dusty structures evident to the east in the original 
HST image.

Figures~\ref{mplots}a and b show the relative amplitudes and
phases of the $m$=1 and $m$=2 Fourier components. These are higher
resolution versions of Figure 9 of BCB, but unlike BCB, the amplitudes
and phases
are now based on deprojected images using the inclination and line of
nodes derived in Section 5. Both Fourier components
show relative amplitudes generally less than 10\% inside 
$r$=16$^{\prime\prime}$. Figure~\ref{mplots}a shows once again the
interesting phase change in the $m$ = 1 component
from $B$ to $V$ to $I$ in the vicinity
of the inner ring (21\rlap{.}$^{\prime\prime}$5 $\leq$ $r$ $\leq$ 
28\rlap{.}$^{\prime\prime}$5). The possible significance of this
change was discussed by BCB and Byrd, Freeman, \& Howard (1993).
We consider it again here in Section 10.4, as well as other details
in the plots. 

\section{Velocity Field and Rotation}

\subsection{Dynamical Parameters and Rotation Curve}

Critical to our interpretation of NGC 4622 is the velocity
field and the rotation of the disk. Figure~\ref{linecont}
shows the continuum and emission maps from the H$\alpha$ Fabry-Perot
interferometry. The emission map shows that the HII regions
in NGC 4622 lie mainly along the two main outer
arms and the southwest half of the inner ring. Little
or no H$\alpha$ emission is detected inside the inner
ring. The H$\alpha$ distribution is asymmetric with
stronger emission in the east outer arm as opposed to the
west outer arm.

The velocity field of NGC 4622 is shown in Figure~\ref{vfield}.
In spite of the low apparent inclination, NGC 4622 shows a 
well-defined line of nodes, the velocity gradient being along the
northeast/southwest direction, consistent with the orientations
of isophotes just outside the main spiral arms. The Fabry-Perot
analysis also provided information on velocity dispersions within
the ionized gas disk. However, the map (not shown here) only displays 
an irregular distribution of small dispersions (less than 20 km s$^{-1}$ 
in general), and only a few small isolated patches having up to 50 
km s$^{-1}$ dispersion. There is no pattern to suggest any serious
disruption of the velocity field perpendicular to the main disk plane.

The kinematic parameters of NGC 4622 were derived assuming only circular 
motions are present and using the iterative method
of Warner, Wright, and Baldwin (1973). All velocity points
were used, weighted according to the cosine of the angle of
each point relative to the line of nodes in the galaxy plane.
The lack of significant ionized gas inside the inner ring
made it difficult to solve for the rotation center, thus
we forced the rotation center in this analysis to be coincident
with the red continuum optical center. This method also did
not constrain the inclination reliably, with any inclination
between 10$^{\circ}$ and 30$^{\circ}$ being possible. This is
not unusual for a low inclination galaxy. Thus, we have used a different 
approach to estimate the kinematic value of this parameter.

The method we use to infer a kinematic inclination is the
Tully-Fisher relation in conjunction with an observed 21-cm
line width. At the time of this writing, we were unable
to find a published 21-cm line profile and width, but
were able to have the galaxy observed at HI specially with the
Parkes 64-m telescope. This spectrum, obtained
by M. Meyer (2002, private communication), will be presented
elsewhere. The main result is that NGC 4622 shows an
asymmetric double-horned line profile having a line center
corresponding to an (optically-defined) systemic velocity
of 4502$\pm$3 km s$^{-1}$ and a line width $W_{20}$ = 157$\pm$5
km s$^{-1}$. We ask what the inclination of NGC 4622 would
have to be to give a distance in the range 40.2$\pm$3.1 Mpc.
We use the $I$-band total magnitude (11.09) derived in this 
paper, and the corrections for inclination, K-dimming, and
Galactic extinction as described by Tully and Pierce (2000).
With equation 7 of Tully and Pierce (2000) and the newly measured
21-cm line width, we obtain a distance to NGC 4622 in the range
40.2$\pm$3.1 Mpc for an inclination in the range 
19\rlap{.}$^{\circ}$3$\pm$1\rlap{.}$^{\circ}$7, excluding
uncertainties in the total magnitude. The expected maximum 
rotation velocity would be 190$\pm$16 km s$^{-1}$.
Additional uncertainties inherent in using this method include the
reliability of using a relation defined entirely by galaxies
more inclined than 45$^{\circ}$, and the complexities of
the Centaurus Cluster itself. Nevertheless, the 21-cm line
width analysis suggests that NGC 4622 is less inclined than
than the value of 26$^{\circ}$ that was derived from the isophote
fitting method.

With an inclination of 19$^{\circ}$ and a rotation center coincident
with the nucleus, iteration for the remaining parameters 
gave a heliocentric systemic radial velocity of 4502 $\pm$ 3 km s$^{-1}$ 
and a kinematic line of nodes position angle of 22$^{\circ}$ $\pm$ 5$^{\circ}$
(J2000). The systemic velocity is in excellent agreement with the 21-cm
value. However, the position angle differs by 35$^{\circ}$ from the
mean photometric major axis position angle derived in Section 
4.1, probably due to the low inclination. The systemic velocity
is larger by more than 100 km s$^{-1}$ than the value given
in RC3 (de Vaucouleurs et al. 1991).

Figure~\ref{ewrotc} shows the implied rotation curve of NGC 4622 
on each side of the minor axis, while Table 2 summarizes the
folded mean rotation curve. The error bars shown
are standard deviations $\sigma$ around the means after
two cycles of 2$\sigma$ rejection. On each side, the rotation
velocity appears to rise with radius. On the northeast side,
the rise is linear and almost solid body. On the southwest side
(the side with the most apparent asymmetry), the rise is more
gentle followed by a rapid change, with the last point reaching
300 km s$^{-1}$. This is much higher than the expected maximum
rotation velocity from the 21-cm line width.

\subsection{Constant Mass-to-Light Ratio Analysis}

To evaluate the meaning of the observed rotation curve, we have
used the $H$-band light distribution to compute the gravitational
potential due to the stars in NGC 4622. Since the bulge is likely
to be rounder than the disk, we have used the decompositions
in Table 1 to subtract the bulge from the $H$-band light
distribution, and then deprojected the residual disk light
according to our orientation parameters. The disk potential 
in the plane of the galaxy was
derived using the method of Quillen, Frogel, and Gonzalez
(1994). The vertical density was assumed to be exponential
with a scaleheight of 325 pc, similar to the scaleheight
of the Galaxy (Gilmore and Reid 1983). The rotation curve
of the disk was derived from the disk potential by computing
a circularly-averaged potential profile as a function of
radius. The rotation curve of the bulge was derived from
the bulge decomposition parameters assuming the bulge is
spherical, using a program from Kalnajs and Hughes (1984).

Figure~\ref{rotc1}a shows the combined rotation curve predicted
from the $H$-band light distribution, assuming mass-to-light
ratios $(M/L_H)_{bulge} = (M/L_H)_{disk}$ = 1.0. This would be 
consistent with an old, single-burst stellar population having
[Fe/H] $\approx$ 0,  based on the evolutionary synthesis models
of Worthey (1994). The predicted combined rotation curve gives 
about the right $V_{c,max}$ if NGC 4622 is at the distance of 
40.2$\pm$3.1 Mpc, but is inconsistent with the observed rotation 
curve. Whereas the predicted rotation curve is
a normal, relatively flat one, with some decline in the
outer parts likely due to the neglect of dark matter,
the observed gaseous rotation curve is rising and does not 
intersect the predicted rotation curve until a radius of 
35\arcs 0 - 45\arcs 0, on both sides of the major axis.

The only way to get the predicted rotation curve to approximate
the observed one is to either reduce the mass-to-light ratio of
the bulge or disk or both, or set this parameter equal to zero for
both the bulge and the disk and consider a pure dark halo model.
Since we have no rotation information inside the bulge-dominated
area, we have little to constrain such a model. Nevertheless,
Figure~\ref{rotc1}b shows a model where we have set the $H$-band mass-to-light
ratio equal to 0.25 for both the bulge and disk and used a "fixed-sigma"
model (Kent 1986) for the halo. The halo model shown has an asymptotic
circular velocity $V_o$ = 190 km s$^{-1}$ and a characteristic radius
$a$ = 10$^{\prime\prime}$. Although the model shows a general rise,
it cannot reproduce the two points at $r$ $>$ 45$^{\prime\prime}$ on
the southwest side. Figure~\ref{rotc1}c shows that a pure halo model 
could account for a rising rotation curve. However, the model shown 
has $V_o$ = 250km s$^{-1}$.

Among Sa spirals, rising rotation curves are not infrequent, especially
for the lower luminosity systems (Rubin et al. 1985). Kent
(1988) shows that an extreme case, NGC 4698, has a rising rotation
curve that is well-fitted by a constant mass-to-light ratio. 
Thus, the disagreement between the observed rising rotation curve and the
one predicted from a constant mass-to-light ratio for NGC 4622 is probably
significant.

\subsection{The Impact of Streaming Motions in the Outer Arms}

The rising rotation curve is an important finding from our study.
It turns out that if the rise is real, we can much more easily explain
the outer leading arms (Byrd, Freeman, \& Buta 2002, hereafter paper II). 
The implication would be that NGC 4622 has a very
important dark matter contribution within the visible disk. However, 
the combination of noncircular motions and a low inclination could distort
the observed rotation curve. There is some evidence for such motions
in the northwest section of the disk where the observed velocities are
higher than expected for pure circular rotation. We ask whether such 
motions (1) are explicable in terms of a density wave; and (2) can 
explain the apparently rising rotation velocities.

First, it is important to note that streaming motions alone cannot resolve
the issue of whether the outer arms are leading or trailing in NGC 4622.
Consider a two-armed spiral that appears to open outward clockwise on 
the sky, as in NGC 4622.
In the rotating reference frame of the spiral pattern, at least in
linear theory, a two-armed trailing density wave rotating counterclockwise,
confined entirely within corotation, and viewed with the west side as
the near side would show the same pattern of streaming motions as a 
two-armed leading density wave rotating clockwise, confined entirely 
outside corotation, and viewed with the east side as the near side.

The main question is whether we expect a substantial rise across the
arms due to the perturbation. To examine this, we use the equations
of linear density wave theory from Rogstad (1971) with a flat
rotation curve set at 200 km s$^{-1}$, implied by the constant
mass-to-light ratio analysis. Even with an assumed density contrast
of 10 (the parameter $c$ in Rogstad 1971), the main effect of the
streaming motions is to cause "wiggles" across each spiral arm in the
sense that, relative to the mean circular velocity, the residual
velocities are positive on the inner edges of the arms and negative
on the outer edges. The pattern seen in NGC 4622 is instead a 
general tendency for the velocities to rise across the arms. Therefore,
this rise is not likely to be an artifact of streaming motions.

\section{The Sense of Winding of the Spiral Arms of NGC 4622}

The information from surface photometry has shown that, while
NGC 4622 has normal bulge/disk parameters for its Hubble type,
its near-infrared photometric structure fails to account for
its observed rotation curve. Even if we did not have any of
this information, we could reliably deduce that NGC 4622 has
leading spiral structure. As noted by Byrd et al. (1989) and
by BCB, either NGC 4622 has a single inner leading arm and two 
normal trailing outer arms (Scenario 1), or a single trailing 
inner arm and two very unusual bright leading outer arms 
(Scenario 2). One of our goals in observing NGC 4622
with HST was to determine if we could {\it prove directly}
which of these two scenarios is correct. Although the first
scenario is clearly more satisfying than the second, since it
is theoretically much easier to make two strong trailing arms than two 
strong leading arms, we have chosen to be open-minded about the 
possibilities. 

The ground-based Fabry-Perot velocity field has shown that gas clouds
on the northeast side of the disk are receding from us relative to
the systemic velocity. If we could determine which side of the disk is
the near side, then we could determine directly which arms lead
in NGC 4622. If Scenario 1 is correct, then NGC 4622
would have to be rotating counterclockwise and the west side 
would be found to be the near side. If Scenario 2 is correct,
then the galaxy would have to be rotating clockwise and the east
side would be found to be the near side.

The best way to determine the near side of a tilted
disk is to look for a reddening and extinction asymmetry across
the line of nodes. If the dust is confined to a thinner plane than
the stellar disk, and especially if a significant bulge is present,
then the bulge is viewed through the dust on the near side and the
dust is viewed through the bulge on the far side, leading to the
reddening and extinction asymmetry. In early studies of this problem (e.g.,
Hubble 1943; de Vaucouleurs 1958), blue light images of highly inclined
spiral galaxies showed the effect unambiguously, and it was concluded
that all spiral arms are trailing. In no case was it ever suggested 
in these early studies that this method could be applied to a galaxy
less inclined than 30$^{\circ}$. In such galaxies it was not possible,
with the old plates, to see the effect well enough for it to
be unambiguous.

In NGC 4622, there are two factors that have a lot to do with our ability
to see this effect. First, as we have shown, the bulge is a significant
fraction of the total luminosity of NGC 4622. Second, the superior
resolution of HST allows us to detect extremely weak dust lanes.
We use two methods to investigate the dust and reddening in NGC 4622.
The first is a $V-I$ color index map, shown in Figure~\ref{vimap}.
With the WFPC2 images obtained, this is the only color index map
we can derive that has a high enough signal-to-noise ratio. The
grayscale in Figure~\ref{vimap} is coded such that red features are
light and blue features are dark. The two
solid white lines in Figure~\ref{vimap} show the position angle of the
kinematic line of nodes, which in the frame of the WFPC2 images
is tipped about 8$^{\circ}$ from vertical. 
The map shows obvious weak, thin dust
lanes concentrated mainly on the east side of the line of nodes
in the region of the inner ring and between the nucleus and
the ring. The west side also shows some dust lanes, but these
appear much weaker by comparison. 

In Figure~\ref{residmap}, we show the $V$-band WFPC2 image of NGC 4622
after subtraction of the $m$=0 Fourier component of the light
distribution (image in upper right panel of Figure~\ref{Fourier}). 
In this map, areas of extinction are clearly seen
east of the line of nodes, with some strong regions of negative
residuals close to the center. That these are generally also zones of
reddening is verified by comparing Figures~\ref{vimap} and ~\ref{residmap}.
Although the bulk of the apparent extinction and reddening is
east of the line of nodes, we can still identify such regions on the
west side of the line of nodes. For example, in Figure~\ref{vimap},
dust patches can be seen to the northwest and southwest within 
5$^{\prime\prime}$ of the nucleus. However, the effects of these appear 
to be muted compared to the east side. Since the extinction and
reddening asymmetries clearly appear to know the line of nodes,
we conclude that the east side of NGC 4622 is the near side of
its disk. With the ground-based velocity field, this would
imply that NGC 4622 is rotating clockwise. {\it In this circumstance,
our Scenario 2 would be the correct one: the two strong outer
arms would be leading while the single inner arm would be trailing.}

Figure~\ref{bhmap} shows a low resolution $B-H$ color index map
of NGC 4622. The line of nodes is indicated by the dashed black
line. In spite of the low resolution, we can still detect
a dust silhouette on the east side, to the left of the bright
foreground star. However, if all we had was this map, we would 
not be able to use the reddening asymmetry method
with confidence. High resolution was clearly needed to reveal
the thin dust lanes on the near side.

A consequence of clockwise rotation of the disk is that not only are
the two outer arms leading, but most of the thin arcs of dust on
the east side would also be leading. Figure~\ref{vimap} shows one thin lane
aligned perpendicular to the line of nodes, connecting a large
dust patch near the line of nodes to a dust lane further out. This
is an interesting peculiarity of the dust distribution, but most of
the lanes simply arc like the spiral arms. Note that although the
Galactic extinction towards NGC 4622 is significant ($A_V \approx$
0.5 mag), the scale of the thin lanes and their curvature like the arms
rules out that they are Galactic in origin. This is verified also
using SkyView (McGlynn 2002) and IRAS 12-100$\mu$m
maps of the field around NGC 4622.

\section{The Feasibility of Detecting a Near Side/Far-Side Asymmetry
in a Low Inclination Galaxy}

It is reasonable to ask if we should really expect to see a near-side/far-side
asymmetry in a galaxy inclined as little as 20$^{\circ}$. First we show that
with modern digital images, the effect can be very obvious in a galaxy inclined as little as
40$^{\circ}$. The Ohio State University Bright Galaxy Survey (Eskridge et al.
2000) includes NGC 2775, one of the lowest inclination galaxies
considered by de Vaucouleurs (1958) for the dust silhouette method.
Even in direct blue light plates, NGC 2775 shows a clear arc-shaped
dust silhouette on its west side (Sandage 1961). Based on his analysis
of other galaxies, de Vaucouleurs suggested that the dust silhouette
method might be applicable to NGC 2775 and that the west side is its
near side. However, he had no rotation curve for NGC 2775 to judge
the sense of winding of its spiral arms. Figure~\ref{n2775} shows 
the OSU $B$-band image and a $B-H$ color index map of NGC 2775, the latter
being coded such that red features are light and blue features are dark.
The $B-H$ map shows a general reddening of the underlying light distribution
on the west side, and several dust arcs, including a complete dust ring
that is enhanced on the west side. Based on this image, whose pixel
resolution is only 1\arcs 5, the west side is obviously the near side.
From ellipse fits to the $B$ and $H$-band isophotes on these images,
we deduce that the inclination of NGC 2775 is only 40$^{\circ}$. 
Like NGC 4622, NGC 2775 is a bulge-dominated galaxy where the dust
silhouette method can be applied in spite of a fairly low inclination.

Does this result for NGC 2775 imply nevertheless that we should
see the effect if it were inclined only 20$^{\circ}$? At groundbased
resolution we expect the effect would be difficult to detect, but
the situation could be different at HST resolution. The best way
to evaluate this is with an HST-resolution model that incorporates the known
bulge and disk properties of NGC 4622, and which makes assumptions
about the dust layer in the galaxy and the likely vertical structure of
the galaxy. The details of the model are described in Appendix A;
here we only discuss the results. We examine the effect for four inclinations:
15$^{\circ}$, 20$^{\circ}$, 25$^{\circ}$, and 30$^{\circ}$. These should
bracket the actual inclination of NGC 4622. Scattering is ignored in
our analysis, but see Elmegreen and Block (1999) for models of
near-side/far-side reddening asymmetries that account for single
scattering, which can reduce the color asymmetry due to the general
background dust layer by 10-15\%. The model includes a general dust
layer and a series of four well-spaced, thin dust rings with widths
comparable to those of the lanes seen on the east side in Figure~\ref{vimap}.

Figure~\ref{modelcmaps} shows model $V-I$ color index maps for the
four inclinations, with a false color code chosen to enhance the
near side/far side reddening asymmetry. The models are oriented so
that the dust rings project into ellipses whose major axes coincide
with the line of nodes of the galaxy. For each model, the near
side corresponds to the left side (or east side on the sky). The
maps show that we can detect a near side/far side reddening asymmetry,
with the chosen optical depth parameters (see Appendix A), even for an inclination as
low as 15$^{\circ}$ for HST resolution. The effect is weak at
15$^{\circ}$ but very obvious at 30$^{\circ}$. 

One way to compare these models with the actual galaxy is with
surface brightness and color asymmetry curves. These do not supercede
Figures~\ref{vimap} and ~\ref{residmap}, but serve to quantify
the asymmetries along two axes. We assume initially that the galaxy
is symmetric and that any residual asymmetry and reddening across the
minor axis is due to our view through the dust layer. This assumption 
is valid at the 0.1 mag level only in the bulge-dominated region of NGC 4622, where isophotes
are mostly centered on the nucleus. However, a non-uniform dust layer as
well as intrinsic asymmetry can affect the asymmetry curves, and this
plays a role at larger radii in NGC 4622.

Figures~\ref{extinction} and ~\ref{reddening} show model surface brightness 
and color asymmetry plots across the kinematic minor axis for the four
inclinations. The differences are 

$$\Delta(V) = \mu_V(east\ side) - \mu_V(west\ side)$$
$$\Delta(V-I) = (\mu_V-\mu_I)(east\ side) - (\mu_V-\mu_I)(west\ side)$$

\noindent
The models were tilted so that the east side would be the near side.
To improve signal-to-noise ratio, we computed average surface
brightnesses along elliptical contours having major axis position
angle along the line of nodes, an axis ratio = $\cos i$, and within a
cone having a half angle of 5$^{\circ}$.
The plots show the clear effects of both the dust rings and the general
dust layer. The general dust layer causes $\Delta V$ and $\Delta(V-I)$
to be generally positive across the line of nodes. The amount of near side/far
side extinction and reddening asymmetry increases with increasing inclination. 
The four dust rings show obvious excess extinction and reddening even at 
$i$ = 15$^{\circ}$. Only the excess reddening due to the two outer dust lanes 
is lost in the noise at $i$ $<$ 30$^{\circ}$.

Figure~\ref{gal20}a shows observed surface brightness and color
asymmetry curves around the kinematic minor axis ($\phi$ = 112$^{\circ}$) 
for NGC 4622. 
Again, to improve signal-to-noise ratio, we computed average surface
brightnesses along elliptical contours having major axis position
angle along the line of nodes, an axis ratio = $\cos 20^{\circ}$, and within a
cone having a half angle of 10$^{\circ}$.
It was necessary to remove the obvious globular clusters from the
images before computing the asymmetry curves, since some of the clusters
are bright enough to perturb the profiles. 

Figure~\ref{gal20}a should be compared with the model asymmetry curves 
for $i$=20$^{\circ}$ in Figures~\ref{extinction} and ~\ref{reddening}. 
In the bulge-dominated region where the isophotes are reasonably well-centered
on the optical nucleus, the extinction and color asymmetries are positive,
as expected if they are due mainly to tilt and the east side is the near side. 
There is a slight excess
of extinction and reddening between 8$^{\prime\prime}$ and 12$^{\prime\prime}$
radius. Beyond 12$^{\prime\prime}$, there are more areas of reddening
and extinction, but also areas of negative color and surface brightness
differences. The negative $\Delta V$ values between 12$^{\prime\prime}$
and 16$^{\prime\prime}$ are due to the inner single arm. At radii approaching 
the optical ring and beyond, the intrinsic
asymmetries in the galaxy's spiral structure impact the asymmetry plots
considerably,
not just near side/far side effects. Nevertheless, even to a radius of
26$^{\prime\prime}$, there is a general tendency for the galaxy to
be redder on the east side. 

These results support our conclusion that the east side of NGC 4622 is the
near side. However, asymmetry curves taken along the  
the kinematic {\it major axis} of NGC 4622 show surface brightness
and reddening asymmetries similar to those seen along the minor axis. 
(No such differences are expected in our symmetric galaxy model.)
Figure~\ref{gal20}b shows the difference between the north and south halves of 
the major axis. Within $r$ = 15$^{\prime\prime}$, there is a slight tendency
for the south half to be redder than the north half, although there is a
more even distribution of positive and negative values of $\Delta(V-I)$
in the bulge-dominated region. The negative values of $\Delta(V-I)$
around $r$ = 9$^{\prime\prime}$ along the major axis are due to
a large obvious dust patch south of the center seen in Figure~\ref{vimap}. 
Although the effects we see are small,
there does appear to be less average reddening from $r$=1$^{\prime\prime}$ to 
5$^{\prime\prime}$ along the major axis compared to the minor axis. Thus, 
the complications along the major axis do not negate our conclusion that
the east side is the near side. Again, we emphasize that these asymmetry
curves do not supercede Figures~\ref{vimap} and ~\ref{residmap}, but 
merely quantify some of the differences seen.

\section{The Globular Cluster System}

We have noted that the spiral arms of NGC 4622 are lined by young 
associations, but in addition to these nearly star-like objects,
the inner region of NGC 4622 is covered by a flurry of other faint 
star-like objects that are very likely to be globular clusters (see
Figure~\ref{onlyclusters}).
This represents the first detection of a rich system
of globular clusters in NGC 4622. In this section, we wish to determine
some of the properties of this system with the main goal being to verify
the distance we are assuming for the galaxy. The associations are discussed
in the next section.

The clusters were first isolated using IRAF routine DAOFIND. 
In order to reliably detect the clusters superposed within the
bright bulge area, the $B$, $V$, and $I$ images were 
flattened by subtracting off the bulge and disk
models of BCB. Since DAOFIND often misses some
sources and finds some spurious sources, we corrected the output catalog
by visual inspection of the images. The final catalogue of sources
included a mix of foreground stars, stellar associations, and globular
clusters. 

Photometry of the sources was performed with IRAF routine PHOT using an
aperture radius of 2 pixels. This allowed us to use the formulae for 
charge transfer efficiency (CTE) 
corrections given by Whitmore, Heyer, and Casertano (1999). Aperture
corrections to 0\rlap{.}$^{\prime\prime}$5 radius were derived from 
foreground stars on the images since the clusters themselves are not
resolved. For WF2-4, values of 0.19, 0.20, 0.20, and 0.24 mag were derived 
for filters F336W, F439W, F555W, and F814W, respectively. For the PC1,
values of 0.42 0.42, 0.42, and 0.58 mag were derived for the same filters.
For Galactic extinction, we use the Schlegel et al. (1998) value
of $E(B-V)$ = 0.139 mag and use Tables 12a and 12b of Holtzman et al.
(1995) to infer corrections
to the natural WFPC2 systems. Since the Galactic reddening is fairly
low, we averaged the implied extinctions in Tables 12a and 12b and
used AF336W=0.658, AF439W=0.578, AF555W=0.441, and AF814W=0.264 mag.
After an additional correction for decontamination, the corrected
natural magnitudes were transformed to the standard systems using
Table 7 and equation 8 of Holtzman et al. (1995).

Figure~\ref{cmdiagram} shows a color-magnitude diagram for all sources 
having $V_o$ $\leq$ 25.0. The absolute magnitude scale is based on a 
distance of 40.2 Mpc. The diagram shows several well-defined regions:

\noindent
(1) A concentration of points having $V_o$ $>$ 21.5 and $-$0.5 $<$
$(V-I)_o$ $<$ 0.7. These are mostly the young associations connected with
the spiral arms and inner ring.

\noindent
(2) A concentration of points having $V_o$ $>$ 22.0 and 0.7 $<$ $(V-I)_o$
$<$ 1.5. These are the sources likely to be globular clusters.

\noindent
(3) Points of any magnitude having $(V-I)_o$ $>$ 1.5, or points of any
color having $V_o$ $<$ 21.0, are likely to be foreground stars,
background galaxies, or heavily reddened sources within the galaxy.

The likely globular clusters are isolated within the box in Figure~\ref{cmdiagram}.
Figure~\ref{xyplot} shows how the objects in the box are distributed within the
WFPC2 field. To evaluate the contribution of field objects to the box, we
show in Figure~\ref{outer} the color-magnitude digram of objects lying more than
65$^{\prime\prime}$ (12.7 kpc) from the center. Even at these large radii, the
box shows a clear concentration of points that are not likely to be merely
field objects. It appears the foreground star contamination in the box is
low or insignificant. The galaxy clearly fills most of the WFPC2 field, and
some likely clusters are found at large radii.

Figure~\ref{colordistribution} shows the distribution of $V-I$ colors for the
250 objects in the box. The solid curve is a gaussian having $<(V-I)_o>$ = 1.04
and dispersion $\sigma$ = 0.19 mag. This is very typical of an old cluster system
in an early-type galaxy (Kundu \& Whitmore 2001). The mean color is consistent with a
mean metallicity of [Fe/H] = $-$0.98 (Kundu \& Whitmore 2001). However,
there may be some dependence of the mean cluster color on radius. If we 
restrict the analysis only to the 27 clusters within the effective radius, 
9\rlap{.}$^{\prime\prime}$38, of the bulge, we get $<(V-I)_o>$ = 1.08 
and a dispersion of 0.14 mag. The slightly redder mean color could indicate
that the metal-rich clusters in NGC 4622 are more centrally concentrated,
as is often seen (see, for example, Larsen, Forbes, \& Brodie 2001).
However, bimodality is not obvious in Figure~\ref{colordistribution},
although a more sophisticated statistical test (such as that described
by Kundu \& Whitmore 2001) might assess this possibility more objectively.

The luminosity function of globular clusters was derived by correcting the observed
luminosity function for both areal incompleteness (due to the limited field
of view of the WFPC2) and for detector incompleteness (due to the variable galaxy
background). The detector incompleteness was evaluated using artificial cluster
experiments. Since none of our photometric models fully accounts for the complex
light distribution, we used the $V$-band WFPC2 image {\it cleaned of all stellar
objects} for the test. The idea is that the cleaned $V$-band image provides the
best background for any artificial cluster tests. The cleaning of the image involved 
a combination of point spread function (psf) fitting and image editing. We created
files with artificial clusters using a psf based on several of the brighter clusters.
We chose 12 intervals of 0.25 mag each from $V_o$ = 22.0 to 25.0. Within each
interval, a random number generator was used to assign $V-I$ colors around a mean
of 1.04 with a dispersion of 0.2 mag (the standard deviation about the mean
color). Also, a random number generator was used to scatter artificial clusters
across most of the WFPC2 field. For each magnitude interval, $\approx$3000
artificial clusters were used to judge completeness as a function of radius
(which correlates approximately with background brightness) in intervals of
5$^{\prime\prime}$ to 65$^{\prime\prime}$ radius.

The artificial clusters were added to the cleaned $V$-band image using IRAF
routine ADDSTAR. IRAF routine DAOFIND was used to locate the added 
clusters, whose coordinates were then matched with the input catalogue. Figure~\ref{completeness}
shows the completeness curves for $V_o$ = 23.00-23.25 and 24.75-25.00. The
incompleteness is a function of both magnitude and background brightness,
especially in the inner few arcseconds, where few clusters are found owing to
the very high background brightness. To minimize the effect of photometric
uncertainties, the analysis was restricted to $V_o$ $\leq$ 25.00.

The areal correction affects only circular annuli whose radii exceed 43$^{\prime\prime}$.
Parts of these annuli are outside the WFPC2 field of view. The observed counts
as a function of magnitude were then corrected for both this areal factor and
the detector incompleteness functions for each magnitude interval. The corrected histogram
of cluster magnitudes is illustrated in Figure~\ref{gclf}, which shows that the
turnover is not reached at the adopted magnitude limit. A gaussian was fitted to
the corrected distribution by fixing the dispersion at $\sigma_{GC}$ = 1.3 mag,
characteristic of the globular cluster universal luminosity function (Kundu \& Whitmore
2001). The turnover magnitude and the number count at this magnitude were then derived
by nonlinear least squares. The turnover magnitude so derived is $<V_o>$ = 25.835$\pm$0.181.
If the intrinsic mean absolute magnitude is $M_V$ = $-$7.41 (Kundu \& Whitmore 2001),
this implies a distance of 44.6$\pm$3.7 Mpc, fully consistent with the Cen30 distance of
40.2$\pm$3.1 Mpc that we have been assuming.

The specific frequency of clusters (number per unit $M_V$ = $-$15 galaxy 
luminosity;
Harris and van den Bergh 1981) was estimated by integrating the gaussian fit
in Figure~\ref{gclf}, which implies the presence of 959 clusters (compared to the 
$\approx$250 brighter than $V_o$ = 25.0 within $r$ = 65$^{\prime\prime}$).
The total $V$-band magnitude estimated by BCB is $V_T$ = 12.41. After correction
for Galactic extinction, $K$-dimming, and inclination, the corrected $V$-band
total magnitude is 11.88. Allowing for the uncertainty in $<V_o>$, the specific
frequency is $S_N$ = 3.4$\pm$ 0.6. This value is comparable, within 
uncertainties, to those for several
large Virgo Es (Kissler-Patig et al. 1997) and to the Sombrero Galaxy M104
(Harris, Harris, \& Harris 1984; Bridges \& Hanes 1992). The referee
suggests that this argues against any very recent major merger events or 
disruption. However, it does not necessarily rule out minor events
involving small galaxies.

The completeness-corrected surface number density distribution of globular clusters
is shown in Figures~\ref{surfdens}a and b. The surface density is plotted
versus $r^{1\over 4}$ and versus $logR$, where $R$ is in kpc based on a distance
of 40.2 Mpc. Figure~\ref{surfdens}a shows that the surface densities approximately
follow an $r^{1\over 4}$ law (solid line), but with a considerably larger
effective radius than the underlying galaxy light distribution (dotted line).
The solid curve in Figure~\ref{surfdens}b shows a King model fit to the surface
densities assuming a tidal radius $r_t$ = 50 kpc. The core radius of the distribution
is 3.1 kpc.

The larger effective radius for the globular clusters as opposed to the
bulge light for NGC 4622 differs from the Sombrero Galaxy, NGC 4594, whose 
globular cluster
surface density distribution declines in a manner similar to the bulge light
(Harris, Harris, \& Harris 1984). Larsen, Forbes, \& Brodie (2001) suggest
that the clusters in that case are mainly associated with the bulge. The lack 
of agreement between effective radii for NGC 4622 suggests that the clusters
are not simply associated with the bulge.

Finally, we ask whether the globular cluster system supports our interpretation
of the east side being the near side of the galaxy. If the clusters are distributed
more spherically than the disk, we should see a near side/far side extinction
and reddening asymmetry similar to what we see in the light distribution. Figure~\ref{eastwest}
shows histograms of the number of clusters as a function of radius, with the solid
histogram including only clusters east of the kinematic line of nodes and the dashed 
histogram including only clusters west of the line of nodes. The plot shows that
the east side has more clusters in the 0$^{\prime\prime}$-5$^{\prime\prime}$ bin
(see also Figure~\ref{onlyclusters}),
while farther out there tends to be more clusters west of the line of nodes.
In the 10$^{\prime\prime}$-15$^{\prime\prime}$ bin, there are few clusters
in the obvious dust lanes on the east side. For 46 east side clusters within a radius
$r$=20$^{\prime\prime}$, the mean $V-I$ color index is 1.077$\pm$0.022, while
for 47 clusters west of the line of nodes within the same radius, the mean $V-I$ color
is 1.057$\pm$0.022. Thus, only a small and insignificant reddening difference
is found between near side and far side clusters.

In summary, we find a rich globular cluster system in NGC 4622 with fairly normal
properties. We do not detect obvious bimodality in its color distribution, which
would imply that no new globular clusters have formed in the last few billion
years.

\section{Star Formation in NGC 4622}

Most of the objects in the region 21.5 $\leq$ $V_o$ $\leq$ 25.0,
$-$0.5 $\leq$ $(V-I)_o$ $\leq$ 0.7 in Figure~\ref{cmdiagram}
are likely to be young stellar associations. This is verified
in the two-color plots in Figure ~\ref{twocolor}. The errors increase
rapidly with increasing magnitude, thus Figure~\ref{twocolor} shows the
plots for two different intervals: 21.5 $\leq$ $V_0$ $\leq$ 23.0 and
23.0 $\leq$ $V_0$ $\leq$ 24.0. Some 
of the excess scatter in these plots is due to the limitations of 
using aperture photometry for the measurements of such associations,
which may have multiple components. Based on comparisons with the 
solar metallicity, Salpeter IMF
models of Bruzual \& Charlot (1996; solid curves in Figure~\ref{twocolor}),
most of the brighter associations are probably less than 10Myr old.
The two reddest objects in the left panels of Figure~\ref{twocolor}
could be affected by a higher than average reddening compared to most of the
other points (see reddening lines in the right panels, for example).
The characteristics of the star formation in NGC 4622 can be summarized
as follows:

\noindent
1. The WFPC2 $U$-band and groundbased H$\alpha$ 
(Figure~\ref{linecont}) images reveal considerable asymmetry in the 
distribution of young associations. These associations are numerous
along the east outer arm and along the northern section of
the west outer arm.

\noindent
2. There is a very thin ridge of star formation along the concave edge of the 
north section of the east outer arm (see Figure~\ref{starform}, feature labeled 1), 
extending from near the line of nodes (nearly vertical) to about due 
north of the nucleus. This ridge looks almost detached from the stellar 
background part of the arm in this region.

\noindent
3. Moving further counterclockwise along this arm, the associations are less offset
from the concave edge, so that by the south part of that arm, some of the associations are
actually on the convex side (Figure~\ref{colorimage}).

\noindent
4. Beginning at about the line of nodes, and going counterclockwise, the star formation
in the west outer arm lies near the convex edge of the arm, completely the opposite
of the east outer arm in the same general area (see Figure~\ref{starform}, feature labeled 2). Further clockwise along this arm,
associations are few and tend to lie either within the arm or near its concave edge (Figure~\ref{colorimage}).

\noindent
5. The star formation within the inner ring is complex. There are virtually no 
blue associations along the northeast half of the ring (region 3 in 
Figure~\ref{starform}). This is also the half that is riddled with dust lanes 
on the proposed near side.

The reddening on the east side of the inner ring thus can partly
explain the one-fold variation in color around the ring found
by BCB (their Figure 11d). However, the asymmetry is not entirely due to 
dust because the part of the ring due north of the nucleus shows little
evidence for dust and no bright young associations (Region 4 of 
Figure~\ref{starform}). Figure~\ref{colorpa}
shows color index versus position angle around the ring based on
the WFPC2 data. These curves were derived by first making a visual
mapping of the ring on the $V$-band WFPC2 image, fitting an ellipse
to its shape, and then circularizing the ring using IRAF routine
IMLINTRAN. The ellipse fit gave a center displaced by $\Delta x$ =
1\arcs 33 \ and $\Delta y$ = $-$2\arcs 14 (in WFPC2 frame), and an axis ratio of 0.91.
On the circularized ring, a series of circular apertures 3\arcs 0 in
diameter was placed around the ring in 1$^{\circ}$ steps. Figure~\ref{colorpa}
shows the approximately sinusoidal variation in color that distinguishes
the inner ring from features seen in other galaxies (Buta \& Combes 1996;
Buta \& Purcell 1998). The youngest (bluest)
associations that appear to be part of the inner single arm are
on a southern portion of the inner ring. These coincide with
bright HII regions in Figure~\ref{linecont}.

\section{Discussion}

\subsection{The Viability of Strong Leading Arms in NGC 4622}

The possibility that two very strong, obvious outer spiral arms in a 
fairly normal-looking galaxy are leading is clearly unprecedented
and questionable even on fairly simple grounds. In their discussion
of density wave theory, Binney and Tremaine (1987) argue that leading
arms are not likely to be seen because they would quickly unwind
and become trailing arms. Indeed, the swing amplifier mechanism
of Toomre (1981) depends on this change, which is accompanied
by a significant amplification of the resulting trailing arms.

The method we have used to deduce the sense of rotation of the 
disk of NGC 4622 is the same one that was used for the generally
accepted conclusion, noted by Binney and Tremaine, that all
spiral arms trail. This conclusion is largely based on the
papers of Hubble (1943) and de Vaucouleurs (1958), who used
samples of one to two dozen galaxies having rotation
data and high quality photographic plates. The assumptions in
determining the near side from an absorption asymmetry are
that the dust layer is confined to a thinner plane than the
disk, and that the bulge is a more three-dimensional
component than is the disk. The galaxies de Vaucouleurs
and Hubble used were highly inclined, and most were
not exceptional examples of spiral galaxies. (In fact, three
of de Vaucouleurs' galaxies are now recognized as flocculent
spirals.) Although high
inclination makes it difficult to see the spiral arms
of a galaxy, we have shown that low inclination does not
preclude using the near-side absorption technique
when the bulge is especially significant and the image
resolution is high. Thus, we believe there is no ambiguity
in the case of NGC 4622. The data favor the east side as
the near side, and the outer arms are leading.

An important question to ask is whether any other aspect of the
spiral structure of NGC 4622 supports this conclusion. For example,
for trailing density waves lying inside corotation, a
star or gas cloud first encounters the wave on its concave side.
A shock would lead to an asymmetric azimuthal profile across
the arm beginning with a sharp peak followed by a gradual 
decline in the direction of rotation (Schweizer 1976). The presence
of sharp dust lanes on the concave sides of spiral arms is often
taken to be the shocks expected, with the young stars appearing further
into the arms. Galaxies like M51, M101, and M83 show this effect
fairly well (Sandage 1961).

In NGC 4622, the east outer arm shows a strong ridge of star
formation on its concave northern edge that is qualitatively similar to
this kind of expectation (feature labeled "1" in Figure~\ref{starform}). 
What appears to be lacking is a clear
concave dust lane; in fact neither of the two outer arms shows
such lanes, which tend to be more evident in late-type spirals.
If the east outer arm lies within corotation, then one could
argue that it would have to be trailing in order for the star
formation ridge to be located on its concave edge. In contrast,
the main star formation connected with the west outer arm (feature
labeled "2" in Figure~\ref{starform})
lies on its outer edge. If it is a trailing arm, then it 
would have to lie outside corotation to show such an effect. 
Since this is contradictory (the outermost portion of the east
arm cannot be inside corotation while the innermost portion of
the west arm, at a smaller radius, is outside), we would have
to conclude that the arms are not trailing. 
The contradiction would be removed if the two outer arms are leading density
waves, because then corotation could be between features "1" and "2"
in Figure~\ref{starform}. This is discussed further in section 10.4.

Toomre (2002) argues that the character of 
the dust lanes in the east section of the inner ring favors that 
the galaxy is rotating counterclockwise. These dust lanes include short 
spurs that are tipped in the same sense as the outer spiral arms, 
i.e., open outward in a clockwise direction, and as we have already noted,
the color index map also shows that most of the small dust features, with one 
likely exception, would be leading if the galaxy is rotating clockwise.
According to Toomre, the short spurs ``look awfully much like other relatively
small-scale results of swing amplification of gas plus dust inhomogeneities
due not only to their own self-gravity but also some of the passing
stars in this galactic shear flow, much like the spiral wakes of Julian
and Toomre (1966), and any such rudimentary features simply have to be
trailing themselves, regardless of any minor merging or interplay that
probably did confuse this galaxy as a whole.'' However, this interpretation,
as straightforward and logical as it may seem, leaves two issues unexplained:
why does the apparent dust distribution know the line of nodes, and
why is the inner arm allowed to open in the opposite sense?

The single inner arm is not likely to be a leading component caught before
it swings and amplifies. Indeed, in the swing amplifier theory of Toomre
(1981), it is a two-armed leading spiral that swings into a much stronger 
two-armed trailing spiral, due to a conspiracy between shear, epicyclic motions,
and self-gravity of the arms. In NGC 4622, the maximum relative amplitudes 
of the $m$ = 1 inner spiral and the $m$ = 2 outer
spiral differ only by a factor of about 2 in $I$ (see Figure~\ref{mplots}, by
comparing the $m$=1 maximum at $r$ = 19$^{\prime\prime}$ with the 
$m$=2 maximum at $r$ = 36$^{\prime\prime}$).
One could therefore ask if it is theoretically more reasonable that
the inner single arm has the leading sense. The feature is weaker and has
little gas except where it overlaps the inner part of the west outer arm; 
it therefore might be less troublesome to be leading than the 
two strong outer spiral arms. There is no question of the
reality of the inner arm  as a genuine spiral feature in NGC 4622. Although
single-armed leading spirals have been shown to result from retrograde
tidal encounters (e.g., Athanassoula 1978; Byrd, Freeman, and Howard
1993; BCB and other references therein), it does not follow that all observed
single arms must be leading (see, e.g., Byrd, Freeman, \& Howard 1994).
For example, the dust silhouette technique combined with a published
H$\alpha$ rotation curve indicates that the single spiral arm seen
in NGC 4378 is trailing (Byrd et al. 1997). Thus, it is just as viable
that the inner arm is trailing, rather than leading.

Elmegreen and Block (1999) have discussed more sophisticated
models of dust that include the effects of scattering. They
note that when the inclination is high enough, the color change
across the line of nodes along the minor axis can be almost
step-like in $V-K$. They also argue that the reddening asymmetry
seen in many galaxies is not due to a preferential dust screen
on one side, but that dust is always present on both sides of
the line of nodes. The fact that we can see such dust on and
across to the far side of the line of nodes is a strong argument
that the observed reddening asymmetry in NGC 4622 is due to tilt.

\subsection{The Influence of a Minor Merger}

As noted by the referee, there is surprisingly little evidence for
disruption in NGC 4622, except for the odd winding of its three
spiral arms. We nevertheless suspect that a minor merger (in
the form of a plunging encounter that left some debris in the center)
could be responsible 
for the unusual structure seen in NGC 4622. This idea is developed more fully
in Paper II. Here we note that several characteristics support
the idea that NGC 4622 has suffered at least one recent minor merger.
The first characteristic is that the galaxy is slightly 
lopsided, with an $m$=1 relative Fourier amplitude of about
25\% just inside the standard isophotal radius (see Figure 9 of BCB).
Rix and Zaritsky (1995) have shown that 30\% of field
spirals in a magnitude-limited sample have significant $m$=1
components in their stellar light distributions. Zaritsky
and Rix (1997) have suggested that these components are largely
caused by tidal interactions and they estimate the merger
rate of small companions. However, as noted by the referee,
the lopsidedness alone may not
be conclusive for NGC 4622 because Jog (2002) has shown that a lopsided 
halo potential can produce similar effects.

A more convincing detail that
suggests a minor merger occurred in NGC 4622 is its short
central dust lane. Unusual dust lanes are often characteristic
of mergers, such as the dust lanes in Centaurus A (Sandage
1961), NGC 1316 (Schweizer 1980), and the "Blackeye Galaxy"
NGC 4826 (Braun et al. 1994; discussed further in the next 
section). The slight change in position
angle of NGC 4622's central dust lane suggests that
the feature is a warped accretion disk. A consequence of
the plunging encounter that may have led to this feature is
that the nucleus of NGC 4622 could have engaged in an
oscillation that might have helped to trigger the galaxy's
lopsidedness (see Miller \& Smith 1992, 1994). In this
regard, B. G. Elmegreen
(2002) has suggested that perhaps the inner parts of NGC 4622
are oscillating with respect to the outer parts. A possible
signature of an oscillating nucleus is a miscentering of
isophotes near the center (Miller \& Smith 1992). In NGC 4622, 
nuclear isophotes are fairly well-centered (see Figure 9a of BCB),
with significant asymmetry appearing only at the beginning
of the inner single arm.

We do not suggest that the particular merger that led to the 
central dust lane in NGC 4622 is necessarily the one that led to
the peculiar leading arms. For example, Byrd, Freeman, \& Howard
(1993) focussed on a small companion galaxy, a likely dwarf E,
lying 1\rlap{.}$^{\prime}$85 east (see Figure 3 of BCB) as a 
possible perturber as well. Nevertheless, the central dust lane
is a very unusual feature to see in a nearly face-on spiral galaxy.
The mass of dust involved could help us to determine how significant
the companion that plunged might have been relative to the mass
of NGC 4622 itself.

\subsection{Other Examples of "Two-Way" Spiral Patterns}

It is important to emphasize that NGC 4622 is not unique.
Other galaxies are known that show features
in common with NGC 4622's spiral pattern. Here we discuss
two other cases having inner single arms and two or more outer
arms winding in opposite senses. We also discuss an unusual
case of a galaxy having two inner arms and two or more
outer arms winding in opposite senses.

\subsubsection{The "Blackeye Galaxy" NGC 4826 (M64)}

The idea that a single inner arm trails and two outer arms lead
is not unprecedented. Walterbos, Braun, \& Kennicutt (1994) discuss
the spiral structure of the "Blackeye" Galaxy NGC 4826 (M64),
a well-known case with counterrotating atomic and ionized gas disks
(Braun et al. 1994). Van Driel \& Buta (1993) had noted a similarity
between NGC 4826 and NGC 4622, in that the inner spiral structure
looks single-armed and winds in the opposite sense from two outer
arms. Walterbos, Braun, \& Kennicutt (1994) used the dust silhouette
method in conjunction with H$\alpha$ and stellar rotation curves
to deduce that the inner spiral structure trails while the two outer arms
lead in NGC 4826. This assertion led to little controversy because
the features seen in NGC 4826 are relatively weak compared to those seen
in NGC 4622. Nevertheless, it would still be difficult to account even
for two weak outer leading arms in conventional density wave
theory. NGC 4826 is more abnormal-looking than NGC 4622 in that
it has much more significant near-side absorption that
gives it its name. It is very likely that an interaction has 
caused the peculiar structure of NGC 4826, because the galaxy shows
counterrotation of its gas disks. The kinematics of atomic
hydrogen in NGC 4622 have not yet been mapped, but HI observations are
scheduled for late 2002.

\subsubsection{A CSRG Galaxy}

In the Catalogue of Southern Ringed Galaxies (CSRG, Buta 1995), the
spiral galaxy ESO 297$-$27, type SA($\underline{\rm r}$s)b, is noted
to have possible two-way spiral structure, involving a single inner
arm winding in the opposite sense to at least two outer spiral arms.
This deduction, based on a small-scale image from the SRC-J southern
sky survey, was not conclusive. However, recently Grouchy and Buta
(2002) imaged the galaxy with a CCD and have been able to verify the
oppositely winding patterns (see Figure~\ref{eso297d27}). The 
observations were made with the CTIO 1.5-m telescope in 2002 August,
and images were obtained with $B$, $V$, $I$, and H$\alpha$ filters.
The galaxy differs from NGC 4622 in that it has a later Hubble type,
is inclined about 55$^{\circ}$, has a disk filled with diffuse
ionized gas, and the relative amplitudes of the spiral arms are
such that the inner arm is stronger than the outer arms. We also 
can identify three outer arms. Like NGC 4622, at least one set of
arms in ESO 297$-$27 must be leading, but unlike NGC 4622, the 
more favorable geometry means that a more conclusive test can be made
as to which arms lead. More information and followup data on this
intriguing system will be provided in Grouchy and Buta (2002).

\subsubsection{The Ringed Spiral Galaxy NGC 3124}

This nearly face-on barred spiral galaxy, type SAB(rs)bc (de Vaucouleurs
et al. 1991), shows an oppositely winding spiral pattern in a red continuum
image obtained by Purcell (1998). Buta (1999) present an $m$ = 2
Fourier image in the $I$-band that shows that the apparent bar of
NGC 3124 can be thought of as a very open two-armed spiral pattern winding
in the opposite sense to a more tightly wound multi-armed outer spiral 
pattern. Like NGC 4622, an inner ring marks the boundary between the
domains of the opposing spirals. The low inclination of NGC 3124, and
its relatively small bulge, have not yet allowed a conclusive determination
of which arms lead. Nevertheless, NGC 3124 provides an intriguing
case of a two-way spiral that, like NGC 4622, seems to involve a
ring as the boundary between the domains of the two opposing sets 
of arms.

\subsection{The Location of Resonances and the Inner Ring of NGC 4622}

The nature of rings in nonbarred galaxies is in general poorly understood. 
Nonbarred galaxies are less likely to have rings than barred galaxies (e.g., 
Buta 1995), but ringed nonbarred galaxies are nevertheless frequent enough
to be significant. Some nonbarred ringed galaxies are likely
to be "ex-barred" galaxies or galaxies with strong ovals (Buta 1999),
but cases like NGC 4622 clearly do not fit into this category.
In several cases (e.g., NGC 7742, de Zeeuw et al. 2002; NGC 4138,
Jore et al. 1996), the ring is counter-rotating with respect
to the main stellar disk.

The inner ring of NGC 4622 is a remarkably complex feature. BCB
noted the offset of the center of the ring from the nucleus of the
galaxy. The lack of any obvious bar-like distortion suggests that
this ring has an unusual origin compared to most rings seen in barred
or weakly-barred galaxies. Based on the study of Byrd, Freeman, 
\& Howard (1993), Buta (1999) suggested that the ring in NGC 4622
is a tidally-driven feature, rather than a bar-driven one. Byrd,
Freeman, \& Howard (1993) suggested that a plunging retrograde
encounter could account for both the mostly stellar inner single
arm and the two outer gaseous arms. However, in their model the 
two outer arms are trailing material arms. BCB suggested that
the ring is a +1:1 resonance with the angular velocity of a companion
at closest approach. However, if the two outer arms are leading,
the nature of the ring becomes more mysterious. Why does it delineate
the boundary between the two opposing senses of spiral structure?
If it is a resonance feature, which resonance might it be? 

Elmegreen, Elmegreen, \& Seiden (1989) have examined smooth spiral
arm amplitude variations in three grand-design spirals (M51, M81,
and M100), and have interpreted the minima in terms of the interference
between inward and outward propagating spiral density waves. When
they link prominent Fourier amplitude minima in each of these spirals 
to the inner
4:1 resonance, they find that the outer spirals terminate near the
outer Lindblad resonance. In Figure 9b of BCB, and also in Figure~\ref{mplots}b
in this paper, the relative $m$=2
Fourier amplitude in the outer spiral arms of NGC 4622 shows a minimum at
$r$ $\approx$ 40$^{\prime\prime}$. If this minimum corresponds to the
inner 4:1 resonance, then we can infer the locations of other
resonances. From a fit of a quadratic to the observed folded rotation
curve (Table 2), this resonance identification implies an inner
Lindblad resonance at $r$ = 21$^{\prime\prime}$, consistent with
the average radius of the inner ring. However, the rising rotation
curve leaves corotation indeterminate, although this resonance would
clearly be beyond the outer boundary of the outer spiral arms, if the
4:1 resonant interpretation of the minimum is correct. 

Another way to locate corotation in a grand design spiral was proposed
by Puerari and Dottori (1997; see also Vera-Villamizar et al. 2001). 
This method uses blue and near-infrared
images to look for a phase shift of star formation relative to a spiral
density wave that is expected to occur across the corotation resonance.
In NGC 4622, we see phase shifts of the type that they predict in the
boundary between the inner single arm and the two outer arms, and in
the middle of the outer arms. Figure~\ref{phaseshifts} shows enlargements
of the $m$ = 2 phases around these regions. Figure~\ref{phaseshifts}a
shows that a strong $B$ to $I$ phase shift occurs at 
$r$ = 21\rlap{.}$^{\prime\prime}$4. This is also where a large
phase shift occurs in the $m$ = 1 component in Figure~\ref{mplots}a.
When we plot this radius on a deprojected image of the galaxy
(see Figure~\ref{depV}), we find that it coincides with the zone where
the inner single arm meets the inner ring; the radius represents a natural
division between the inner and outer spirals in NGC 4622. 
In the outer arms (Figure~\ref{phaseshifts}b), a much smaller
phase shift occurs in the opposite sense at $r$ = 36\rlap{.}$^{\prime\prime}$4.
This radius is shown as the larger circle in Figure~\ref{depV}, and
it passes between regions 1 and 2 shown in Figure~\ref{starform}.
Comparing the plots with Figure 2 of Puerari \& Dottori (1997),
we find that the phase change in the outer arms, although very small,
is consistent with a two-armed leading pattern (as noted already in
section 10.1). However, the small size of the phase shift, as
well as the rising rotation curve, complicates this interpretation.
Nevertheless, on the northeast side, where the star formation differences
are most obvious, the observed rotation curve still favors a slightly
declining angular velocity, which might be enough to cause the observed
phase shift.

Referring again to Figure 2 of Puerari \& Dottori (1997), the much larger
phase change in Figure~\ref{phaseshifts}a is at first sight consistent with the
inner arm also being leading. However, the applicability of the
Puerari \& Dottori method to this zone is unclear, since their
method assumes the presence only of a single two-armed density wave winding
either clockwise or counterclockwise, not nested, oppositely winding
patterns. The two outer arms intersect the inner arm roughly
north and south of the nucleus, and since these arms are bluer
than the inner arm, they impact the Fourier phases in the
transition zone. 

These results do not really clarify the nature of the inner ring.
We believe this can best be answered only with a 
detailed dynamical model of the galaxy. In paper II, we will
interpret the phase change at $r$ = 
21\rlap{.}$^{\prime\prime}$4 as a corotation radius for a pair of
$m$=1 periodic orbits, in the form of two nested ellipses with maximum radii
180$^{\circ}$ apart. While paper II will create the arm structure
via a plunging encounter, the central regions will evolve toward the
survival for such periodic orbits.

\section{Conclusions}

Our main conclusions from this study are:

\noindent
1. The two strong outer spiral arms in NGC 4622 appear to have the
leading sense. This is based on the standard dust silhouette method 
in conjunction with radial velocities used to infer spiral arm winding 
sense in the much earlier studies of Hubble (1943) and de Vaucouleurs
(1958). In this paper, the principal evidence for this conclusion is based on
the observed velocity field in Figure~\ref{vfield}, 
the $V-I$ color index map in Figure~\ref{vimap},
and the residual intensity map in Figure~\ref{residmap}.

\noindent
2. Isophote shapes suggest an inclination of 26$^{\circ}$. However, the
H$\alpha$ kinematics as well as the 21-cm line profile width favor
a lower inclination, closer to 20$^{\circ}$.

\noindent
3. NGC 4622 has a significant bulge component that allows the dust
silhouette method to be used reliably even if the galaxy's
inclination is only about 20$^{\circ}$. The reddening asymmetry
across the line of nodes is so obvious that it almost argues for
the higher inclination to be correct. 

\noindent
4. There is clear reddening on both sides of the line of nodes, so
there is no need to assume that the reddening asymmetry seen in the
galaxy is purely intrinsic and has nothing to do with tilt. We see the
expected effect that far-side dust lanes are muted compared to near-side
lanes.

\noindent
5. A model of the dust distribution in a tilted symmetric galaxy
was used to show the feasibility of detecting, at HST resolution,
a near side/far side
reddening asymmetry to an inclination as low as 15$^{\circ}$, given the
observed bulge and disk parameters of NGC 4622.
Observed asymmetry curves along the kinematic major and minor
axes show some of the expected effects, but are complicated by details
along each axis. 

6. The distribution of star formation in the outer arms weakly
supports the leading sense of the arms, based on the Puerari
\& Dottori (1997) method of locating corotation. The effect
found (a phase shift in the location of star-forming
regions relative to the background stellar arms) is small,
but is at least consistent with the dust asymmetry results.
The reliability of the method, however, is questionable in the
presence of a rising rotation curve. Nevertheless, 
the observed rotation curve on the northeast side
still favors a slightly declining angular velocity with radius,
which may be enough allow the phase shift to be seen.

\noindent
7. Toomre has argued independently that the small-scale aspects of
the dust lanes on the east side of the inner ring strongly favor
a counterclockwise rotation and trailing sense for the outer arms.
However, this does not explain why the reddening asymmetry appears
to know the line of nodes, nor does it account for the inner single
spiral arm.

\noindent
8. The ionized gas in NGC 4622 mainly follows the two outer stellar arms
and the southwest portion of the inner ring. Thus, it is unlikely
that these gaseous arms represent a separate system counterrotating
relative to the stars, as is sometimes observed in other galaxies.
The observed kinematics of the HII regions is such that the rotation 
curve is rising across the outer arms. This rise is not an artifact of 
streaming motions. In contrast to the ionized gas, the near-infrared light
distribution predicts, under the assumption of a constant mass-to-light
ratio, a normal relatively flat rotation curve. The implication is that
NGC 4622 may be dark matter dominated even in the inner parts of the
visible disk.

\noindent
9. NGC 4622 has a significant system of old globular clusters that extends
well beyond the main optical region. The specific frequency of this
system is 3.4$\pm$0.6 and its mean color is $<(V-I)_o>$ = 1.04 $\pm$ 0.19.
The inferred turnover in the globular cluster luminosity function confirms
the membership of the galaxy in the Centaurus Cluster.

\noindent
10. NGC 4622 is not unique as a "two-way" spiral. At least three other
examples are known. These include, NGC 4826, where two
weak outer arms have been suggested to be leading; ESO 297$-$27, which
shows a bright single inner arm winding opposite to three weaker outer
arms; and NGC 3124, an inner-ringed spiral where a 
bar-like feature can be interpreted as an open two-armed spiral winding
opposite to the outer arms. These cases suggest that leading spirals
may not be as rare or impossible as once thought. More examples
will no doubt be found as greater awareness of the phenomenon is
recognized.

Further studies of NGC 4622 are needed to better assess
exactly what is going on, particularly observations of
its stellar and atomic gas kinematics. If the galaxy has an extended
HI disk, this may provide information on possible warping.
The galaxy also needs to be checked
for counter-rotation in the inner regions, an aspect that may be
common among early-type disk galaxies. Haynes et al.
(2000) describe kinematic studies of four early-type galaxies
and find evidence in each case for a decoupling of the ionized gas and
stellar disks, sometimes throughout the disk or in the
central area. Minor mergers are likely to be responsible for
these unusual components. In NGC 4622, we have used the stellar light
distribution to assess the near side, and the ionized
gas to assess the kinematics. If the ionized gas disk
is coupled to the stellar disk morphologically, as
it appears to be in NGC 4622 (that is, the ionized clouds
for the most part follow the stellar arms), then this should be a
fair way to determine the sense of winding of the spiral
arms in NGC 4622. 

Table 3 provides a summary of all of the main parameters of NGC
4622 derived in this paper.



\acknowledgments

We thank M. Meyer and S. D. Ryder for getting us a 21-cm line profile
of NGC 4622 on short notice, and especially M. Meyer for deriving some
of the parameters of the profile. We also thank Z. Levay and the Hubble
Heritage Team for making the fine color image of NGC 4622 which we use
in Figure 1 of this paper. We acknowledge helpful comments from A.
Toomre, B. G. Elmegreen, and especially from an anonymous referee.
This work was supported by NASA/STScI
Grant GO 8707 to the University of Alabama, and NSF Grant 9802918
to Bevill State College in Fayette. This research made use
of the NASA/IPAC Extragalactic Database (NED), which is operated 
by the Jet Propulsion Laboratory, California Institute of Technology,
under contract with the National Aeronautics and Space Administration.
The Digitized Sky Surveys were produced at the Space Telescope Science
Institute under U. S. Government grant NAG W-2166.
Funding for the OSU Bright Spiral Galaxy Survey was provided 
by grants from The National Science Foundation (grants AST-9217716 
and AST-9617006), with additional funding by the Ohio State University. 


\appendix

\section{Model of the Near Side/Far Side Reddening Asymmetry
in a Low Inclination Galaxy}

For the disk component, we use a double exponential 
having volume luminosity density ($L_{\odot} pc^{-3}$)

$$E^{II}(R,z)=E^{II}_0 e^{-{R\over h_R}}\ e^{-{z\over h_z}}$$

\noindent
where $h_R$ is the radial scalelength and $h_z$ is the vertical
scaleheight. For the bulge component, we use the Young (1976) asymptotic 
volume density for the $r^{1\over 4}$ law for a spherical galaxy:

$$E^I(r)=E^I_0{e^{-{bj}}\over 2j^3}({\pi\over 8bj})^{1\over 2}$$

\noindent
where $j$=$(r/r_e)^{1\over 4}$ and $b$ is a constant, 7.669. Here $r_e$ is the
effective radius of the bulge in projection. The parameters $h_R$,
$r_e$, $E^I_0$, and $E^{II}_0$ are based on the bulge/disk decompositions.
The parameter $h_z$ is assumed to be 325pc, the same as for the Galaxy
(Gilmore and Reid 1983).

In our model galaxy, we assume there is a dust layer having vertical
half-thickness $h_d$ = 0.3$h_z$. There are two parts to this dust layer.
The extended part is a uniform layer having face-on optical depth $\tau_{0d}(V)$ =
0.3 and $\tau_{0d}(I)$ = 0.18. Within this layer there are four rings each having a width of
12 pix (=0\arcs 12) centered at radii of 1\arcs 0, 2\arcs 0, 4\arcs 0, and
6\arcs 0. Each ring is assumed to have $\tau_{0r}$ = 3$\tau_{0d}$.
(These optical depths are based roughly on the analysis of overlapping
galaxy pairs by White \& Keel 1992).
With these features, we integrate the bulge and disk light in three parts each:
the part in front of the dust layer, the part mixed with the dust
layer, and the part behind the dust layer. For a line of sight passing
through coordinates ($x_0$, $y_0$) in the disk plane, the cylindrical
radius for a point at $z$ relative to the plane is 
$R(z)=\sqrt{x_0^2+(y_0+z\tan i)^2}$, where $i$ is the assumed inclination.
Then the three contributions for the disk component are given by

$$I^{II}(x_0,y_0)=\int_{h_d}^{\inf} E^{II}(R(z),z)\sec i\ dz +
\int_{-h_d}^{h_d} E^{II}(R(z),z) \sec i\ e^{-{{\tau_{0}(h_d-z)\sec i\over 2h_d}}}dz +$$
$$\int_{-\inf}^{-h_d} E^{II}(R(z),z)\sec i\ e^{-{\tau_{0} \sec i}} dz$$

\noindent
For the bulge component surface brightness at ($x_0,y_0$), we use the radius
$r=\sqrt{x_0^2+(y_0+z\tan i)^2+z^2}$. Then the three contributions for the
bulge component are given by

$$I^{I}(x_0,y_0)=\int_{h_d}^{\inf} E^{I}(r(z))\sec i\ dz +
\int_{-h_d}^{h_d} E^{I}(r(z)) \sec i\ e^{-{{\tau_{0}(h_d-z)\sec i\over 2h_d}}}dz +$$
$$\int_{-\inf}^{-h_d} E^{I}(r(z))\sec i\ e^{-{\tau_{0} \sec i}} dz$$

\noindent
The final model involves summing the contributions for the bulge and disk
components in both $V$ and $I$, matching the resolution by smoothing each
model image with a gaussian whose full width at half maximum is consistent
with the stars on the original images, and then adding noise consistent 
with the observed images. 




\section{Discovery of a Supernova in the WFPC2 Images}

The WFPC2 images we have used in this paper, which were taken
from 2001 May 25.57- May25.91 (UT), reveal a likely supernova
located 91.9 arcsec slightly east of north of the nucleus (see 
Figure~\ref{sn2001jx}). The object is seen on all four filters used for 
our WFPC2 exposures. The coordinates of the object are
RA(J2000) = 12h 42m 40.04s, DEC(J2000) = $-$40$^{\circ}$ 43$^{\prime}$ 12\rlap{.}$^{\prime\prime}$54.
There is nothing at this position in the field of the galaxy seen
on the Digitized Sky Survey. At the adopted distance of 40.2 Mpc,
this position projects 18 kpc from the nucleus, well outside the
visible disk. Since there is no evidence of any massive star
formation in this region, it is possible that the object is a
Type Ia supernova. Its location so far from the center prevented it
from being noticed when the images were first obtained. The object
has been designated SN 2001jx by the IAU Central Bureau of Astronomical
Telegrams (IAU Circ. 7833).

The image of the possible supernova is saturated on all but the
F336W image, which gives U=17.5 (May 25.6 UT). The apparent U-B color index
is +0.3 or greater. After correcting for Galactic
extinction, and referring to the U-band light curve of SN 1981b
in NGC 4536 (Buta and Turner 1983), we estimate
that our images were taken about 15-20 days past maximum light
if the object was a Type Ia supernova.

\clearpage

\centerline{REFERENCES}

\noindent
Athanassoula, E. 1978, \aap, 69, 395

\noindent
Binney, J. and Tremaine, S. 1987, Galactic Dynamics, Princeton, Princeton Univ. Press

\noindent
Braun, R., Walterbos, R. M., Kennicutt, R. C., \& Tacconi, L. J. 1994,
\apj, 420, 558

\noindent
Bridges, T. J. \& Hanes, D. A. 1992, \aj, 103, 800

\noindent
Bruzual, G. \& Charlot, S. 1996, AAS CD-ROM, Vol. 7

\noindent
Buta, R. 1995, \apjs, 96, 39 (CSRG)

\noindent
Buta, R. 1999, Astrophys. \& Space Sci., 269-270, 79

\noindent
Buta, R., Byrd, G. G., \& Freeman, T. 2002, IAU Circ. 7833

\noindent
Buta, R. \& Combes, F. 1996, Fundamentals of Cosmic Physics, 17, 95

\noindent
Buta, R., Crocker, D. A., and Byrd, G. G. 1992, \aj, 103, 1526 (BCB)

\noindent
Buta, R. \& Purcell, G. B. 1998, \aj, 115, 484 (BP98)

\noindent
Buta, R. \& Turner, A. 1983, \pasp, 95, 72

\noindent
Buta, R., Treuthardt, P. M., Byrd, G. G., \& Crocker, D. A. 2001, \aj, 120, 1289

\noindent
Byrd, G., Freeman, T., \& Buta, R. 2002, in preparation (paper II)

\noindent
Byrd, G., Freeman, T., \& Howard, S. 1993, Astron. J., 105, 477

\noindent
Byrd, G., Freeman, T., \& Howard, S. 1994, Astron. J., 108, 2078

\noindent
Byrd, G. G., Thomasson, M., Donner, K. J., Sundelius, B., Huang, T.-Y., \& Valtonen, M. J. 1989, Celest. Mech., 45, 31 

\noindent
Byrd, G. G., Purcell, G. B., Buta, R. J., McCormick, D., \& Freeman, T.
1997, in Star Formation Near and Far, S. S. Holt and L. G. Mundy, eds.,
AIP Conf. Ser. 393, p. 283

\noindent
Carollo, C. M., Stiavelli, M., \& Mack, J. 1998, \aj, 116, 68

\noindent
Carollo, C. M., Stiavelli, M., Seigar, M., de Zeeuw, P. T.,
\& Dejonghe, H. 2002, \aj, 123, 159

\noindent
de Vaucouleurs, G. 1958, \apj, 127, 487

\noindent
de Vaucouleurs, G., de Vaucouleurs, A., Corwin, H., Buta, R.,
Paturel, G., \& Fouqu\'e, P. 1991, Third Reference Catalogue
of Bright Galaxies, Springer-Verlag, New York

\noindent
de Zeeuw, P. T. et al. 2002, \mnras, 329, 513

\noindent
Elmegreen, B. G. 2002, private communication

\noindent
Elmegreen, B. G. \& Block, D. L. 1999, \mnras, 303, 133

\noindent
Elmegreen, B. G., Elmegreen, D. M., \& Seiden, P. E. 1989, \apj, 343, 602

\noindent
Eskridge, P. et al. 2000, \aj, 119, 536

\noindent
Faber, S. M. et al. 1997, \aj, 114, 1771

\noindent
Freeman, K. C. 1970, \apj, 160, 811

\noindent
Garcia, A. M. 1993, \aaps, 100, 47

\noindent
Gilmore, G. \& Reid, N. 1983, \mnras, 202, 1025

\noindent
Grillmair, C., Faber, S. M., Lauer, T. R., Hester, J. J., Lynds, C. R.,
O'Neill, E. J., \& Scowen, P. A. 1997, \aj, 113, 225

\noindent
Grouchy, R. \& Buta, R. 2002, in preparation

\noindent
Harris, W. E. \& van den Bergh, S. 1981, \aj, 86, 1627

\noindent
Harris, W. E., Harris, H. C., \& Harris, G. L. H. 1984, \aj, 89, 216

\noindent
Haynes, M. P., Jore, K. P., Barrett, E. A., Broeils, A. H., \&
Murray, B. M. 2000, \aj, 120, 703

\noindent
Holtzman, J. A., Burrows, C. J., Casertano, S., Hester, J. J.,
Trauger, J. T., Watson, A. M., \& Worthey, G. 1995, \pasp, 107, 
1065

\noindent
Hubble, E. 1943, \apj, 97, 112

\noindent
Jedrzejewski, R. 1987, \mnras, 226, 747

\noindent
Jog, C. J. 2002, \aap, 391, 471

\noindent
Jore, K. P., Broeils, A. H., \& Haynes, M. P. 1996, \aj, 112, 438

\noindent
Joyce, R. R. 1992, ASP Conference Series 23, S. B. Howell, ed.,
p. 258

\noindent
Julian, W. H. \& Toomre, A. 1966, \apj, 146, 810

\noindent
Kalnajs, A. \& Hughes, S. M. 1984, private communication

\noindent
Kent, S. M. 1986, \aj, 91, 1301

\noindent
Kent, S. M. 1988, \aj, 96, 514

\noindent
Kissler-Patig, M., Kohle, S., Hilker, M., Richtler, T., Infante, L.,
\& Quintana, H. 1997, \aap, 319, 470

\noindent
Kormendy, J. 1987, in Structure and Dynamics of Elliptical Galaxies,
T. de Zeeuw, ed., Dordrecht, Reidel, p. 17

\noindent
Kundu, A. \& Whitmore, B. C. 2001, \aj, 122, 1251

\noindent
Larsen, S. S., Forbes, D. A., \& Brodie, J. P. 2001, \mnras, 327, 1116

\noindent
Lauer, T. et al. 1995, \aj, 110, 2622

\noindent
Lucey, J. R., Currie, M. J., \& Dickens, R. J. 1986, \mnras, 221, 453

\noindent
McGlynn, T. A. 2002, SkyView: The Internet's Virtual Telescope,
http://skys.gsfc.nasa.gov/

\noindent
Miller, R. H. \& Smith, B. F. 1992, \apj, 393, 508

\noindent
Miller, R. H. \& Smith, B. F. 1994, Cel. Mech. \& Dyn. Astr., 59, 161

\noindent
Mollenhoff, C. \& Heidt, J. 2001, \aap, 368, 16

\noindent
Puerari, I. \& Dottori, H. 1997, \apj, 476, L73

\noindent
Purcell, G. B. 1998, PhD Thesis, Univ. of Alabama

\noindent
Quillen, A. C., Frogel, J. A., and Gonz\'alez, R. A. 1994, \apj, 437,
162

\noindent
Rix, H.-W. \& Zaritsky, D. 1995, \apj, 447, 82

\noindent
Rogstad, D. H. 1971, \aap, 13, 108

\noindent
Rubin, V. C., Burstein, D., Ford, W. K., \& Thonnard, N. 1985, \apj, 289,
81

\noindent
Sandage, A. 1961, The Hubble Atlas
of Galaxies, Carnegie Institution of Washington Publication No. 618

\noindent
Schlegel, D. J., Finkbeiner, D. P., \& Davis, M. 1998, \apj, 500, 525

\noindent
Schommer, R. A., Bothun, G. D., Williams, T. B., \& Mould, J. R.
1993, \aj, 105, 97

\noindent
Schweizer, F. 1976, \apjs, 31, 313

\noindent
Schweizer, F. 1980, \apj, 237, 303

\noindent
Scott, J. S. 1995, Master's Thesis, Univ. of Alabama 

\noindent
S\'ersic, J. 1968, Atlas de Galaxias Australes, Observatorio Astron\'omica de C\'ordoba

\noindent
Simien, F. \& de Vaucouleurs, G. 1986, \apj, 302, 564

\noindent
Tonry, J. L., Dressler, A., Blakeslee, J. P., Ajhar, E. A.,
Fletcher. A. B., Luppino, G. A., Metzger, M. R., \& Moore, C. B.
2001, \apj, 546, 681

\noindent
Toomre, A. 1981, in The Structure and Evolution of Normal Galaxies, S. M. Fall \& D. Lynden-Bell, eds., Cambridge Univ. Press, Cambridge, p. 111

\noindent
Toomre, A. 2002, private communication

\noindent
Tully, R. B. \& Pierce, M. J. 2000, \apj, 533, 744

\noindent
Vera-Villimizar, N., Dottori, H., Puerari, I., \& de Carvalho, R. 2001,
\apj, 547, 187

\noindent
Walterbos, R. A. M., Braun, R., \& Kennicutt, R. C. 1994, \aj, 107, 184

\noindent
Warner, P. J., Wright, M. C. H., \& Baldwin, J. E. 1973, \mnras, 163,
163

\noindent
White, R. E. \& Keel, W. C. 1992, Nature, 359, 129

\noindent
Whitmore, B., Heyer, I., \& Casertano, S. 1999, \pasp, 111, 1559

\noindent
Worthey, G. 1994, \apjs, 95, 107

\noindent
van Driel, W. \& Buta, R. 1993, PASJ, 45, 47

\noindent
Young, P. J. 1976, \aj, 81, 807

\noindent
Zaritsky, D. \& Rix, H.-W. 1997, \apj, 477, 118

\clearpage

\begin{deluxetable}{lcccc}
\tabletypesize{\scriptsize}
\tablewidth{0pc}
\tablecaption{Bulge/Disk Solutions for NGC 4622}
\tablehead{
\colhead{Profile} & 
\colhead{$a^I$} &
\colhead{$b^I$} &
\colhead{$a^{II}$} &
\colhead{$b^{II}$} 
}
\startdata
major axis $B$ & 14.3375 & 4.7239 & 22.3846 & 0.06963 \\
major axis $I$ & 11.9377 & 4.6395 & 19.9959 & 0.07070 \\
major axis $H$ &  9.6429 & (4.7585) & 18.1477 & 0.07330 \\
minor axis $B$ & 14.4436 & 4.7261 & 22.2001 & 0.07837 \\
minor axis $I$ & 11.5837 & 4.9445 & 19.7284 & 0.07808 \\
minor axis $H$ &  9.7237 & (4.7585) & 17.9286 & 0.07945 \\
\enddata
\end{deluxetable}

\clearpage

\begin{deluxetable}{ccrcc}
\tabletypesize{\scriptsize}
\tablewidth{0pc}
\tablecaption{Folded Mean Rotation Curve of NGC 4622 for $i$=19$^{\circ}$}
\tablehead{
\colhead{radius} & 
\colhead{$V_c$} &
\colhead{Number of} &
\colhead{$\sigma$} &
\colhead{mean error} \\
\colhead{(arcsec)} &
\colhead{km s$^{-1}$} &
\colhead{velocity points} &
\colhead{km s$^{-1}$} &
\colhead{km s$^{-1}$} 
}
\startdata
      17.5  &      131.4  &     117   &     24.6    &    2.55     \\
      22.5  &      131.6  &    2574   &     38.0    &    0.85     \\ 
      27.5  &      156.2  &    3895   &     39.3    &    0.69     \\
      32.5  &      152.0  &    4296   &     40.3    &    0.68     \\ 
      37.5  &      171.1  &    3994   &     29.5    &    0.51     \\ 
      42.5  &      183.6  &    2852   &     30.1    &    0.62     \\ 
      47.5  &      232.5  &    1462   &     24.4    &    0.70     \\ 
      52.5  &      299.8  &      78   &     24.9    &    2.94     \\ 
\enddata
\end{deluxetable}

\begin{deluxetable}{ll}
\tabletypesize{\scriptsize}
\tablewidth{0pc}
\tablecaption{Summary of Derived Parameters of NGC 4622}
\tablehead{
\colhead{Parameter} & 
\colhead{Value} 
}
\startdata
Distance assumed & 40.2$\pm$3.1 Mpc (Tully \& Pierce 2000) \\
GCLF distance & 44.6$\pm$3.7 Mpc (this paper) \\
Photometric inclination & 26$^{\circ}$$\pm$4$^{\circ}$ \\ 
Optical systemic velocity & 4502$\pm$ 3 km s$^{-1}$ \\
Kinematic inclination\tablenotemark{a} & 19\rlap{.}$^{\circ}$3$\pm$1\rlap{.}$^{\circ}$7 \\
Bulge effective radius & 9\rlap{.}$^{\prime\prime}$38$\pm$0\rlap{.}$^{\prime\prime}$56 \\
Disk effective radius & 26\rlap{.}$^{\prime\prime}$0$\pm$0\rlap{.}$^{\prime\prime}$2 \\
$B(0)_c$ (Freeman 1970) & 21.81 mag arcsec$^{-2}$ \\
Total magnitude $B_T$ & 13.44 \\
Total magnitude $I_T$ & 11.09 \\
Total magnitude $H_T$ & 9.16 \\
Absolute magnitude $M_B^{b,k,i}$ & $-$20.3 \\
Relative bulge luminosity $k_I(B)$ & 0.39 \\
Relative bulge luminosity $k_I(I)$ & 0.51 \\
Relative bulge luminosity $k_I(H)$ & 0.58 \\
GC specific frequency & 3.4$\pm$0.6 \\
GC $<(V-I)_o>$ & 1.04$\pm$0.19 \\
\enddata
\tablenotetext{a}{Based on observed 21-cm line width and assumed distance
of 40.2 Mpc (see text)}
\end{deluxetable}

\clearpage


\centerline{\bf Figure Captions}

\figcaption[buta_fig01.ps]{
\label{colorimage}
Hubble Heritage color image based on the F336W, F439W, F555W, and F814W 
WFPC2 images of NGC 4622. In this and all other WFPC2 images in this article,
north is oriented 30$^{\circ}$ clockwise from the top vertical,
while east is oriented 30$^{\circ}$ clockwise from the left
horizontal. The field shown is 1\rlap{.}$^{\prime}$57 $\times$ 
1\rlap{.}$^{\prime}$38.}

\figcaption[buta_fig02.ps]{
\label{center}
The central 5$^{\prime\prime}$$\times$5$^{\prime\prime}$ of NGC 4622 in the $V$-band
(F555W) image (left) and in $V-I$ (right), showing the strong central dust lane. North is to the
upper right and east is to the upper left, as in Figure 1. The color index
map is coded such that redder features are light and bluer features are
dark.}

\figcaption[buta_fig03.ps]{
\label{deepB}
Left: $B$-band image of NGC 4622 obtained with the CTIO 1.5-m telescope.
Right: the same image block-averaged 8$\times$8 pixels and displayed
to reveal the very faint outer isophotes. Each field is 
5\rlap{.}$^{\prime}$0 square. In this and all other groundbased images
in this article, north is to the top and east is to the left.}

\figcaption[buta_fig04.ps]{
\label{qpa}
Plots of minor-to-major axis ratio $q$ and position angle $\phi$ versus
semimajor axis radius $a$ of low resolution outer
isophotes of NGC 4622, based on ellipse fits. Filled circles are
for $B$ while open circles are for $I$. The position angle
$\phi$ is measured eastward from north.}

\figcaption[buta_fig05.ps]{
\label{decomp}
Folded major and minor axis profiles of NGC 4622 showing the combined bulge and
disk model in $B$ and $I$. In the lower panels, filled circles refer to
groundbased data while crosses refer to WFPC2 data.}

\figcaption[buta_fig06.ps]{
\label{lowresprofs}
Folded major axis profiles of NGC 4622 in the $H$-band and in $B$ and $I$
matched to the resolution of the $H$-band image. Bulge, disk, and total
profiles are shown based on standard decompositions.} 

\figcaption[buta_fig07.ps]{
\label{nukers}
Ellipse fit surface brightness and color indices (both corrected for 
Galactic extinction), ellipticity ($\epsilon$), and position angle ($\phi$)
profiles within $r$ = 16\arcs 0. The solid curve superposed on the $I$-band
profile is a fit of a "Nuker" law
to the central surface brightnesses. The position angle
$\phi$ is measured eastward from north, and the radius scale in pc
is based on a distance of 40.2 Mpc.}

\figcaption[buta_fig08.ps]{
\label{qpanuc}
Results of ellipse fits to isochromes of the central dust lane. The 
isochromes fitted range from $V-I$ = 1.45 to 1.72. The parameter
$q$ is the minor-to-major axis ratio. The position angle
$\phi$ is measured eastward from north.}

\figcaption[buta_fig09.ps]{
\label{Fourier}
Fourier decomposition (sky plane) of the WFPC2 $I$-band stellar background light distribution of NGC 4622. Upper left:
sum of $m$=0-6 terms. Upper right: $m$=0 image. Lower left: $m$=1 image. Lower
right: $m$=2 image. North is to the upper right and east is to the upper
left, as in Figure 1. Each frame covers a field of 1\rlap{.}$^{\prime}$50
$\times$ 1\rlap{.}$^{\prime}$43.} 

\figcaption[buta_fig10.ps]{
\label{onlyclusters}
$I$-band WFPC2 image of NGC 4622 after subtraction of the $m$=0-6 Fourier components. This removes
the bulge, disk, and smooth background light of the spiral arms to reveal
mostly the young and old cluster systems in NGC 4622. Orientation of this
image is the same as for Figure 1. The field shown is 
1\rlap{.}$^{\prime}$54
$\times$ 1\rlap{.}$^{\prime}$58. 
}

\figcaption[buta_fig11a.ps,buta_fig11b.ps]{
\label{mplots}
Relative Fourier amplitudes (intensity units) and phases for the (a) $m$=1
component and (b) the $m$=2 component. These are based on deprojected
images assuming a major axis position angle of 22$^{\circ}$ and an inclination
of 19\rlap{.}$^{\circ}$3. The phases are measured counterclockwise
from the line of nodes, which was oriented vertically in the deprojected
images.} 

\figcaption[buta_fig12.ps]{
\label{linecont}
Left: red continuum image of NGC 4622. Right: H$\alpha$ line emission map of NGC 4622. Both are from Fabry-Perot interferometry. The field shown in each map
is 1\rlap{.}$^{\prime}$96 square. North is to the top and east is to the left
in these groundbased images.}

\figcaption[buta_fig13.ps]{
\label{vfield}
Radial velocity field of NGC 4622 from Fabry-Perot interferometry, color-coded
so that reddish regions correspond to 4550 km s$^{-1}$ and bluish regions
to 4450 km s$^{-1}$. Orientation and field same as in Figure 12.}

\figcaption[buta_fig14.ps]{
\label{ewrotc}
Rotation curve of NGC 4622 based on an inclination of 19$^{\circ}$ and
a line of nodes position angle of 22$^{\circ}$.}

\figcaption[buta_fig15a.ps,buta_fig15b.ps,buta_fig15c.ps]{
\label{rotc1}
(a) A comparison between the rotation curve observed and the rotation
curve inferred from the $H$-band light distribution, based on
a mass-to-light ratio of 1.0 for both the bulge and the disk.
(b) A comparison between the rotation curve observed and the rotation
curve inferred from the $H$-band light distribution, based on
a mass-to-light ratio of 0.25 for both the bulge and disk 
and a ``fixed-$\sigma$" halo component (Kent 1986) having the parameters 
indicated. (c)
A comparison between the rotation curve observed and the rotation
curve inferred from a pure "fixed-$\sigma$" halo model having the
parameters indicated.}

\figcaption[buta_fig16.ps]{
\label{vimap}
WFPC2 $V-I$ color index map of NGC 4622. North is to the upper right
and east is to the upper left, as in Figure 1. The solid lines show the
kinematic line of nodes for comparison, corresponding to an actual
position angle of 22$^{\circ}$. The map is coded such that redder
regions are light and bluer regions are dark. It shows that thin 
dust lanes are confined mainly on the east side of the line of 
nodes. The field shown is 1\rlap{.}$^{\prime}$47 square.}

\figcaption[buta_fig17.ps]{
\label{residmap}
$V$-band WFPC2 map after subtraction of an $m$=0 Fourier image. Orientation
and field are the same as in Figure 16. The solid white line is the kinematic
line of nodes. The image shows again that the dust is mostly seen
east of the line of nodes, but also shows evidence for weaker dust lanes
west of the line of nodes.}

\figcaption[buta_fig18.ps]{
\label{bhmap}
A low resolution $B-H$ color index map of NGC 4622. The black ragged
line shows the position angle of the kinematic line of nodes. The field
shown is 2\rlap{.}$^{\prime}$43 square. North is at the top and east is
to the left in this groundbased image.}

\figcaption[buta_fig19.ps]{
\label{n2775}
Left: $B$-band image of NGC 2775 from the OSU Bright Galaxy Survey. 
Right: $B-H$ color index map, coded such that redder regions are
light and bluer regions are dark. Although the inclination is only
40$^{\circ}$, this image conclusively shows that the west side of 
NGC 2775 is the near side. The field shown is 6\rlap{.}$^{\prime}$4 
square. North is to the top and east is to the left.}

\figcaption[buta_fig20.ps]{
\label{modelcmaps}
Model color index maps (excluding scattering)
for four inclinations (upper left: 15$^{\circ}$, upper right: 20$^{\circ}$,
lower left: 25$^{\circ}$, lower right: 30$^{\circ}$), with false color code 
to enhance the reddening asymmetry across the line of nodes (oriented
at $-$7\rlap{.}$^{\circ}$5 in these images, the same as for the WFPC2
images of NGC 4622). Four dust rings were included in the model,
the innermost one being most prominent in these maps. The noise level
is approximately matched to that of the WFPC2 $V$-band image of
NGC 4622. The assumed inclination is indicated in each panel.}

\figcaption[buta_fig21.ps]{
\label{extinction}
Plots of surface brightness asymmetry (in magnitudes) along the minor axis (east side minus
west side) for the four
models in Figure 20. The profiles are based on averages
along contours having a major axis position angle along the line of nodes
and an axis ratio = $\cos i$, within a cone having a half angle of
5$^{\circ}$. The short vertical lines refer to the positions of the
model dust rings, which have a face-on optical depth $\tau_o$
three times that of the background dust layer.}

\figcaption[buta_fig22.ps]{
\label{reddening}
Plots of color asymmetry (in magnitudes) along the minor axis (again east side minus west side)
for the four models in Figure 20. The profiles are based on averages
along contours having a major axis position angle along the line of nodes
and an axis ratio = $\cos i$, within a cone having a half angle of
5$^{\circ}$.}

\figcaption[buta_fig23a.ps,buta_fig23b.ps]{
\label{gal20}
Plots of WFPC2 surface brightness and color asymmetry along the kinematic
minor (a) and major (b) axes of NGC 4622. The profiles are based on averages
along contours having a major axis position angle along the line of nodes
and an axis ratio = $\cos 20^{\circ}$, within a cone having a half angle of
10$^{\circ}$. The short vertical lines in (a) point to dust features seen in the
images. The differences are in the sense east minus west for the minor axis,
and north minus south for the major axis.}

\figcaption[buta_fig24.ps]{
\label{cmdiagram}
Color-magnitude diagram of point or near-point sources in the WFPC2 field of
NGC 4622. The dotted box isolates the objects that are likely to be
globular clusters. The absolute magnitude scale is based on a distance
of 40.2 Mpc.}

\figcaption[buta_fig25.ps]{
\label{xyplot}
The distribution of the 250 objects lying within the dotted box in Figure 24.
Crosses refer to WF3, filled circles to WF4, open circles to WF2, and
plus symbols to PC1. Coordinates are relative to the nucleus, and
the plot has the same orientation as the image in Figure 1.}

\figcaption[buta_fig26.ps]{
\label{outer}
Color-magnitude diagram of sources having $r$ $>$ 65$^{\prime\prime}$ within the
WFPC2 field. The dotted box is the same as shown in Figure 24, and probably includes
mostly globular clusters.}

\figcaption[buta_fig27.ps]{
\label{colordistribution}
The distribution of $(V-I)_o$ colors of the objects within the dotted box in
Figure 24. The solid curve is a gaussian fitted to the histogram. The mean,
1.04, and dispersion, 0.19, are typical of old globular cluster systems seen
in other galaxies (Kundu \& Whitmore 2001).}

\figcaption[buta_fig28.ps]{
\label{completeness}
Completeness curves for two magnitude intervals to show how the detectability of the
globular clusters depends on both magnitude and background brightness.}

\figcaption[buta_fig29.ps]{
\label{gclf}
$V$-band luminosity function of globular clusters in NGC 4622, corrected for both
areal and detector incompleteness. The solid curve is a gaussian fit to the histogram
having a fixed dispersion of $\sigma$ = 1.3 mag.}

\figcaption[buta_fig30a.ps,buta_fig30b.ps]{
\label{surfdens}
a. Surface density of globular clusters in NGC 4622 versus $r^{1\over 4}$ in projection,
where $r$ is the radius in arcseconds. The solid line is a weighted fit of an $r^{1\over 4}$ law
to the observed points. The dotted line shows the fit forced to have the same slope
as the background starlight. b. same as a but the curve now represents a
fit of a King model to the surface densities. In this case, the radius $R$
is in kpc.}

\figcaption[buta_fig31.ps]{
\label{eastwest}
Histograms of the numbers of globular clusters east (solid line) and west (dotted line) of
the kinematic line of nodes.}

\figcaption[buta_fig32.ps]{
\label{twocolor}
Two-color plots of those sources in Figure 24 having 21.5 $\leq$ $V_o$ $\leq$ 23.0 (left panels) and
23.0 $<$ $V_o$ $\leq$ 24.0 (right panels) in the color range $-1$ $\leq$ $(V-I)_o$ $\leq$ 0.7. The
solid curves in each plot are evolutionary synthesis models (for
solar metallicity and a Salpeter IMF) from Bruzual \& Charlot
(1996; see also Buta et al. 2001). Several points on the curves
in the left panels are labeled by the cluster age in years. 
Reddening lines are shown for a visual extinction $A_V$ = 1 mag in the 
right panels. These plots confirm that most of the sources in the
region to the left of the dotted box in Figure 24 are young stellar 
associations.}

\figcaption[buta_fig33.ps]{
\label{starform}
Closeup of northeast part of NGC 4622 (F555W filter) showing star-forming 
ridges. Features labeled 1-4 are discussed in the text. The area shown has
dimensions 0\rlap{.}$^{\prime}$99 $\times$ 0\rlap{.}$^{\prime}$57. North
is to the upper right and east is to the upper left, as in Figure 1.}

\figcaption[buta_fig34.ps]{
\label{colorpa}
Color index versus position angle around the inner ring of NGC 4622,
based on WFPC2 images. The position angle is measured relative to
true north.}

\figcaption[buta_fig35.ps]{
\label{eso297d27}
Groundbased CCD image (sum of $B$, $V$, and $I$-band images) of ESO 297$-$27,
a newly identified "two-way" spiral (Grouchy \& Buta 2002). The image
has been deprojected from a 55$^{\circ}$ assumed inclination, to better
display the galaxy's structure. In this case, a single inner arm
opens outward clockwise, while up to three weaker outer arms open
outward counterclockwise. The field shown in 1\rlap{.}$^{\prime}$89
square. Field stars have been removed.}

\figcaption[buta_fig36.ps]{
\label{phaseshifts}
Enlargements of the behavior of the phase of the $m$ = 2 component
of NGC 4622 around two likely significant radii: (a) the region of
the transition from the single inner arm to the two outer arms,
and (b) a region in the outer arms where star formation changes
from being on the convex side of the arms to the concave side.}

\figcaption[buta_fig37.ps]{
\label{depV}
Deprojected $V$-band WFPC2 image showing the locations of the
transition radii highlighted in Figure 36. The field covers a
region 1\rlap{.}$^{\prime}$63 $\times$ 1\rlap{.}$^{\prime}$63
in area, and is oriented such that the kinematic line of nodes
is vertical.}

\figcaption[buta_fig38.ps]{
\label{sn2001jx}
HST WFPC2 F555W image of NGC 4622, rotated so that north is at the top 
and east is to the left. The arrow points to a supernova, now known as
SN 2001jx (IAU Circ. 7833), that was
present during the WFPC2 observations. The arrow has a length of 
21\rlap{.}$^{\prime\prime}$5.}





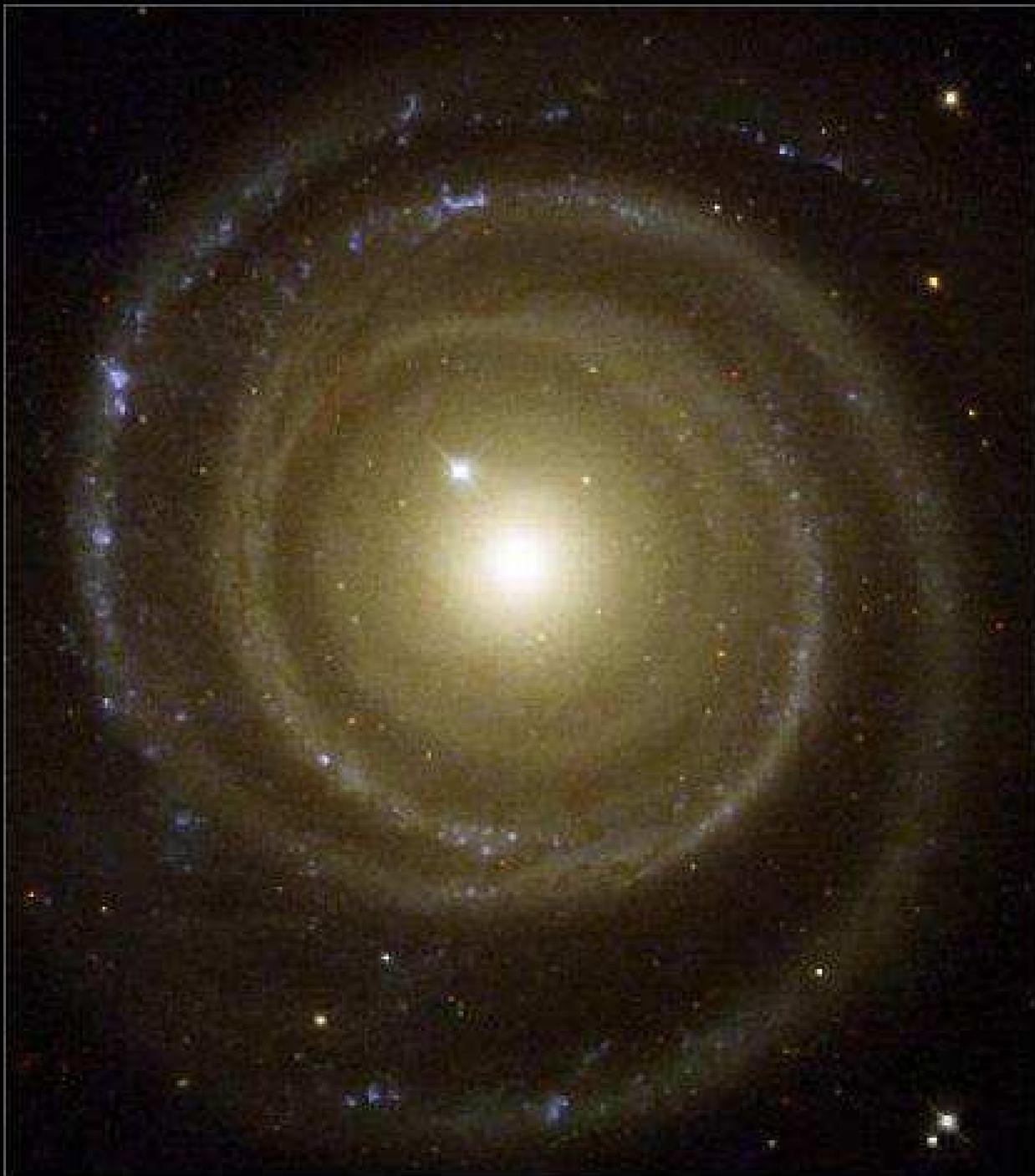
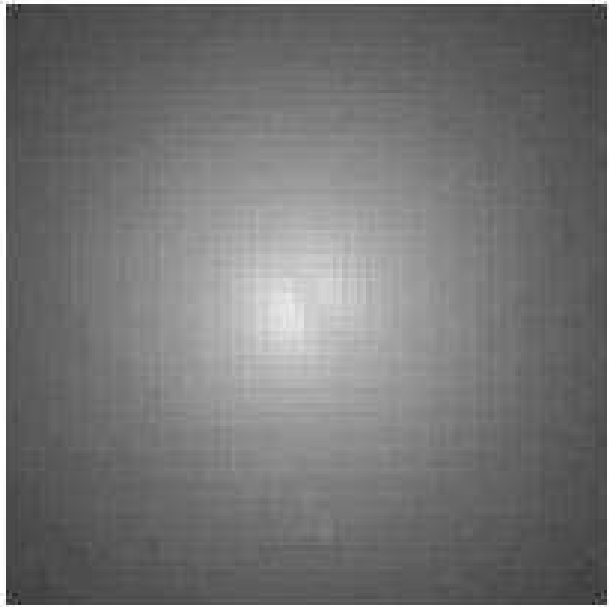
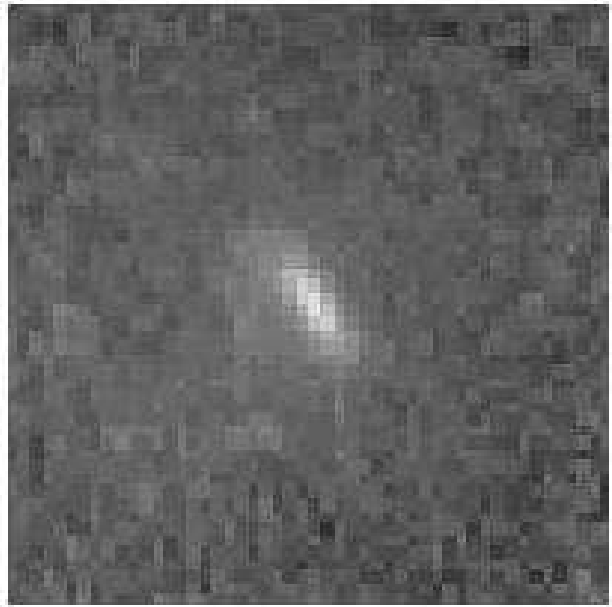
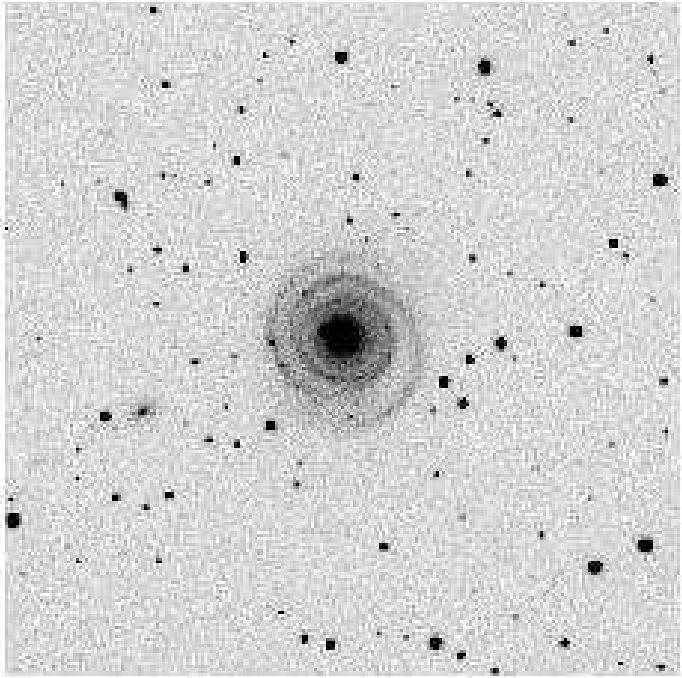
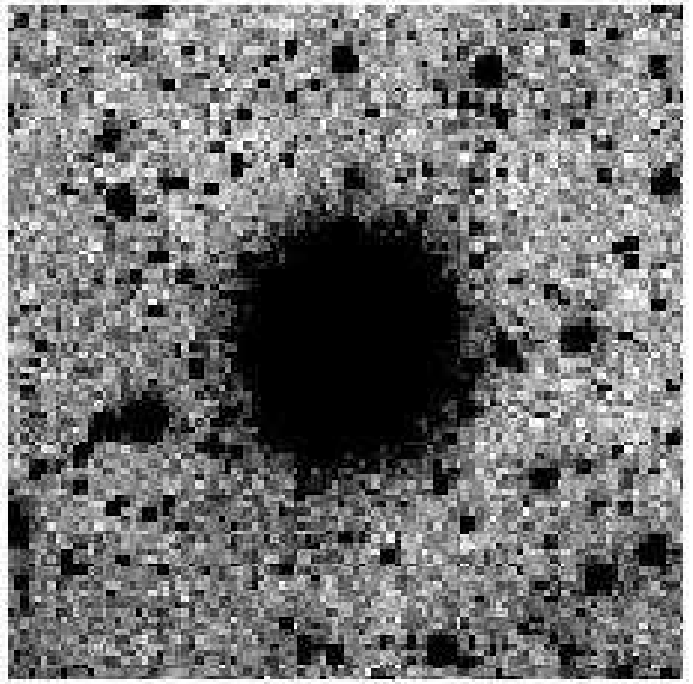
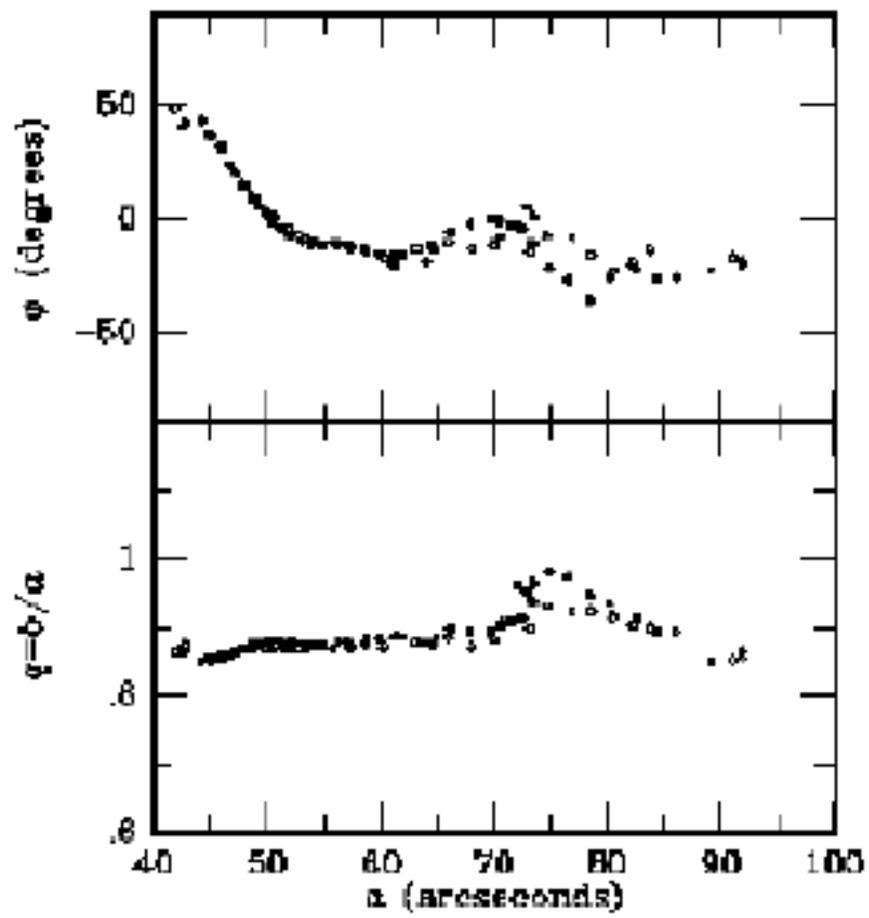
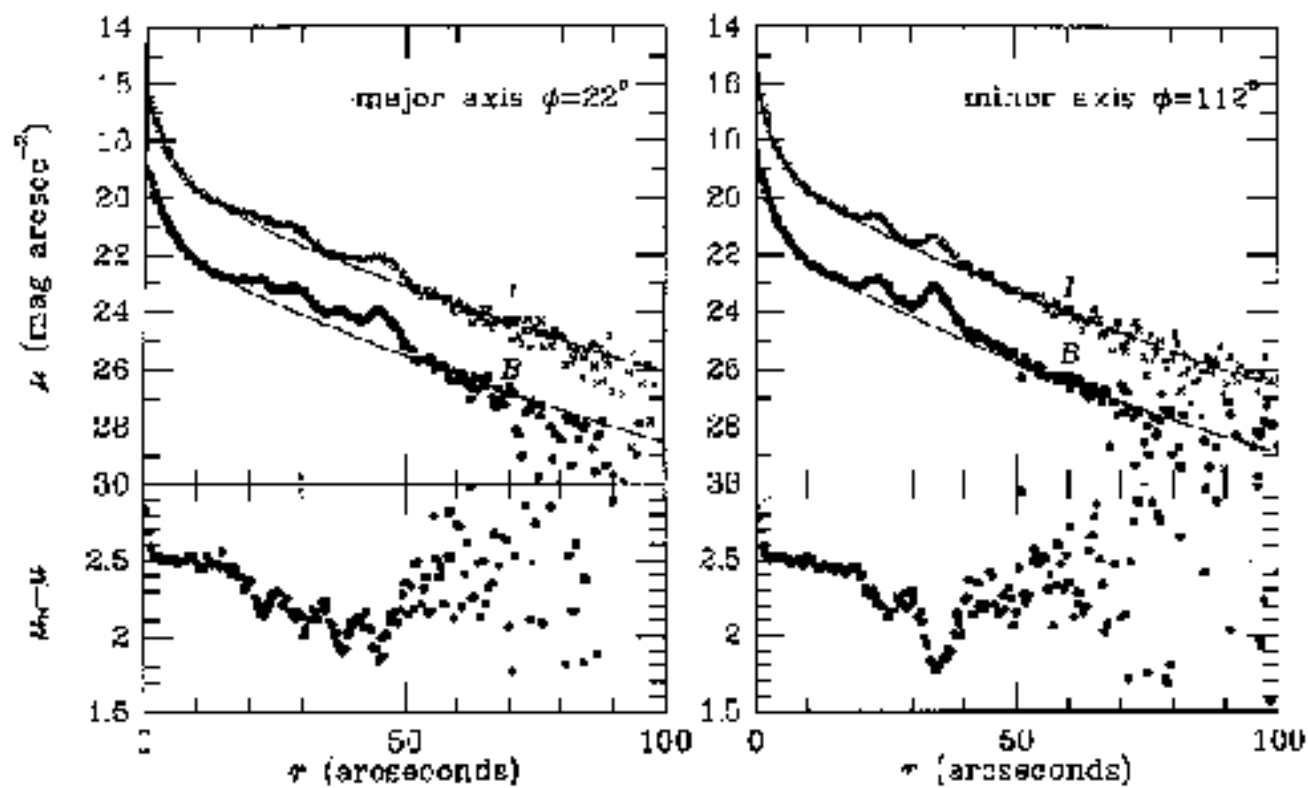
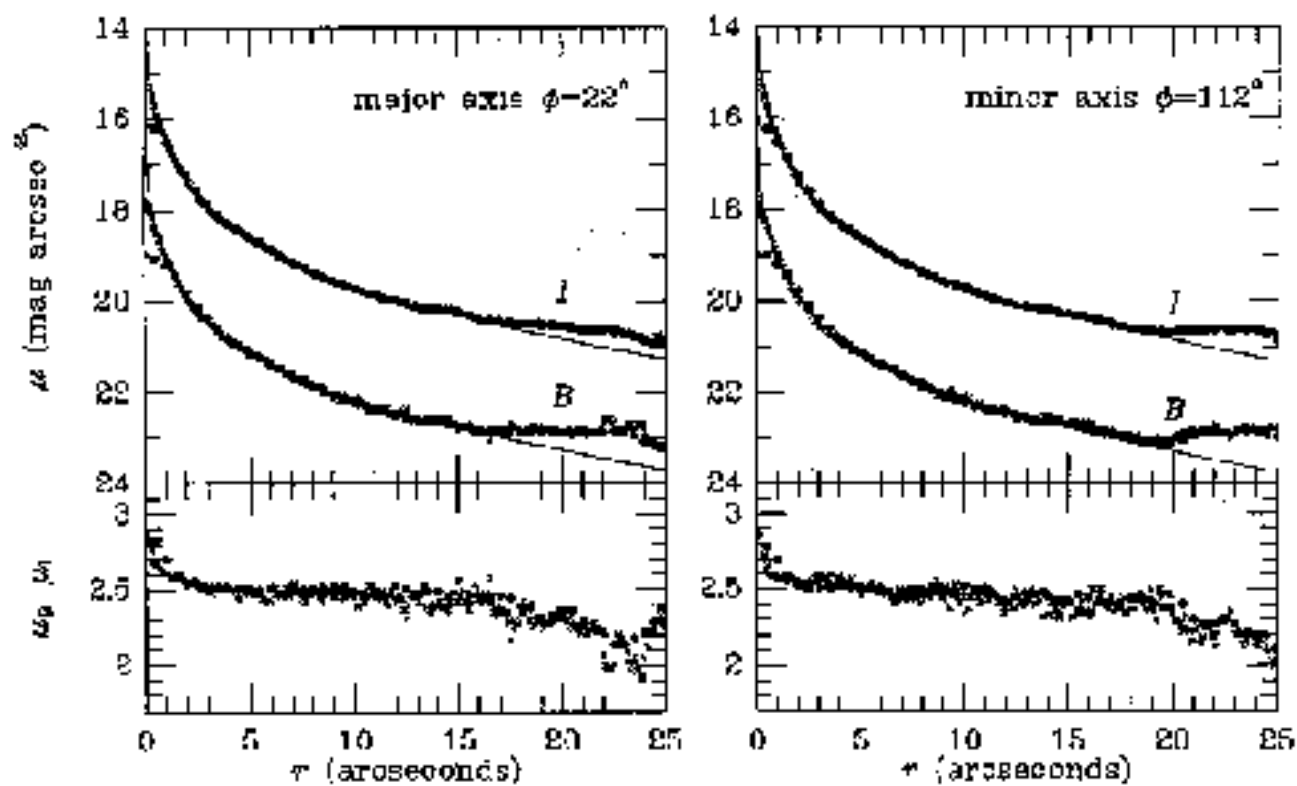
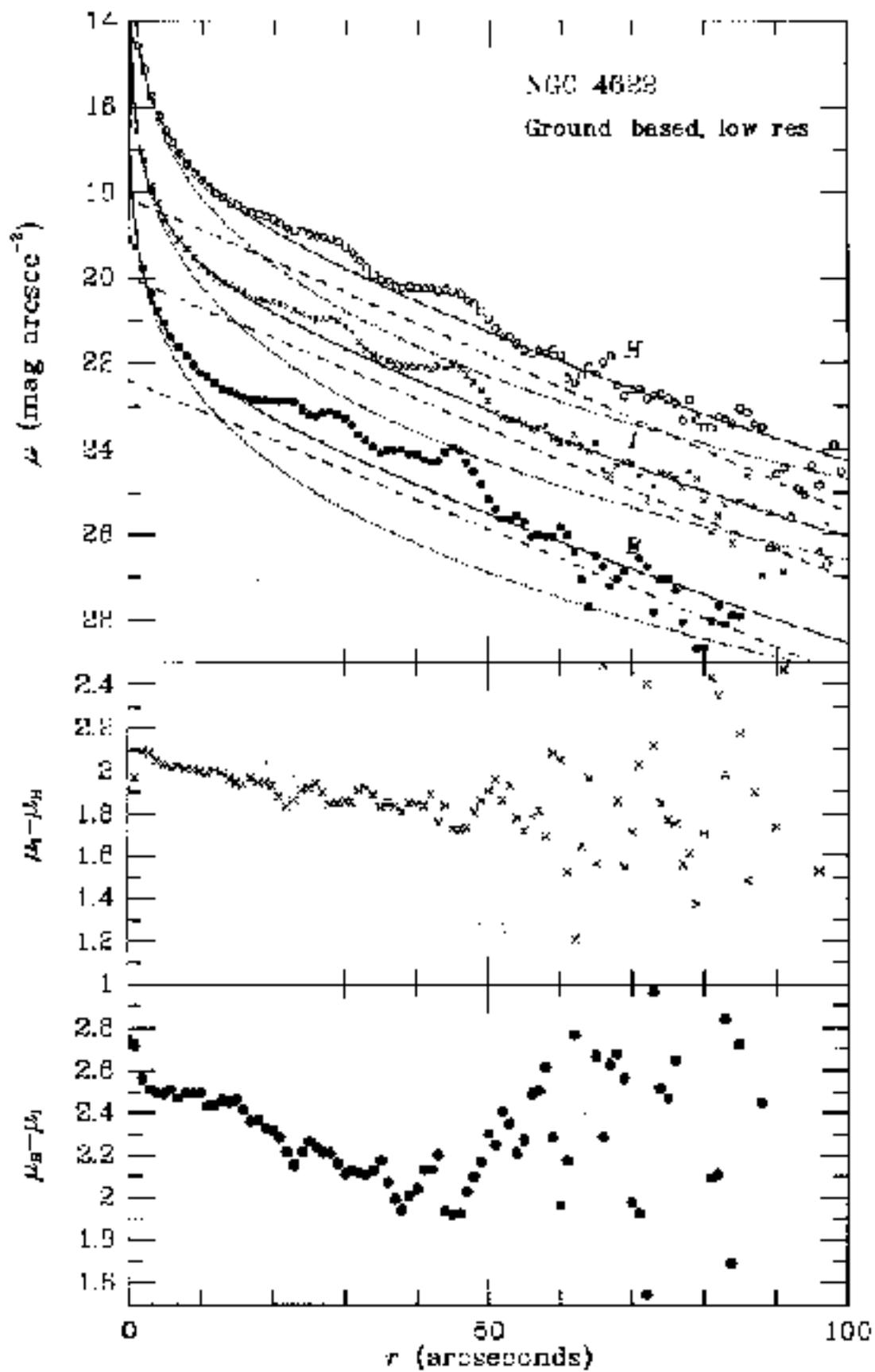
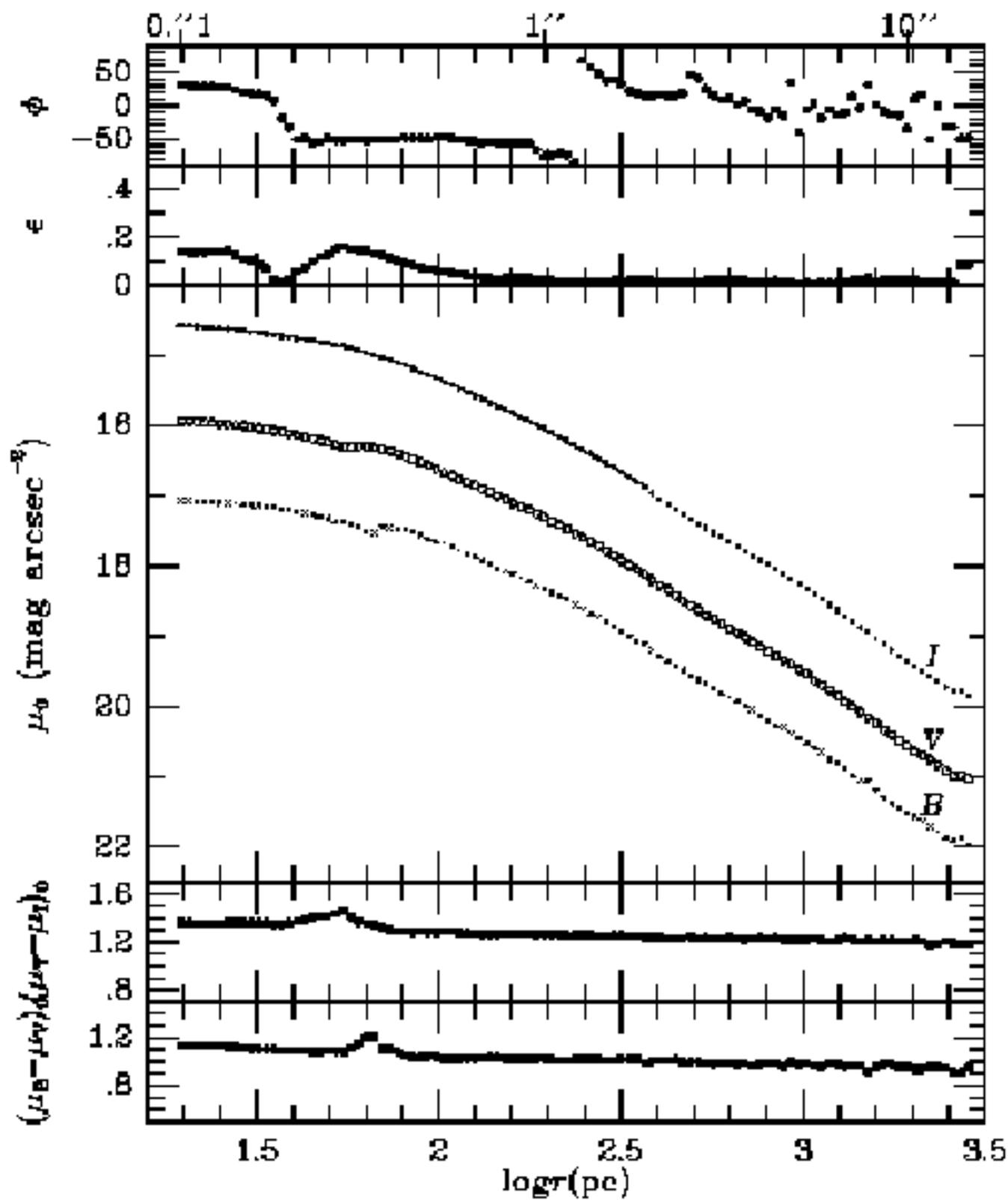
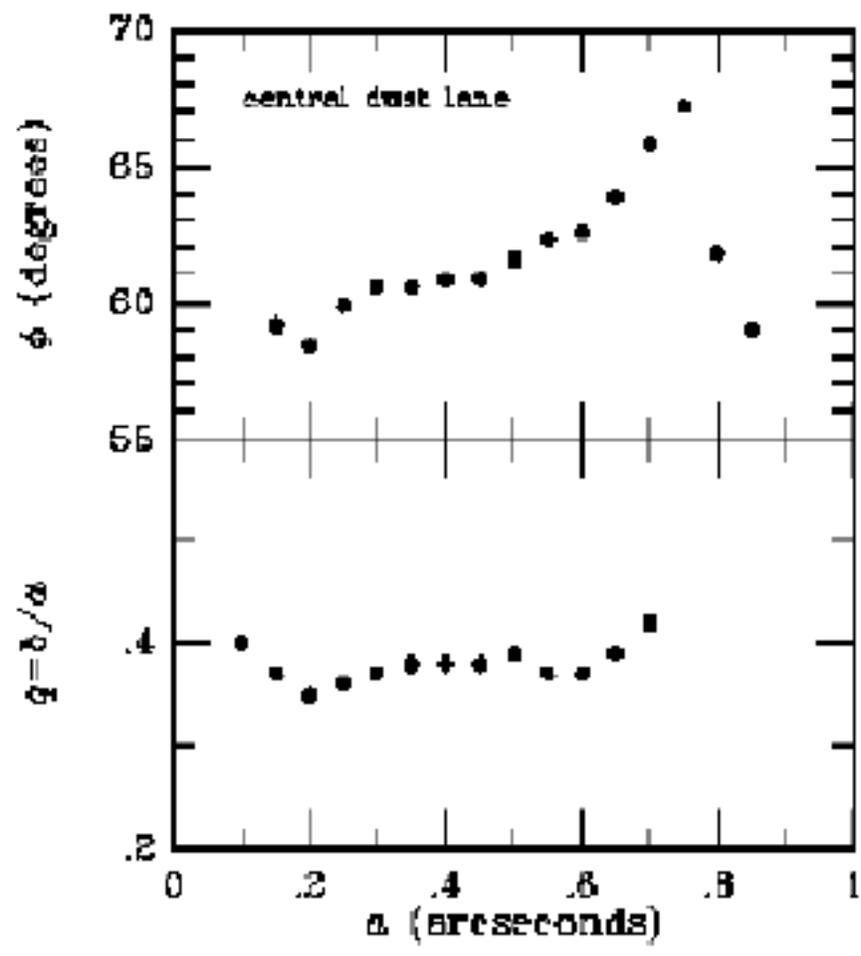
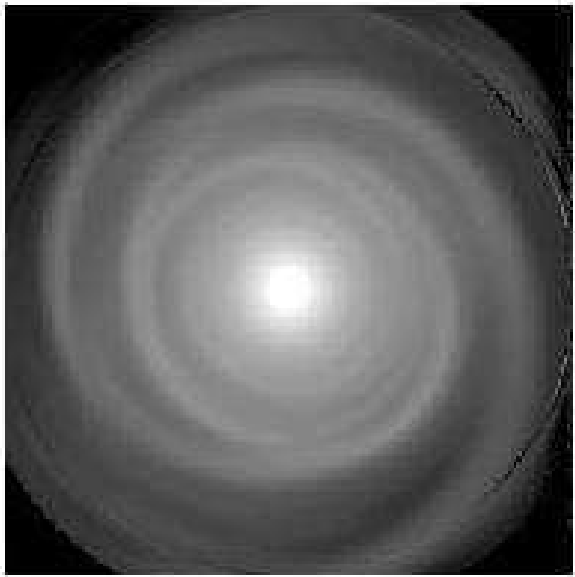
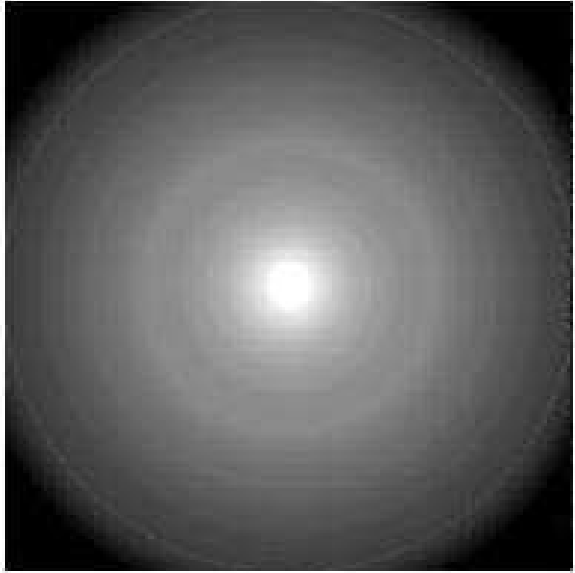
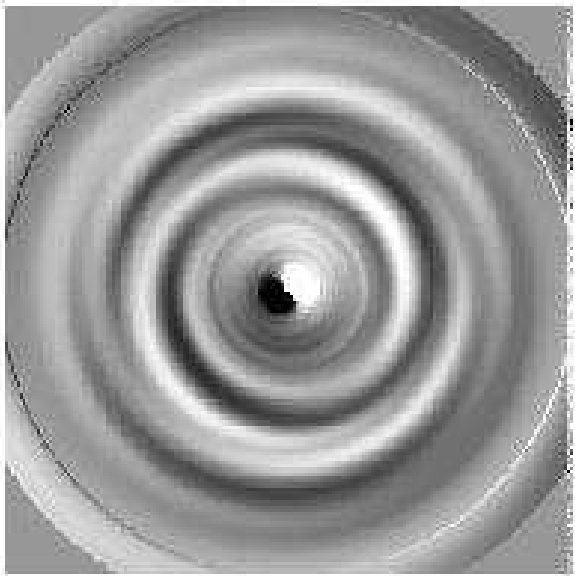
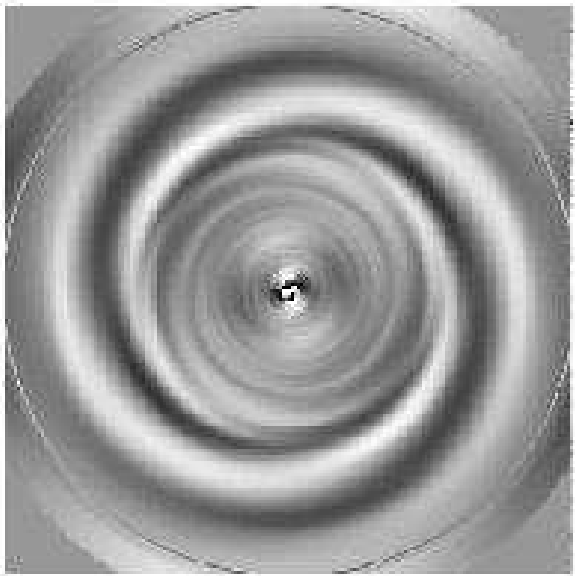
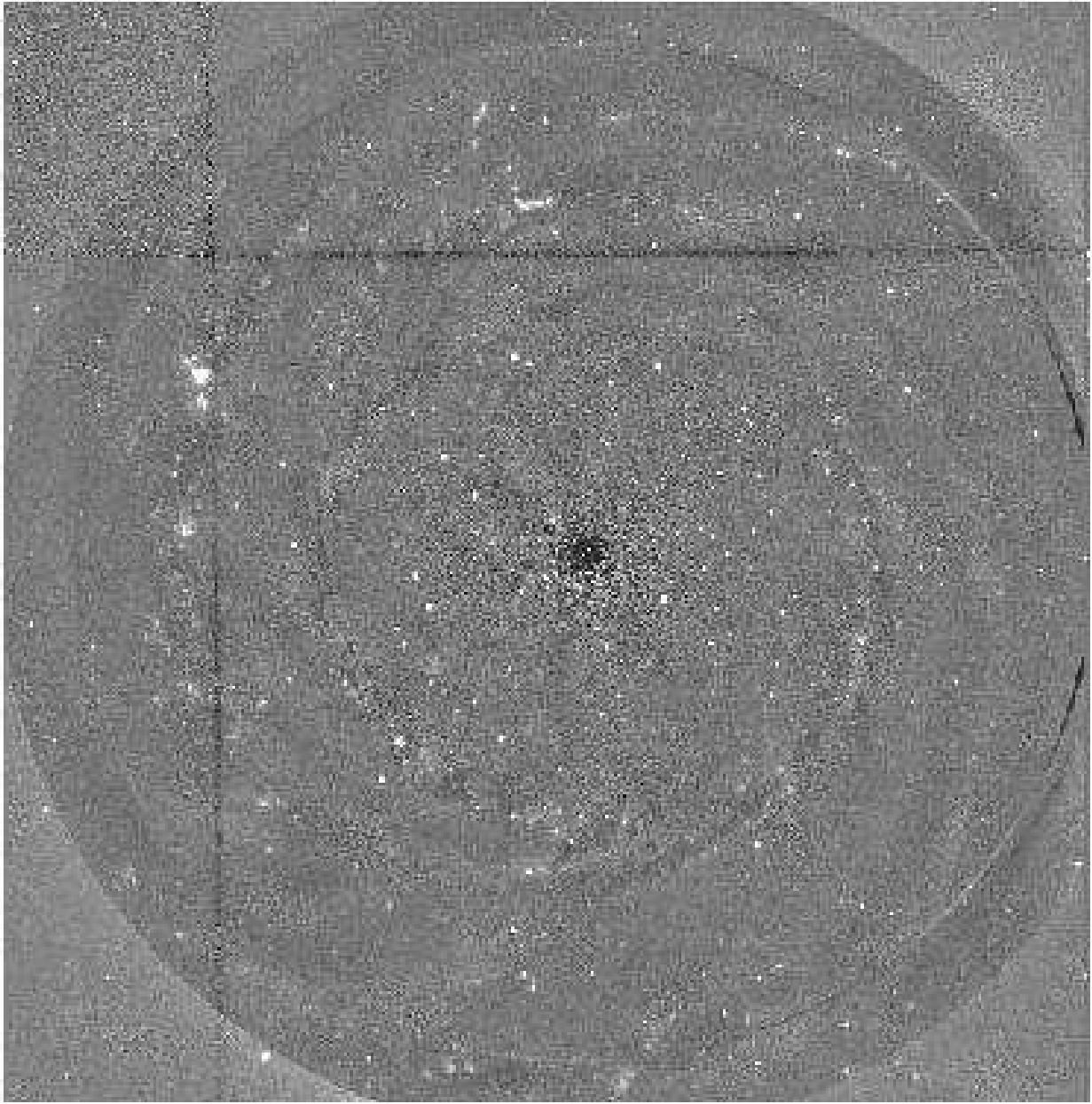
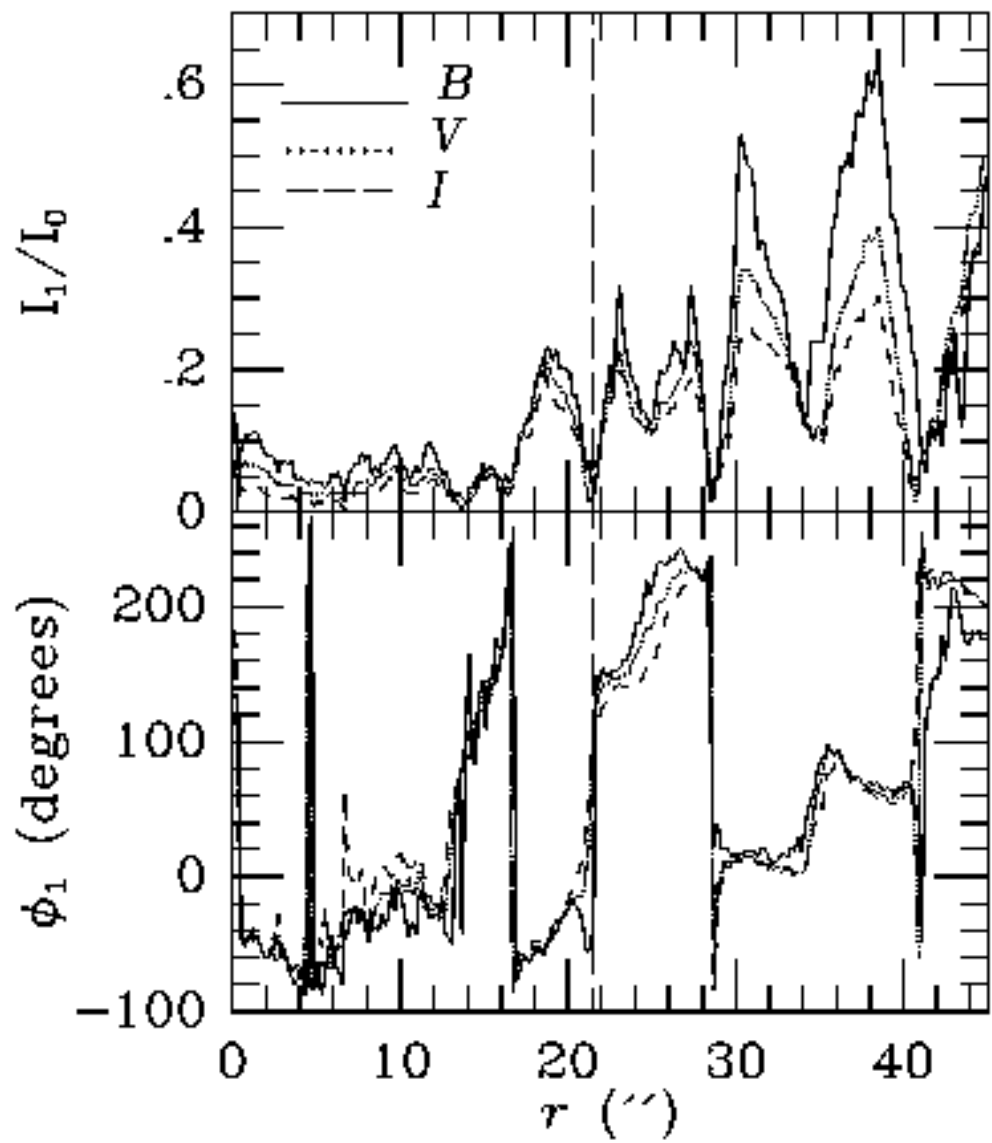
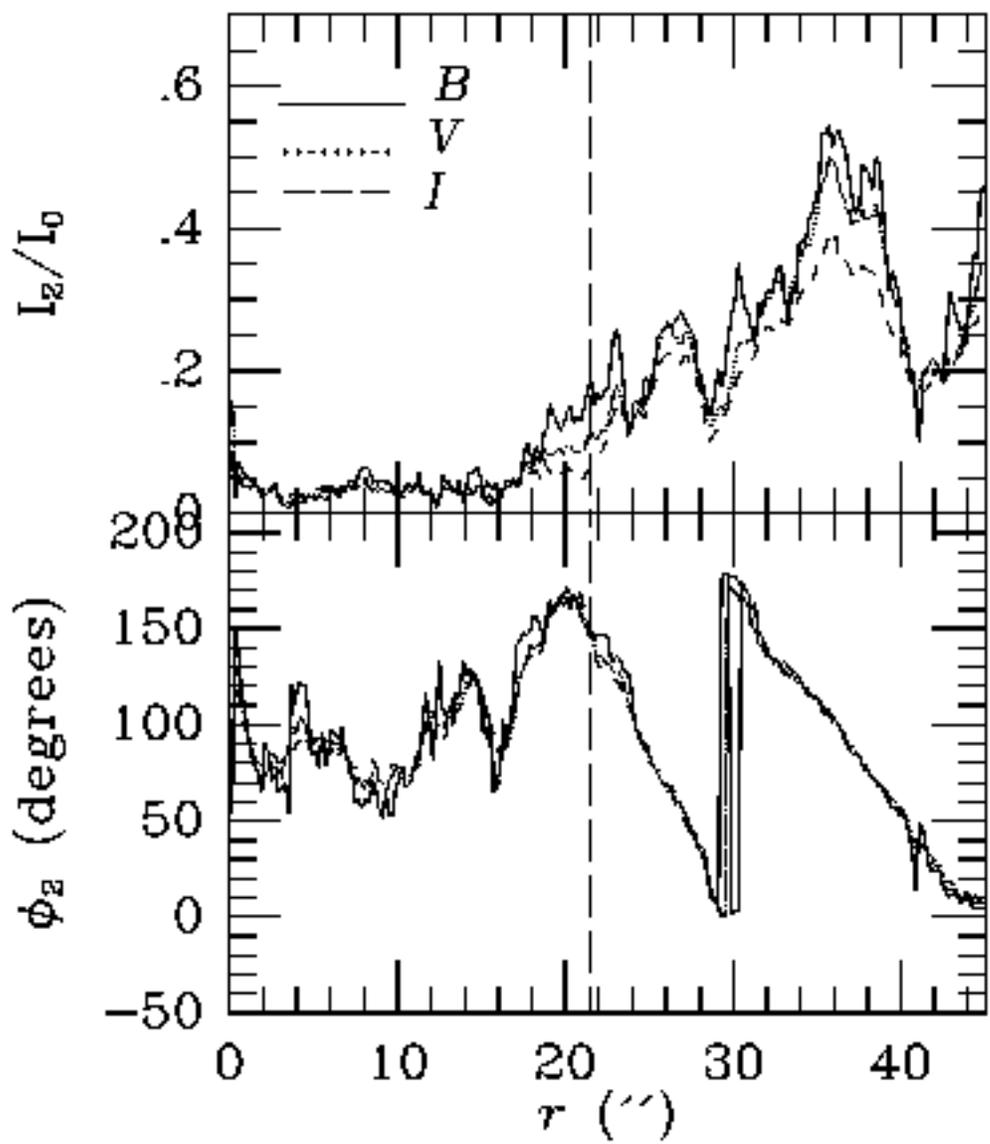
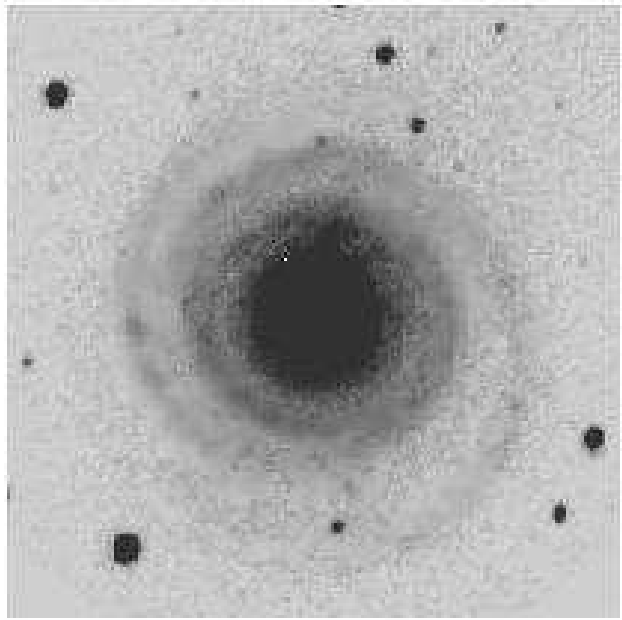
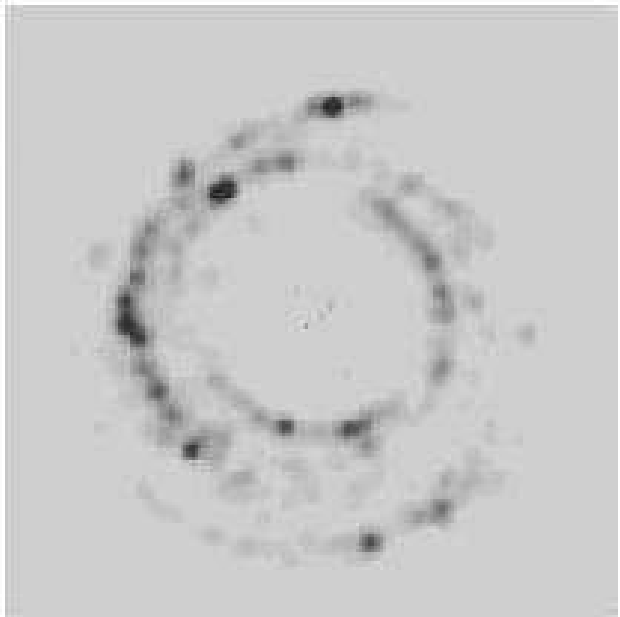
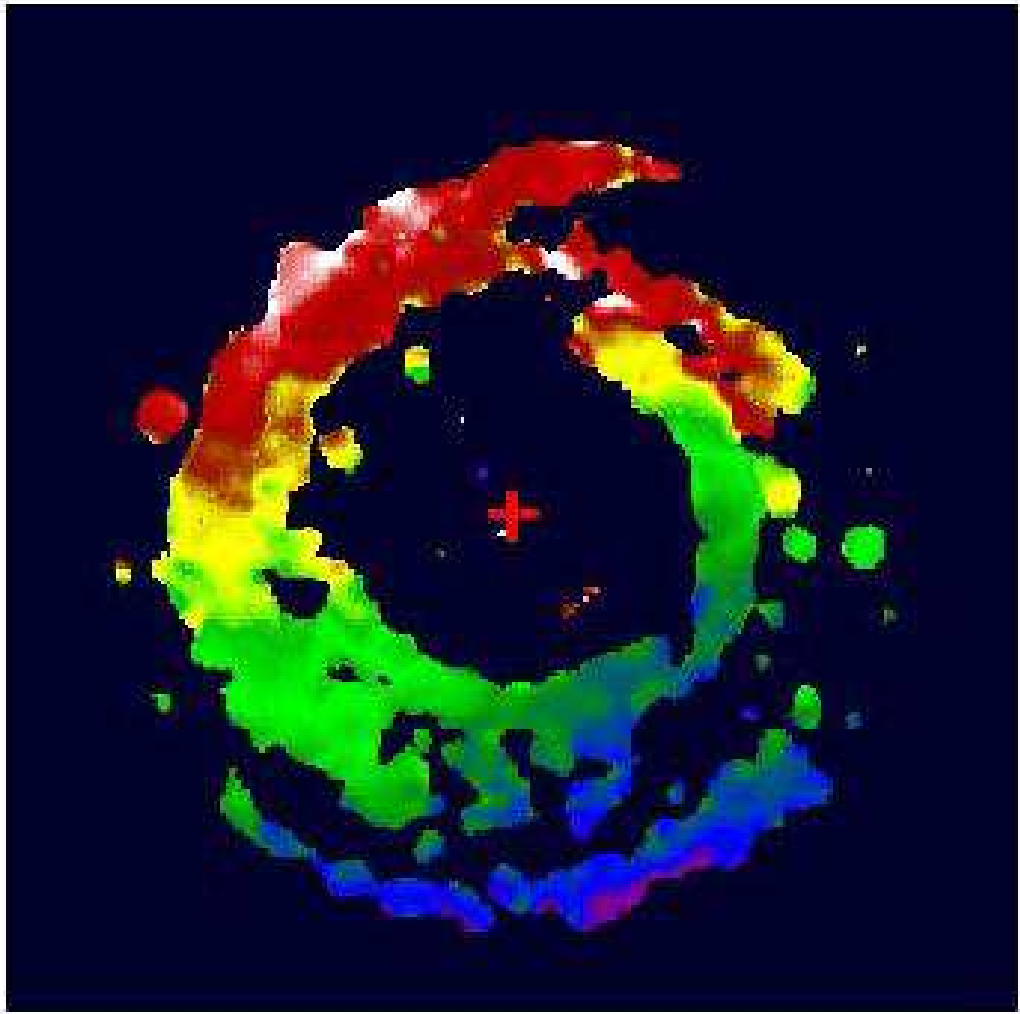
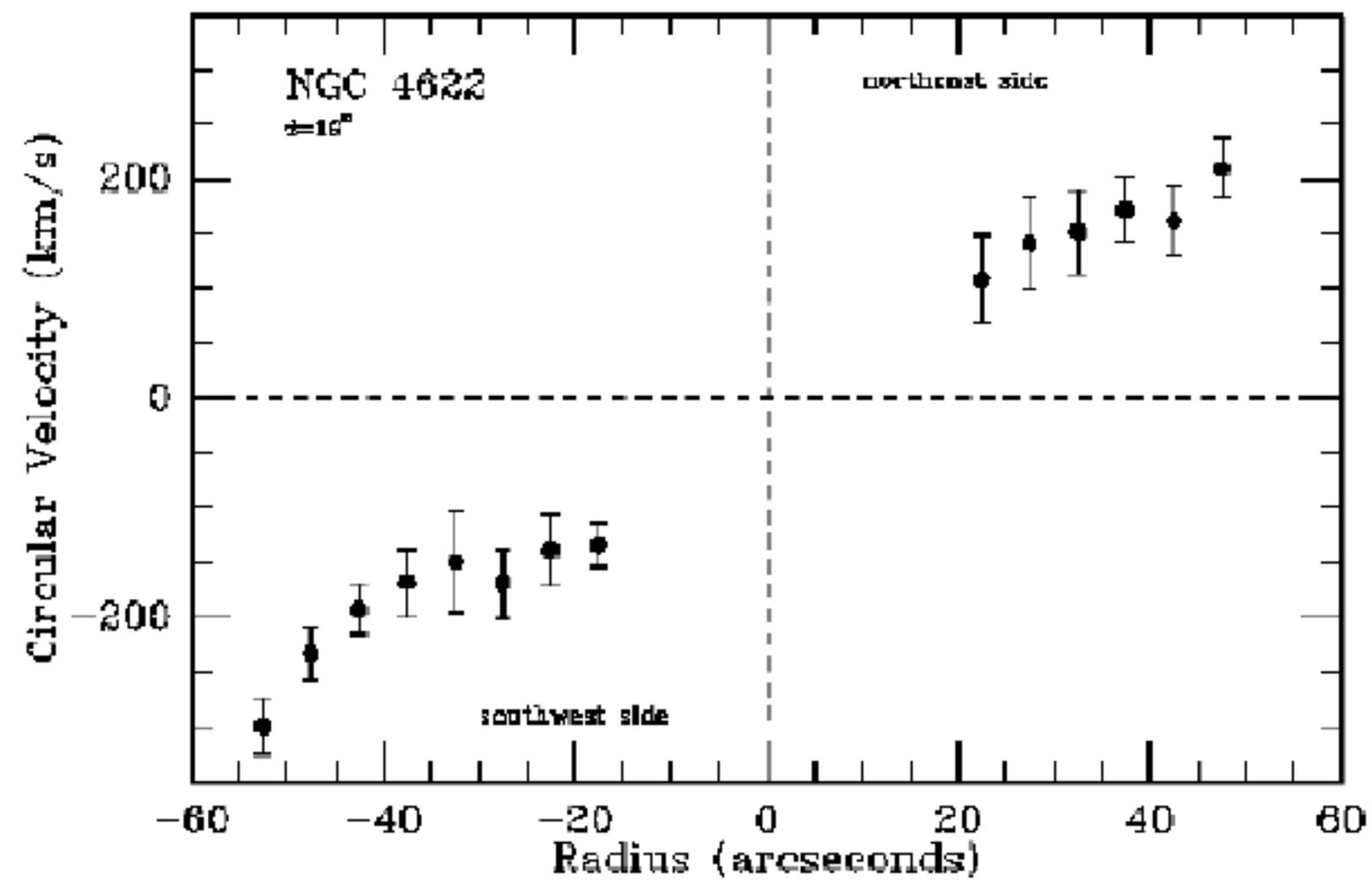
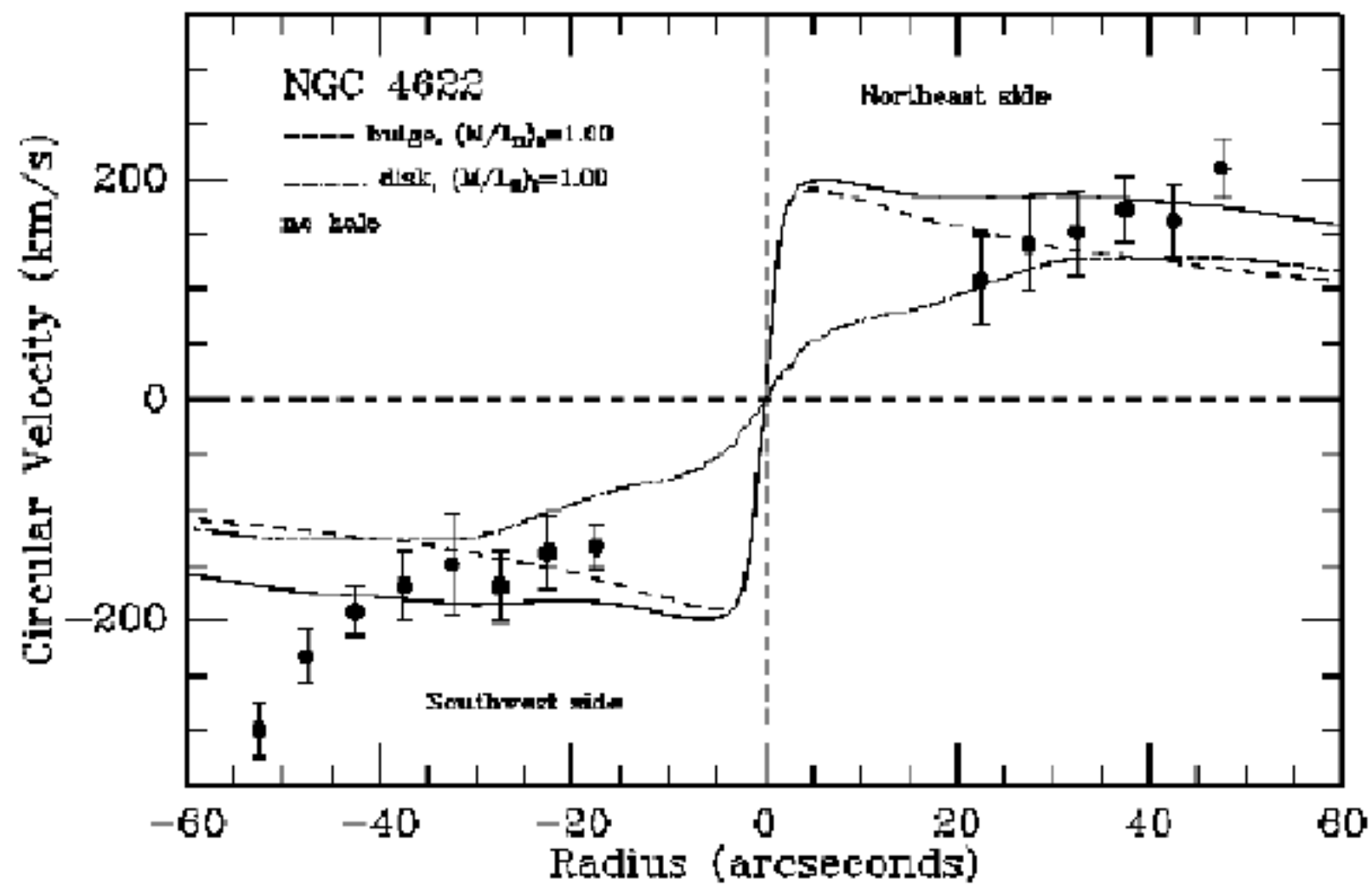
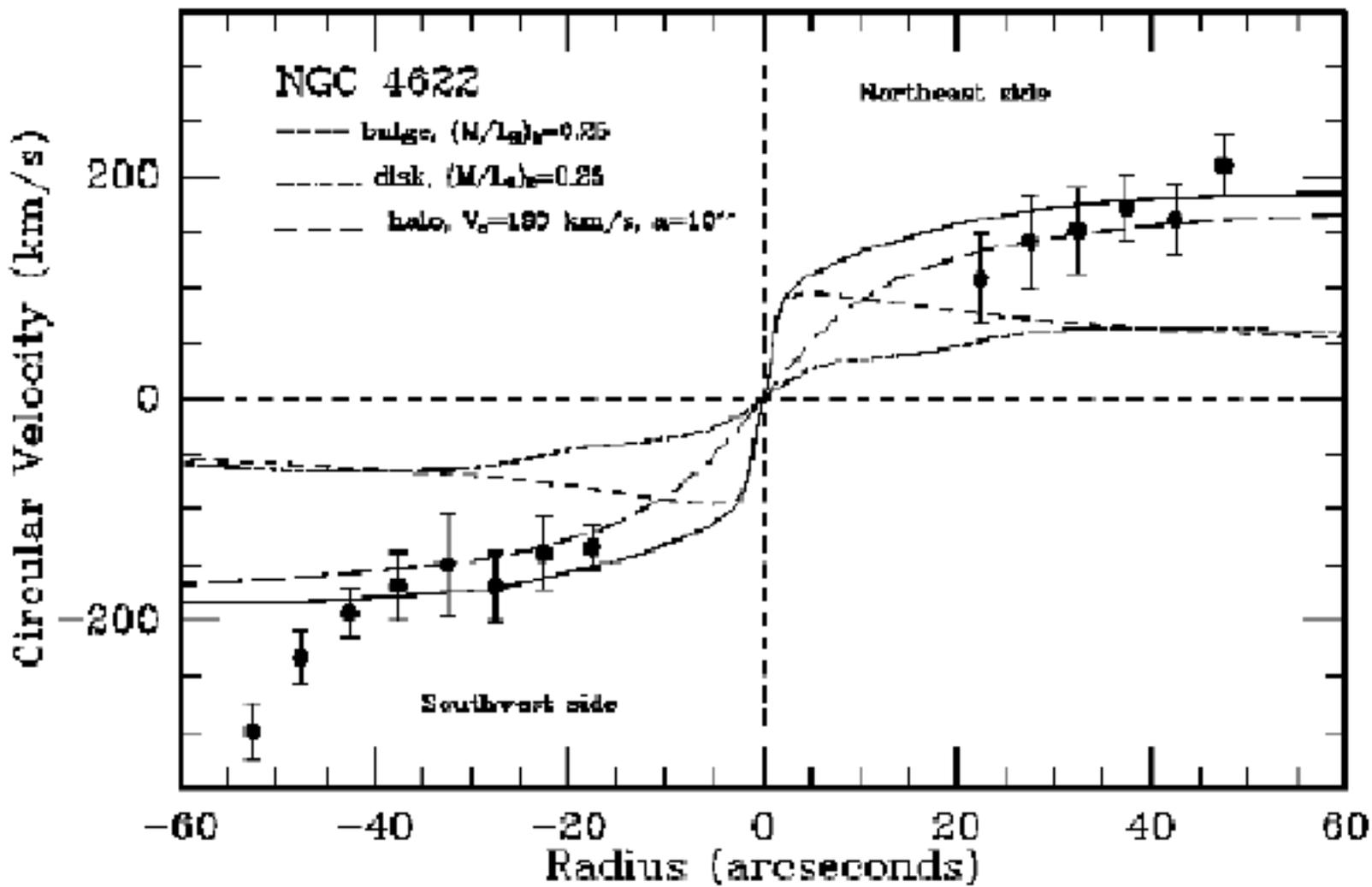
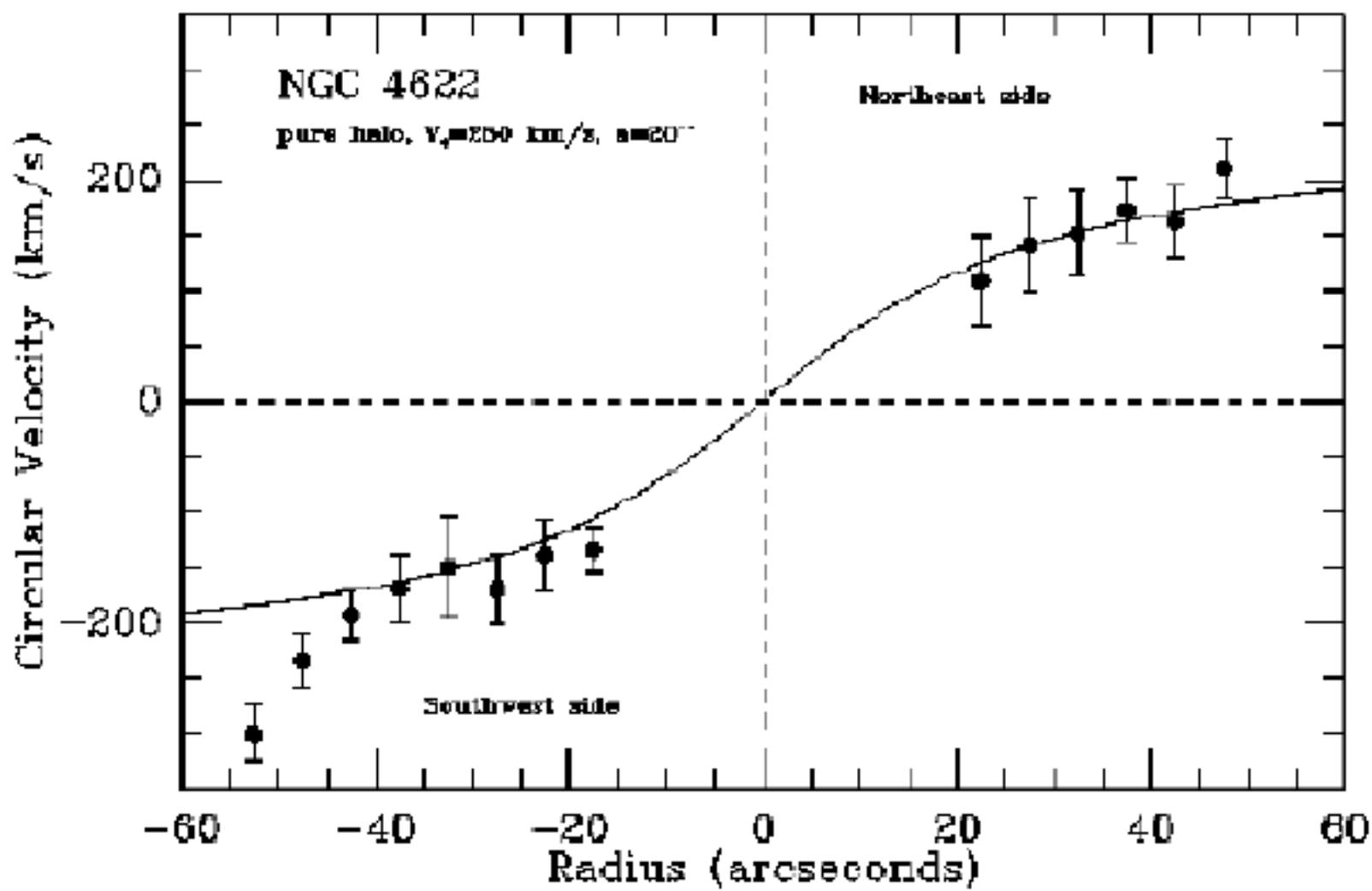
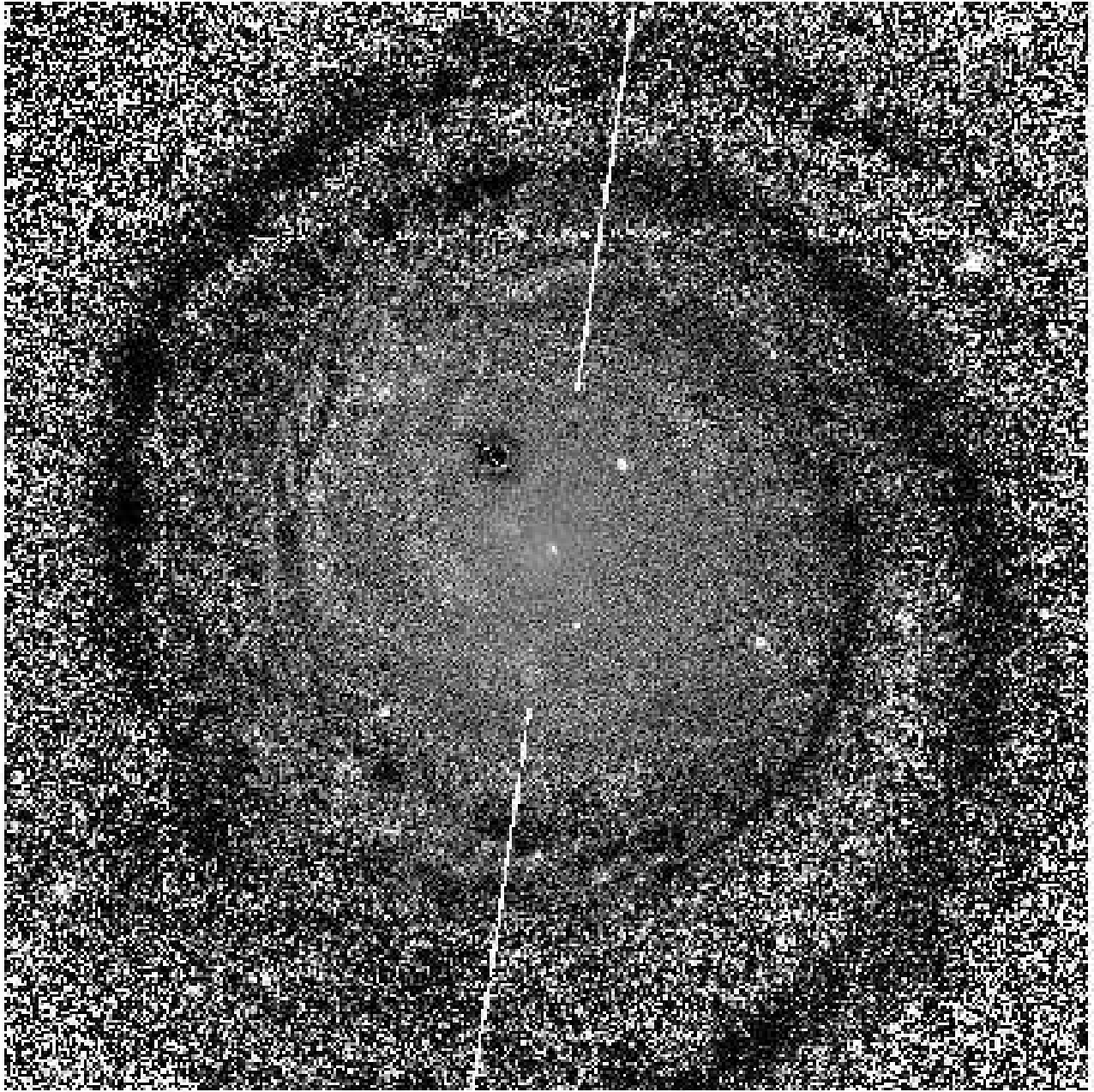
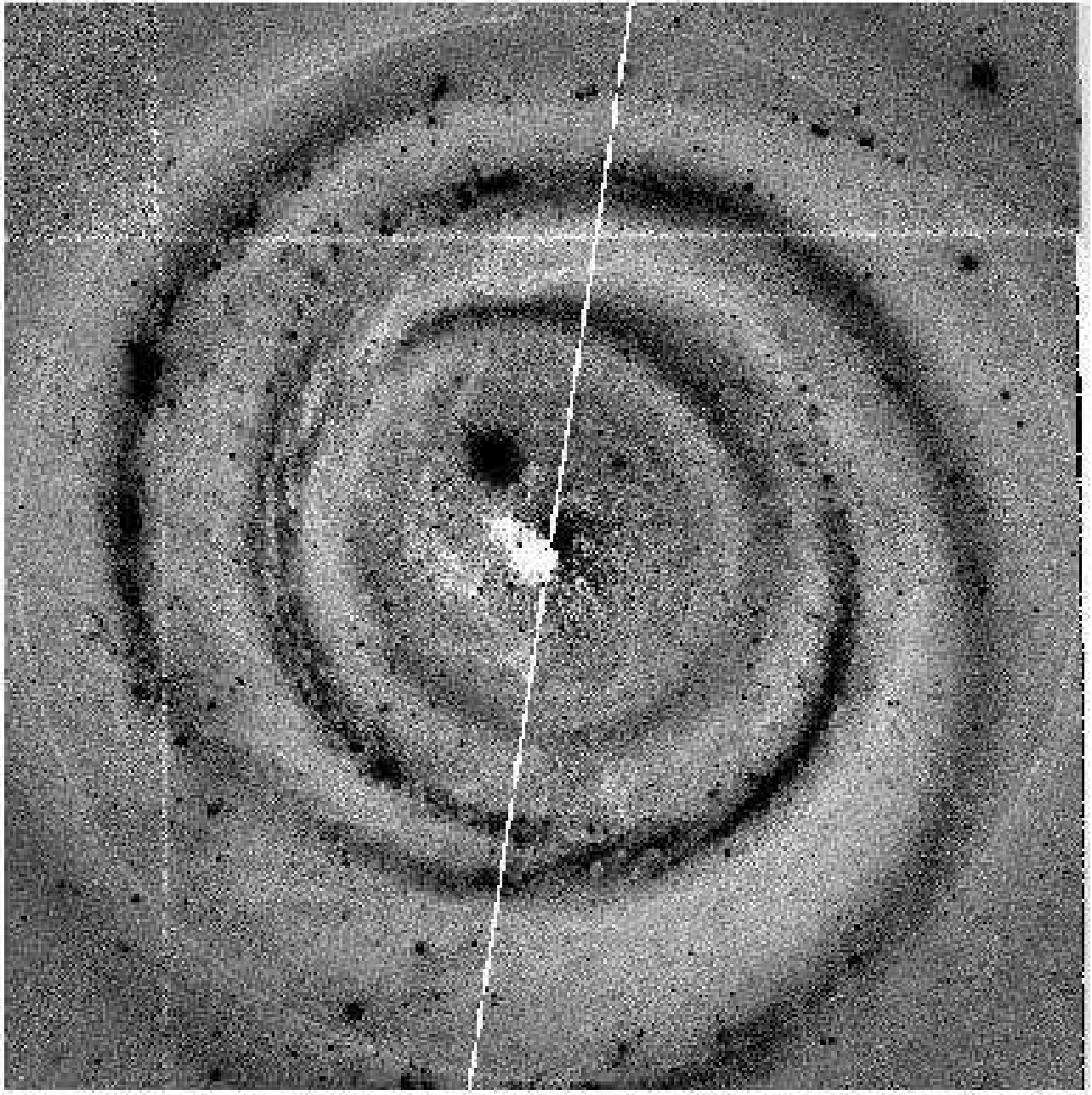
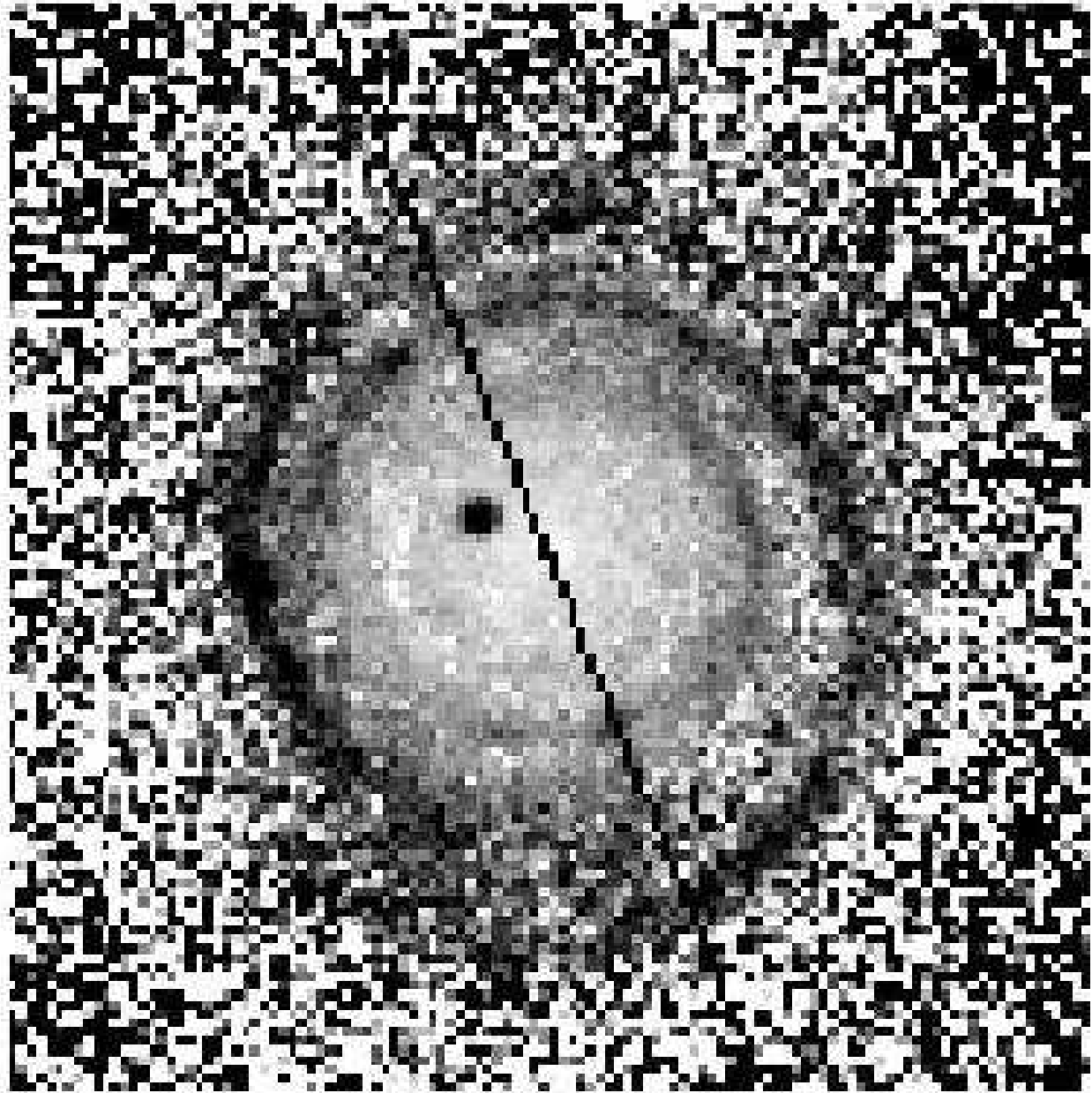
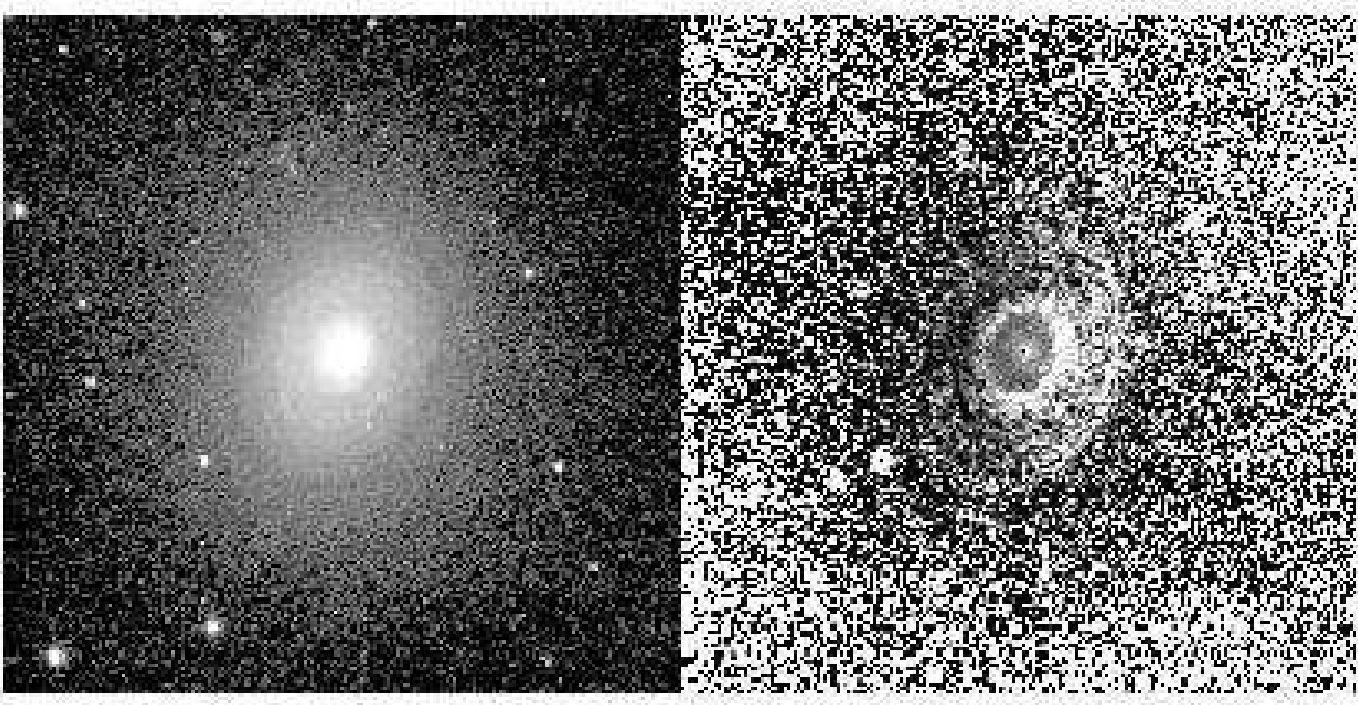
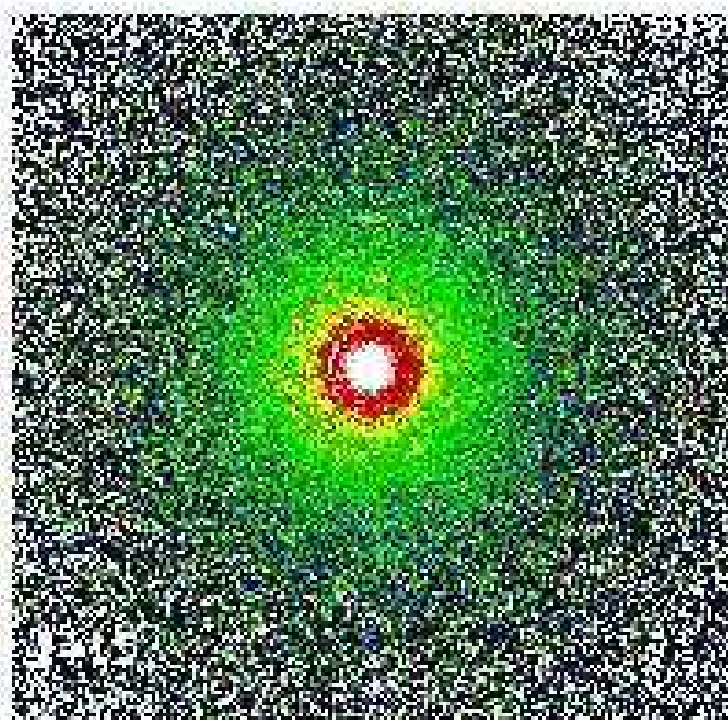
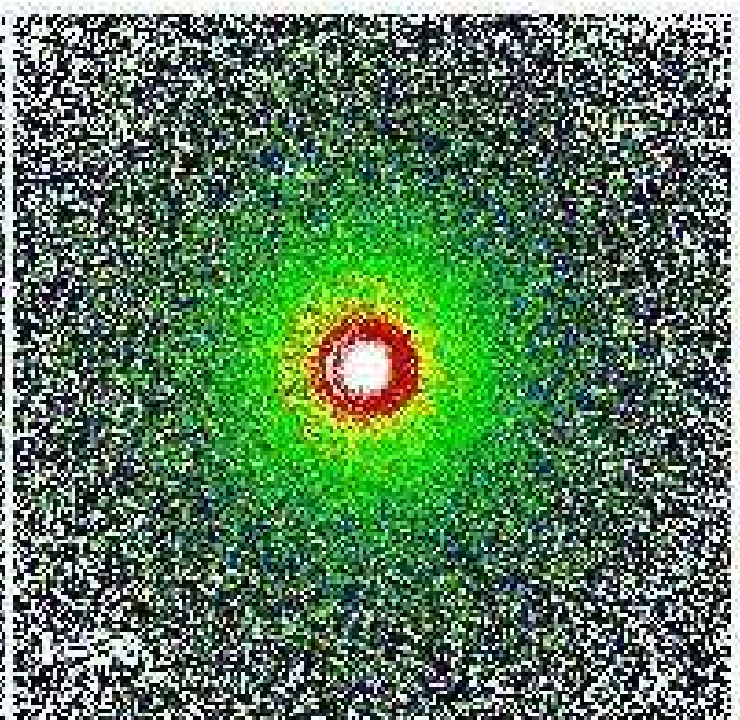
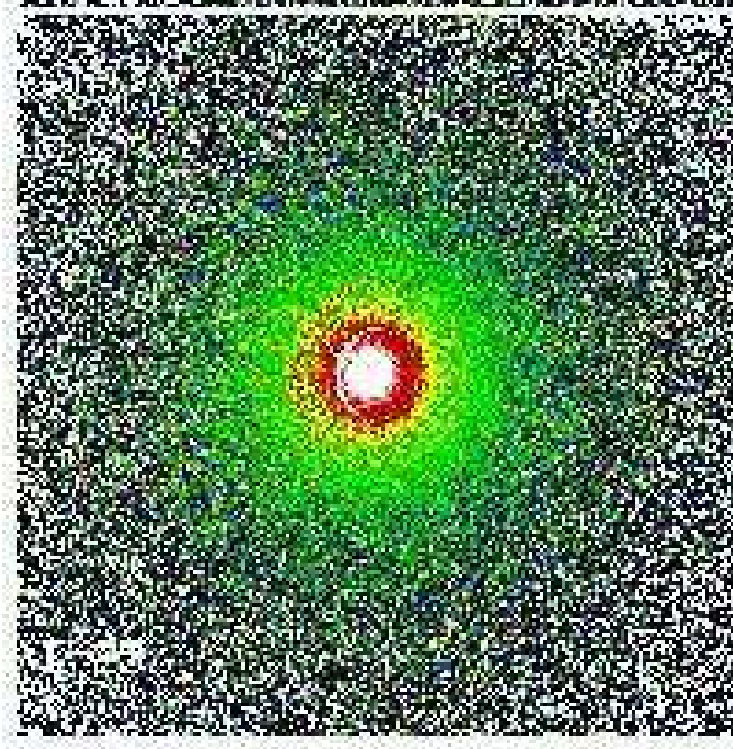
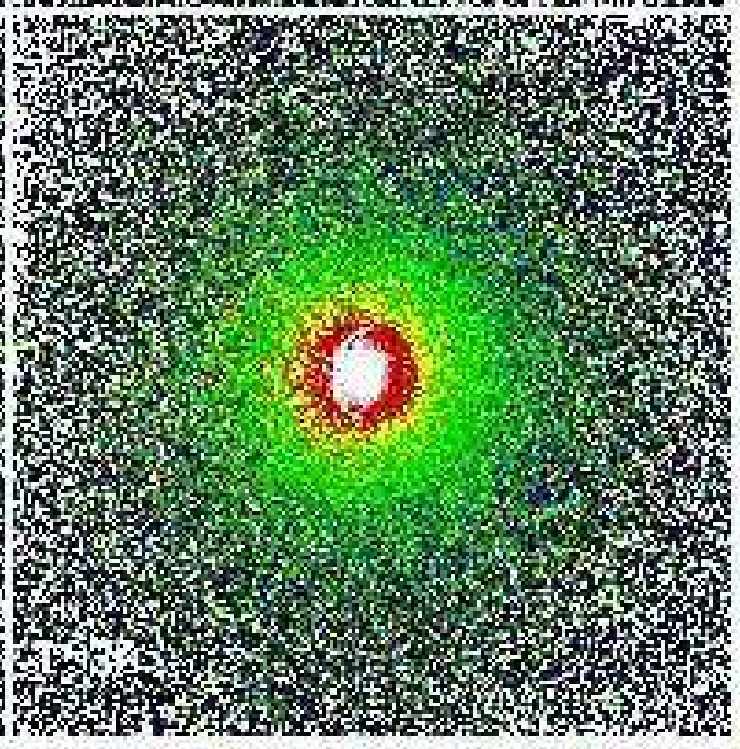
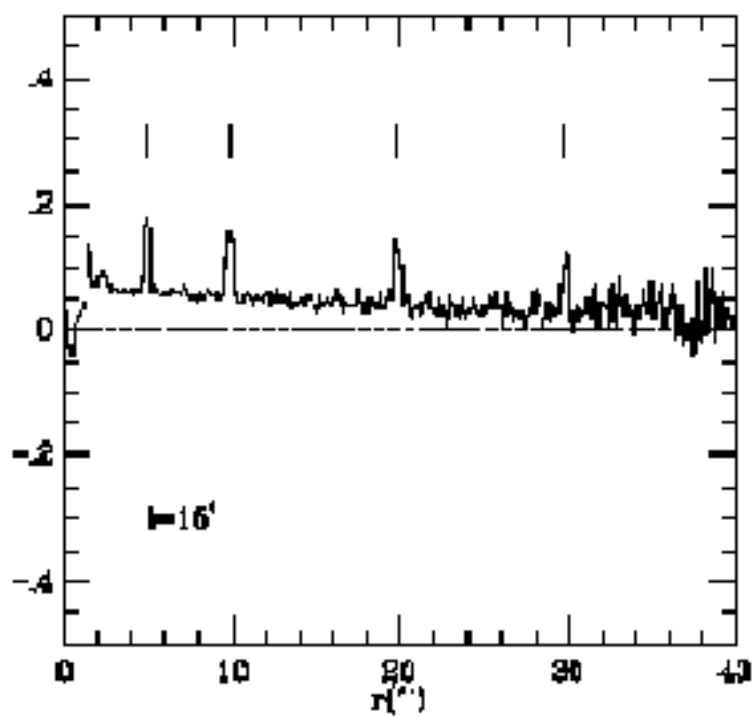
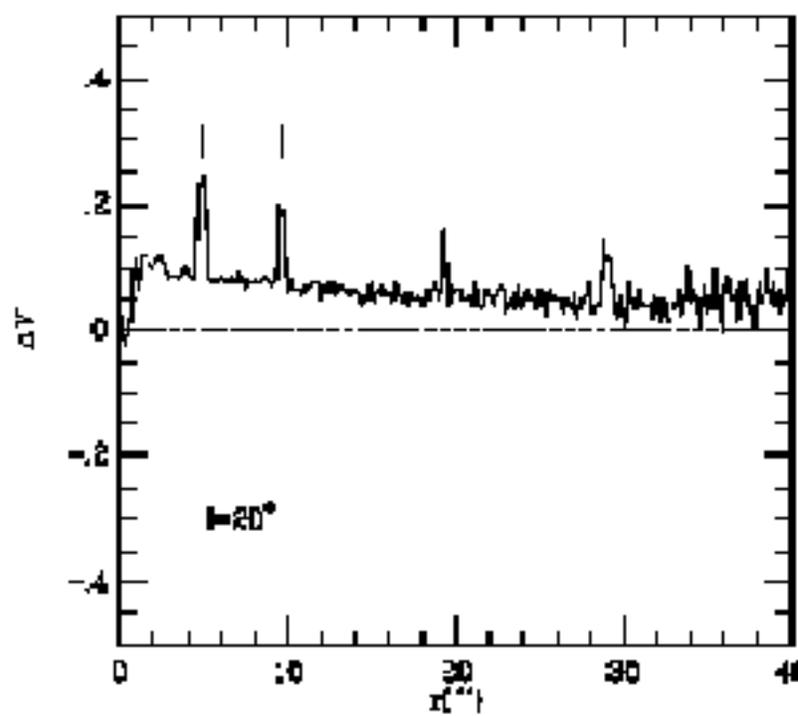
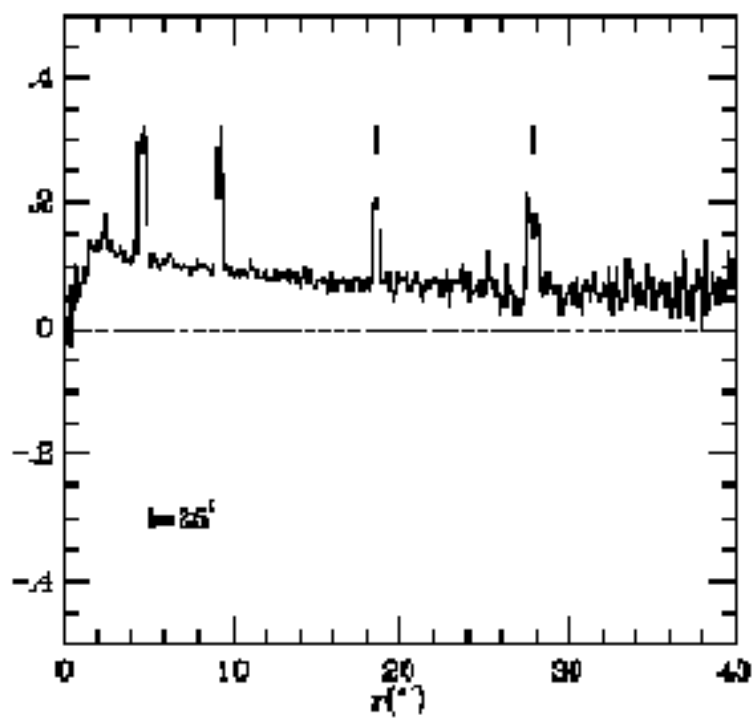
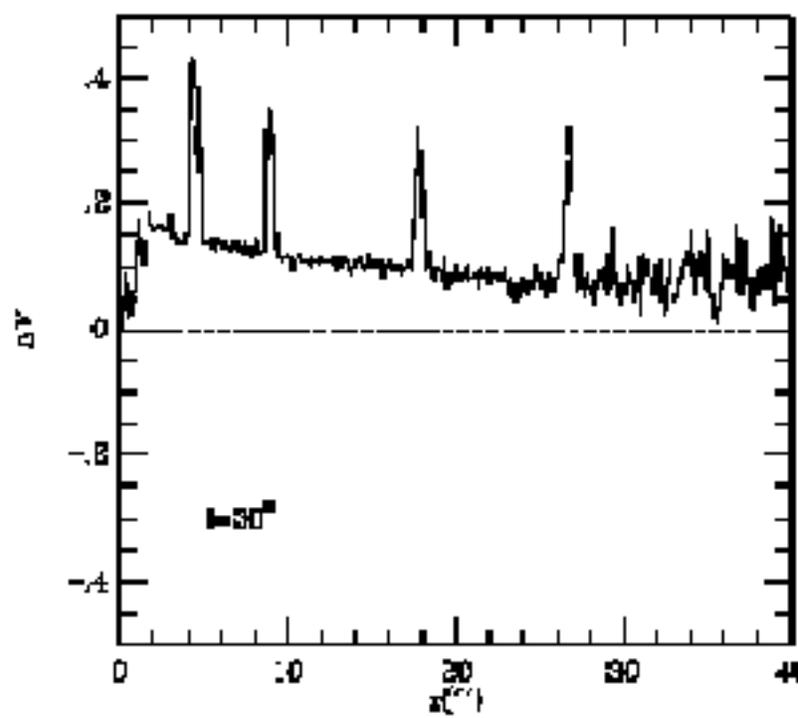
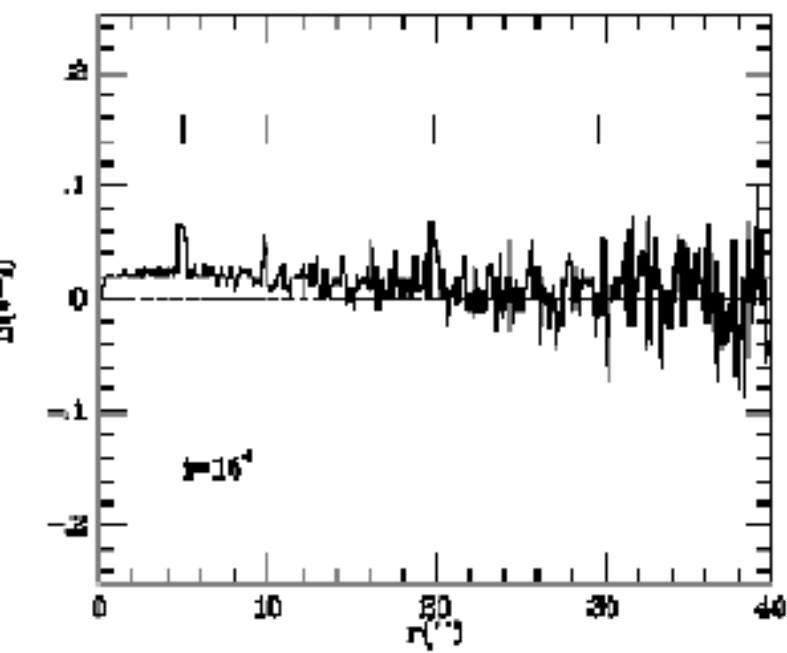
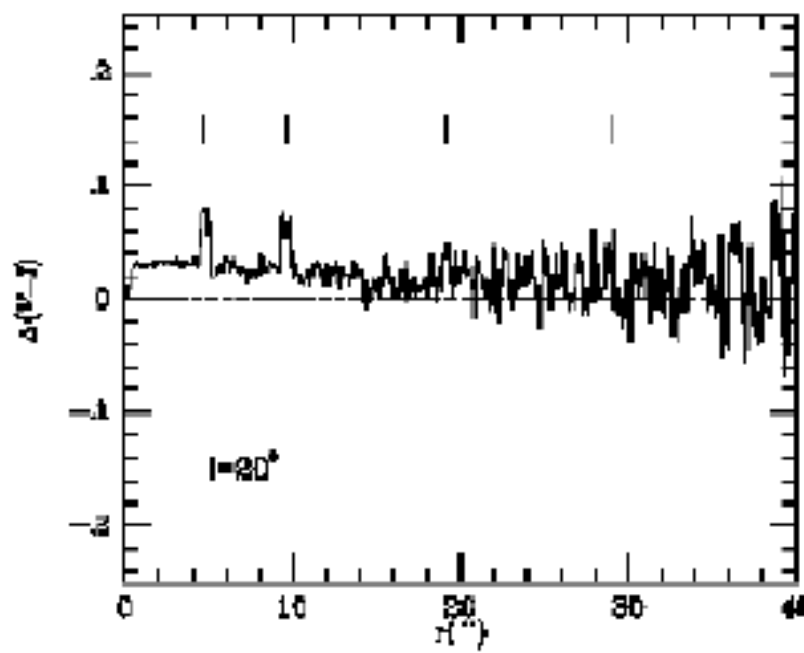
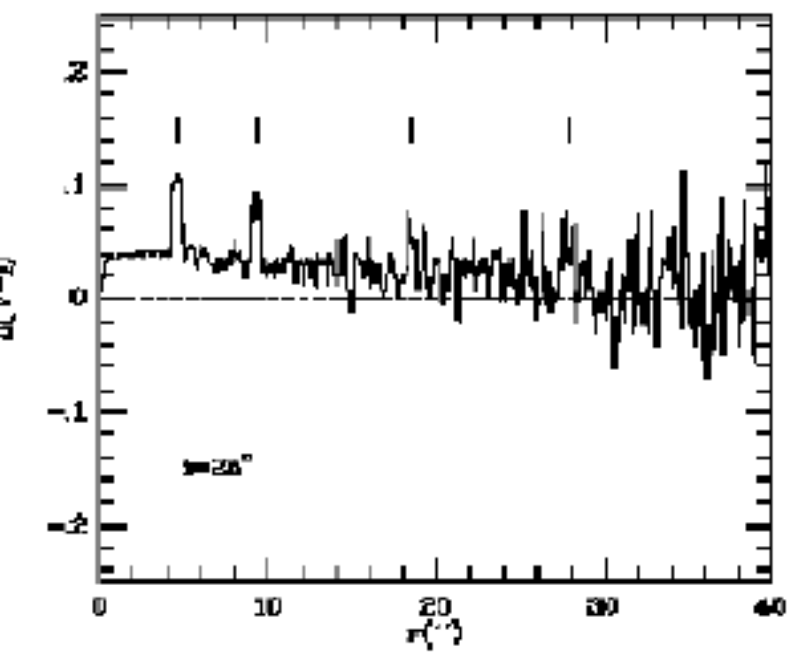
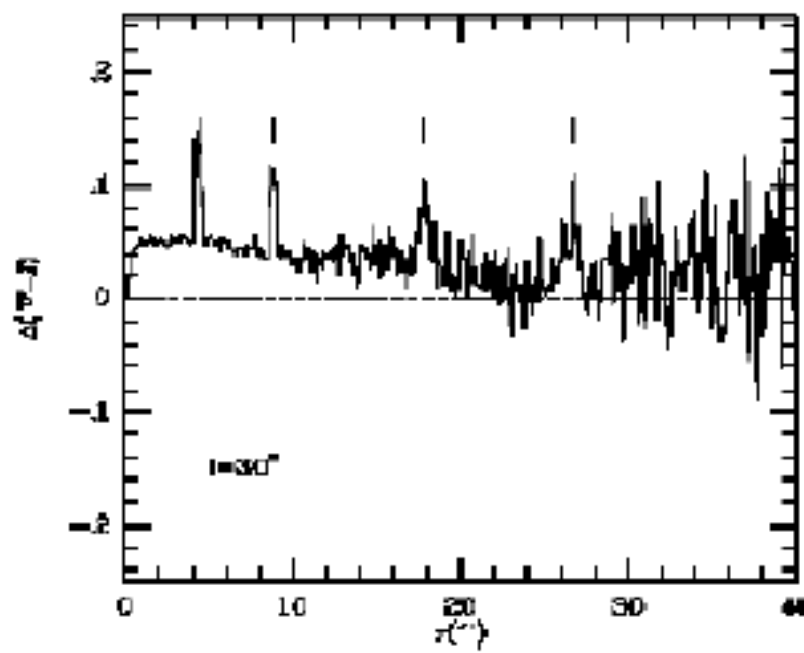
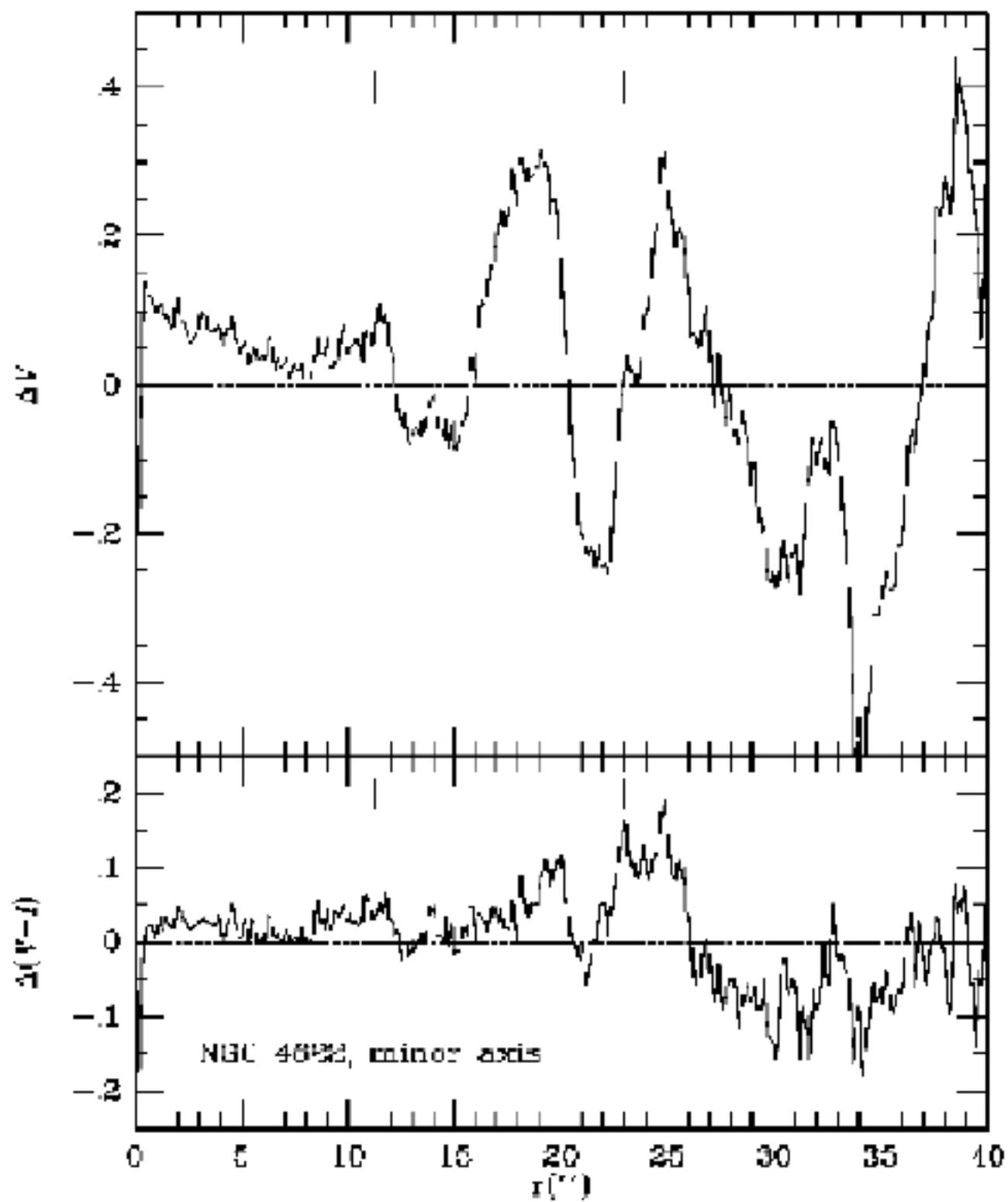
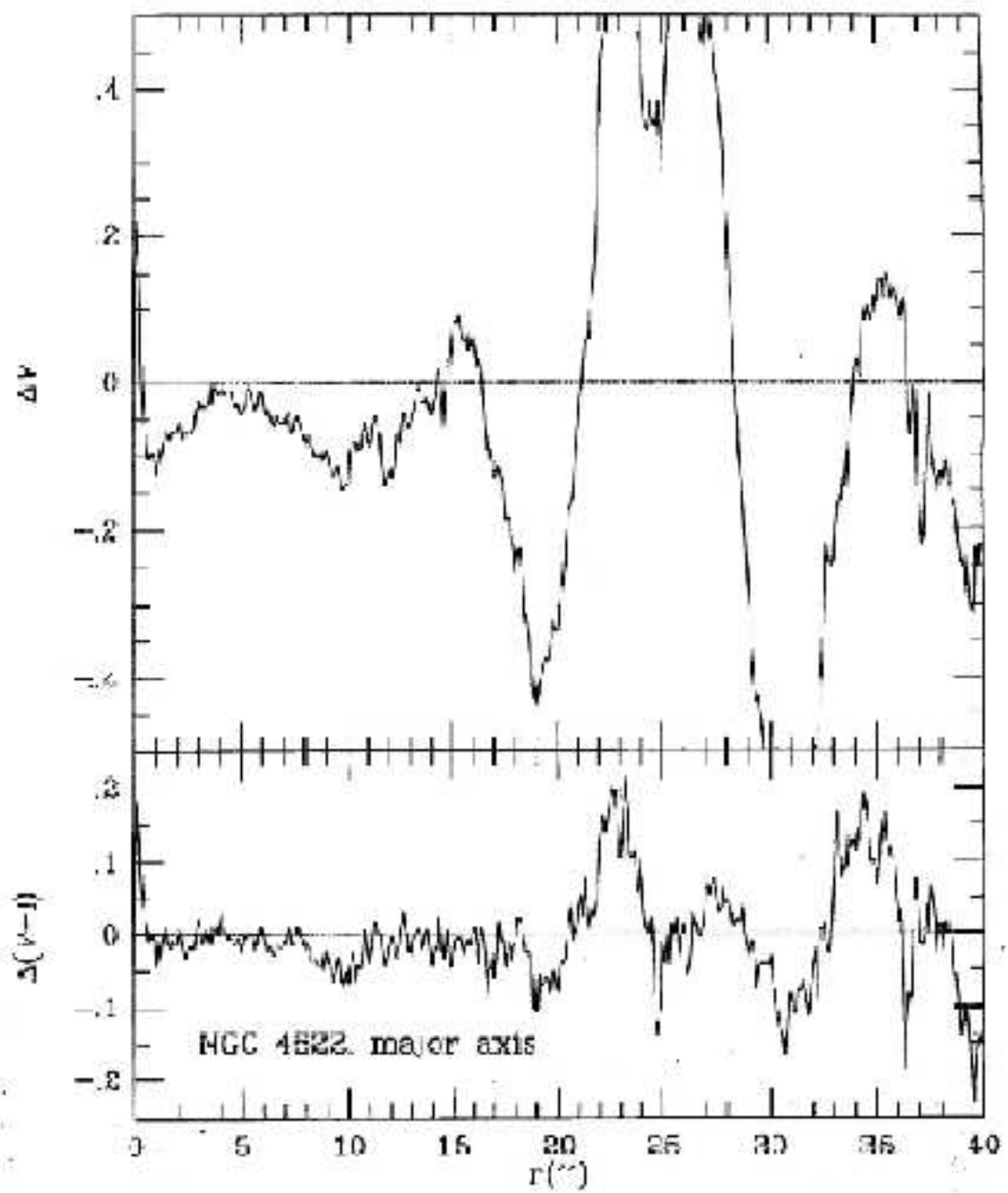
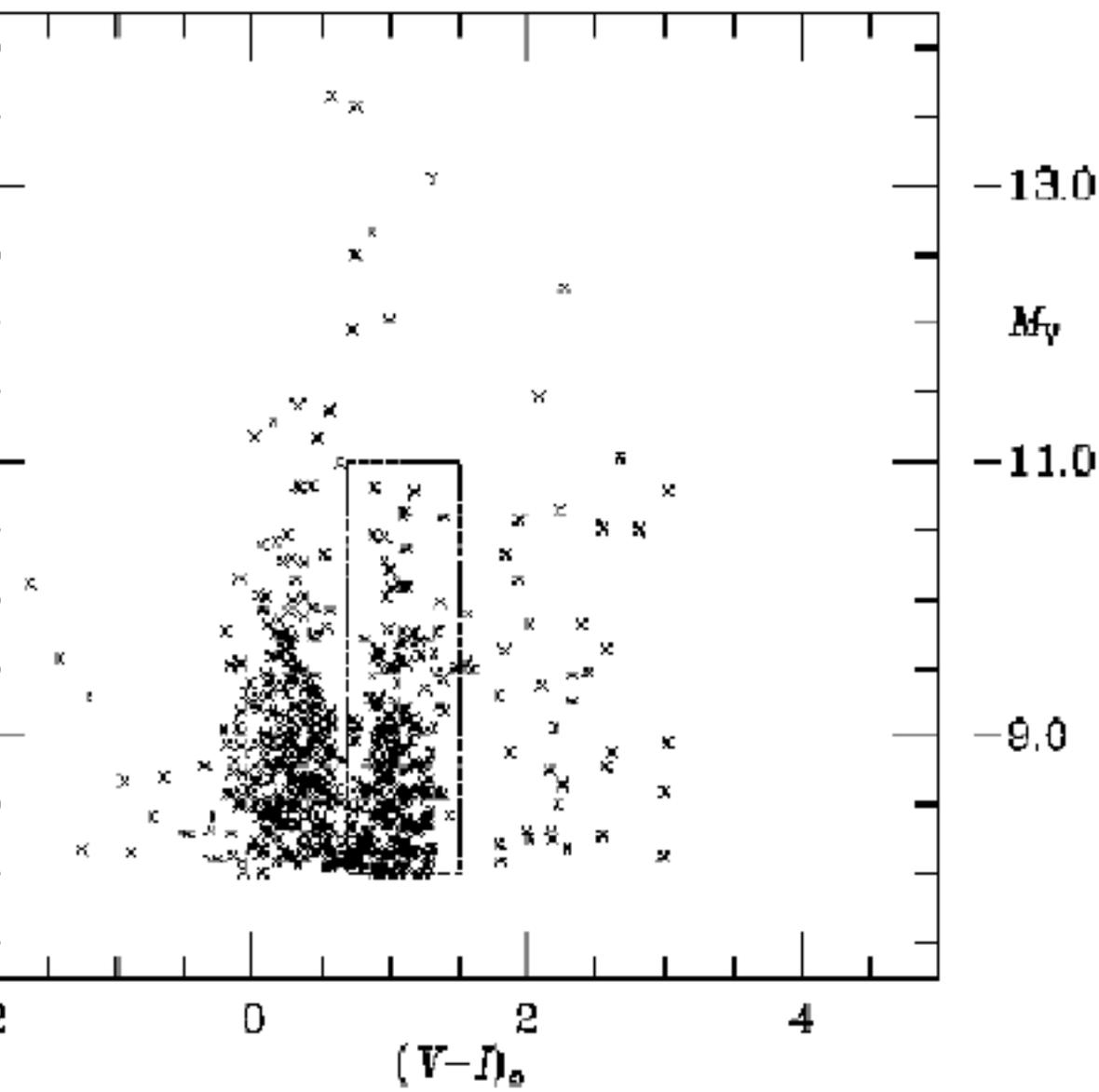
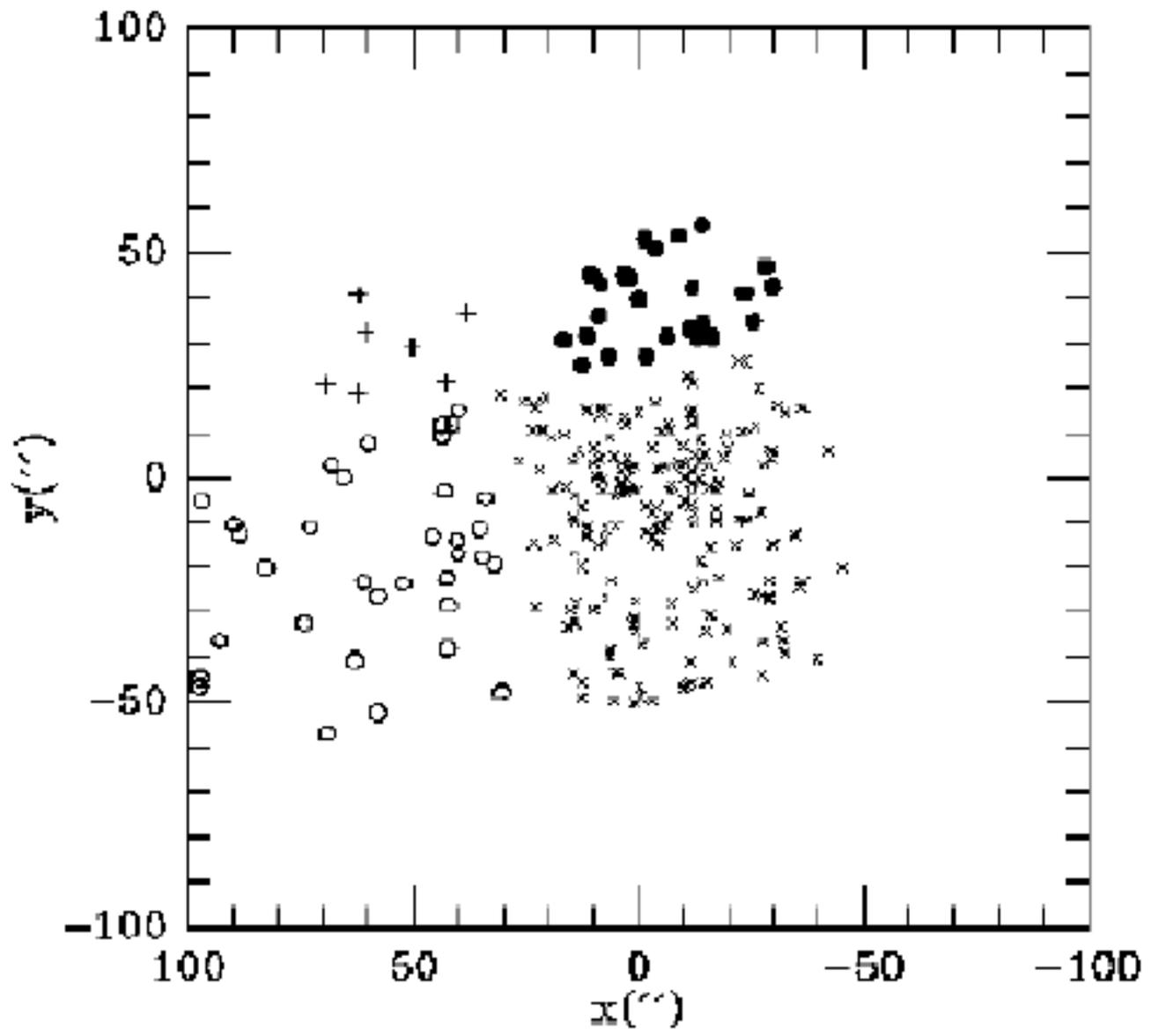
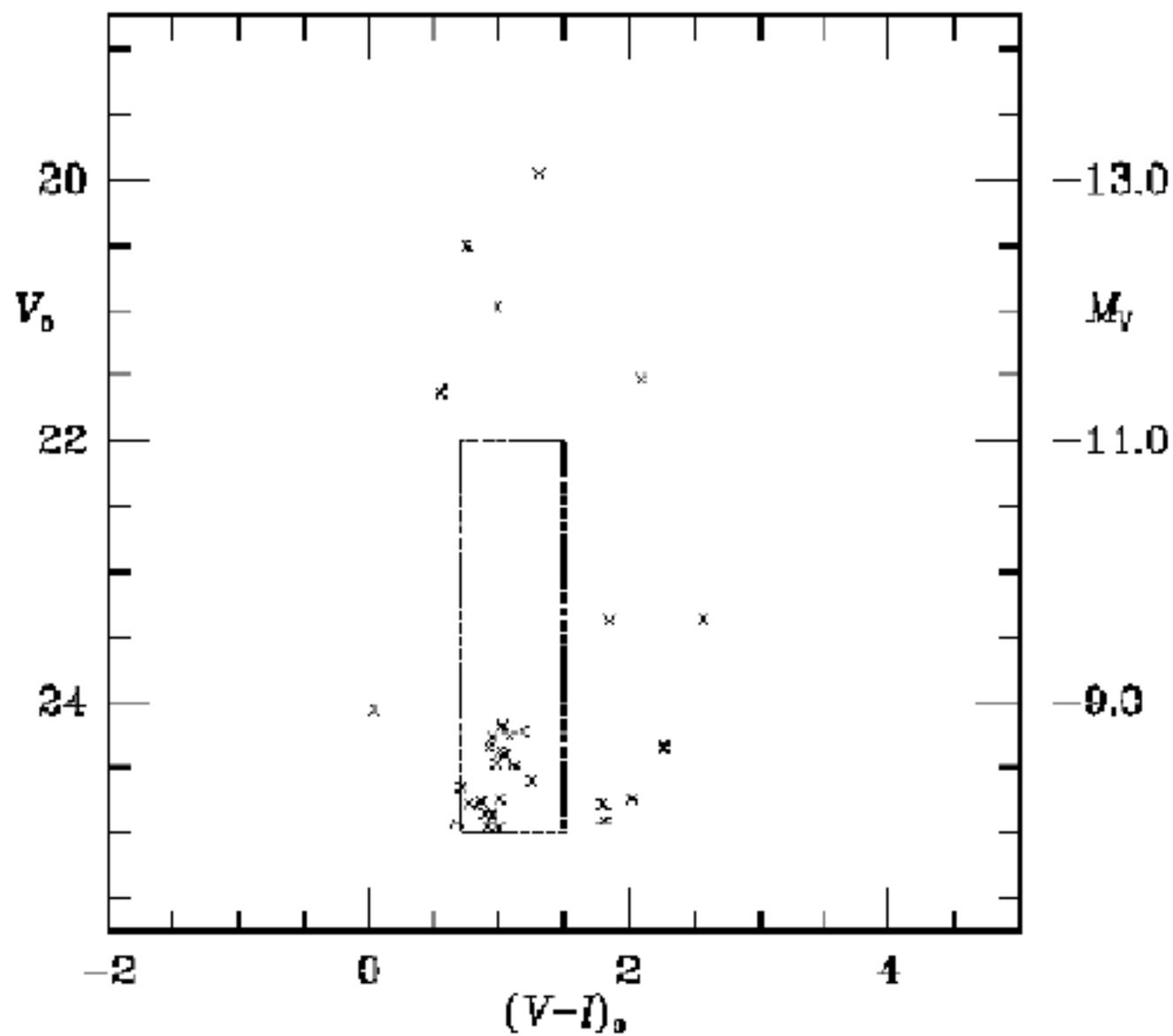
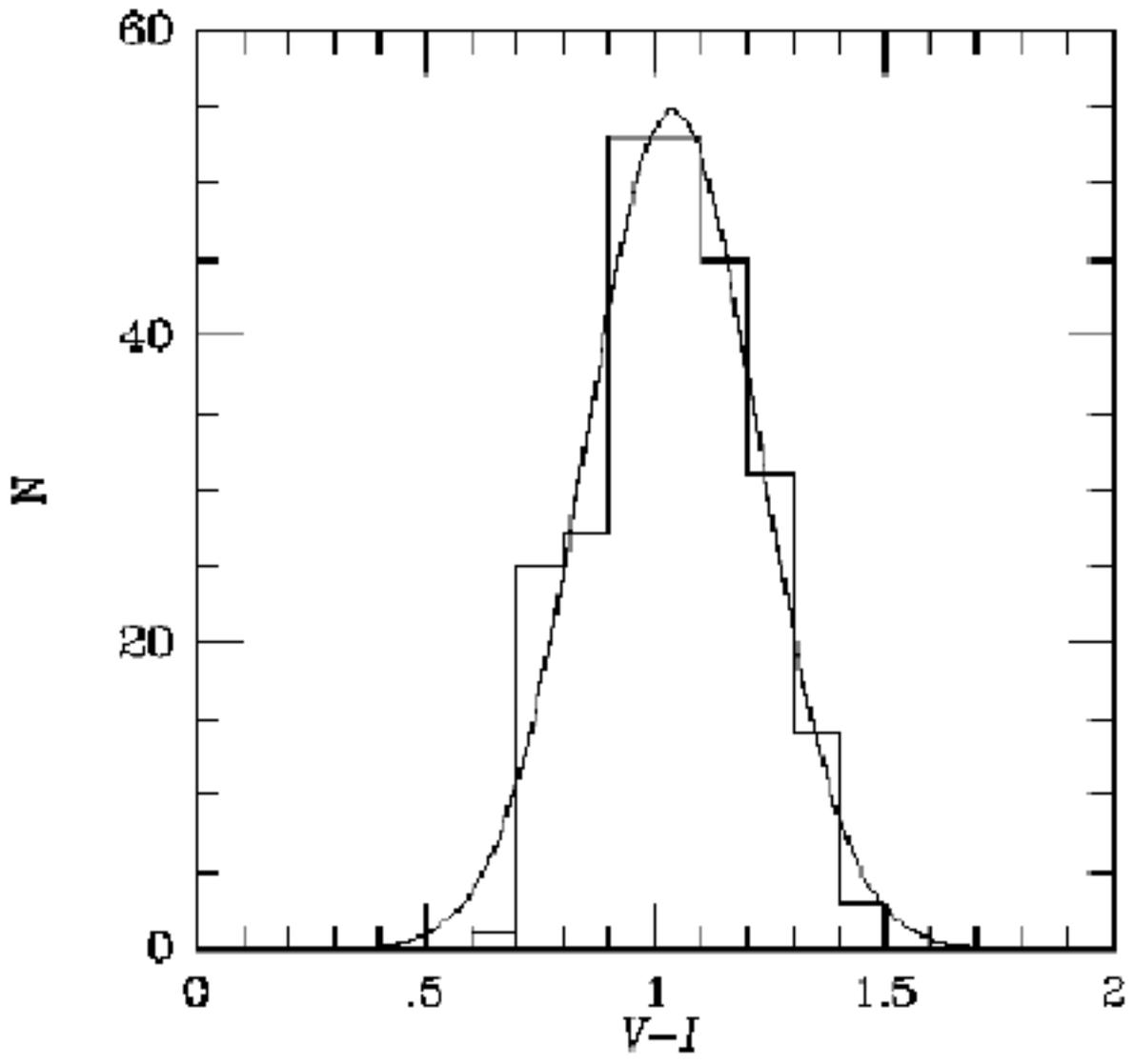
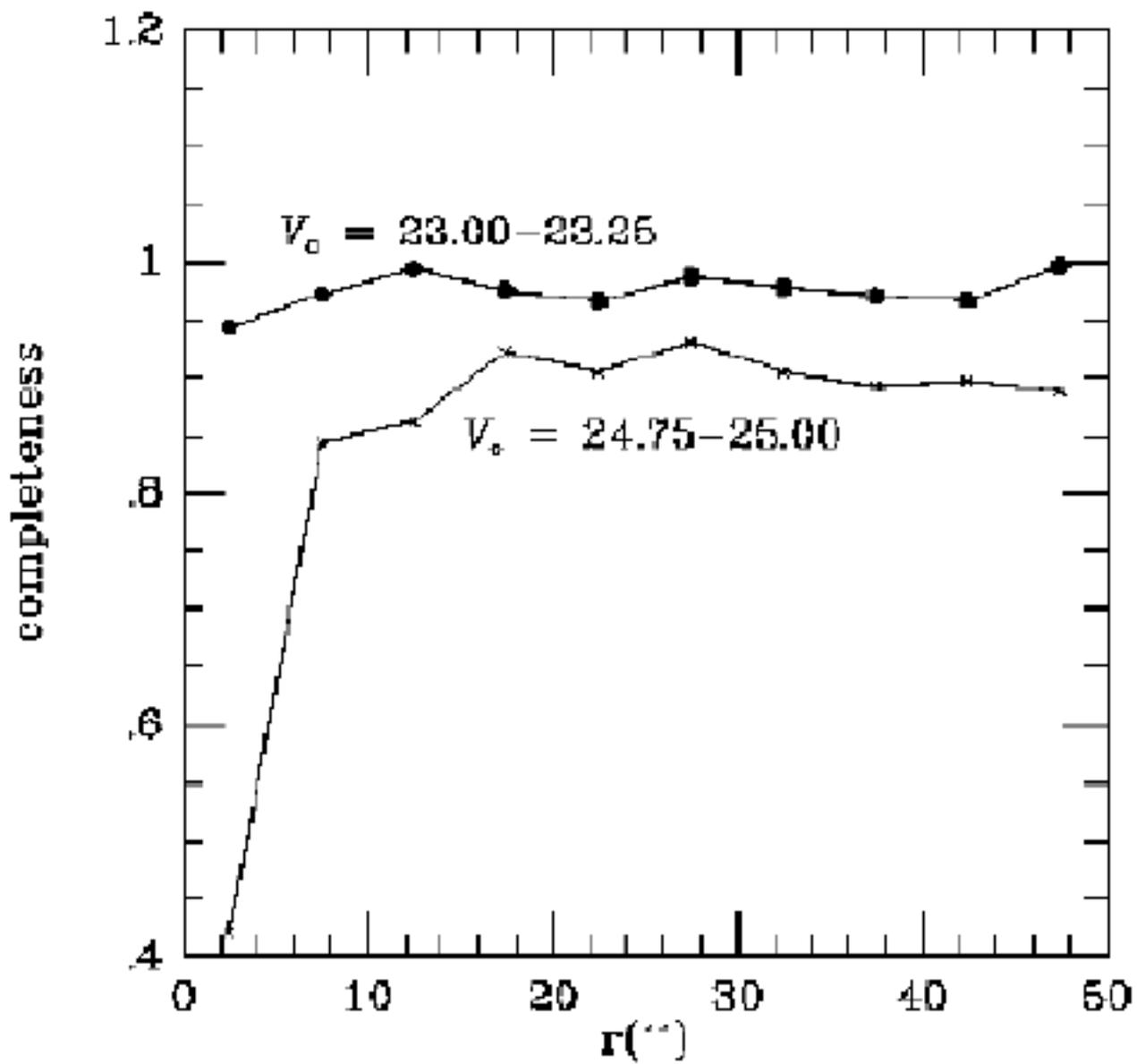
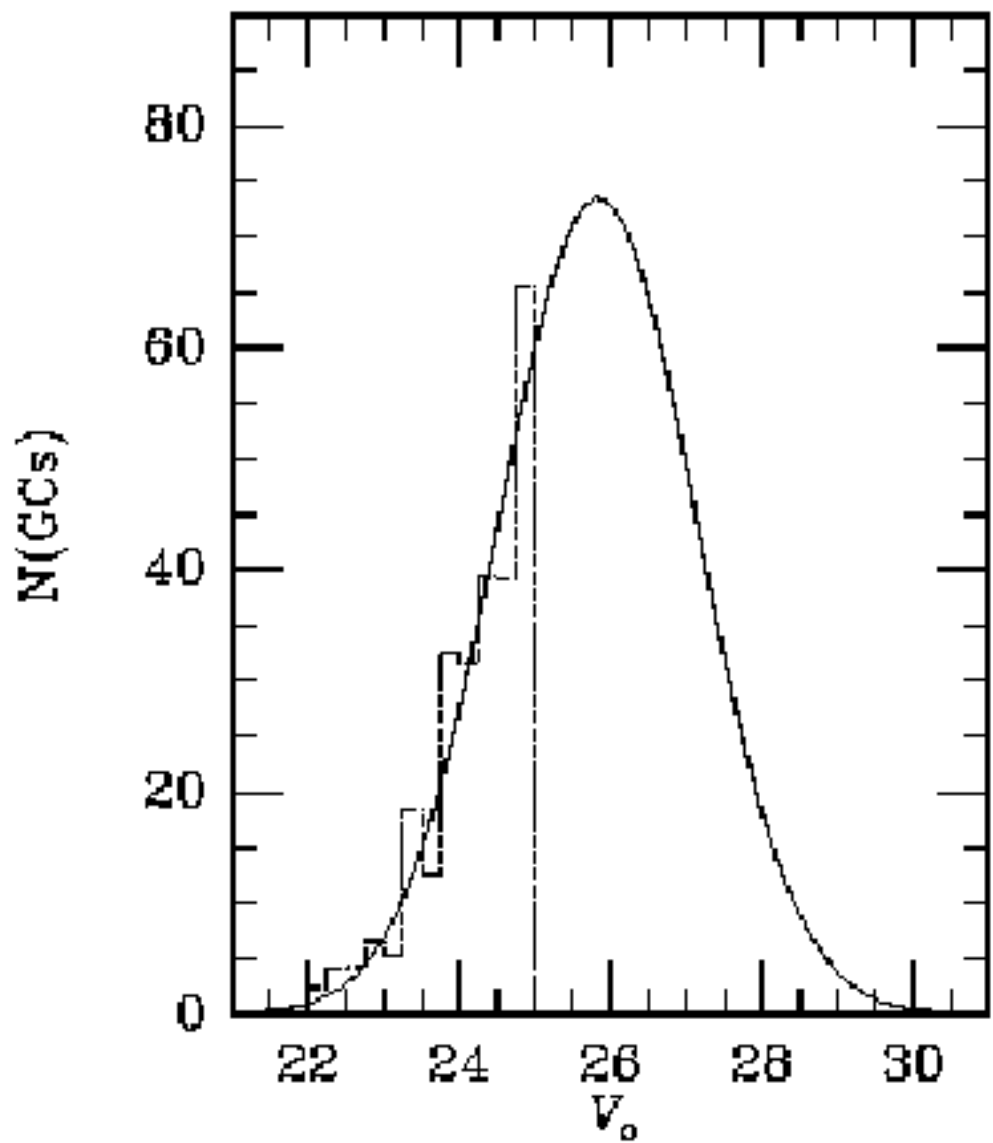
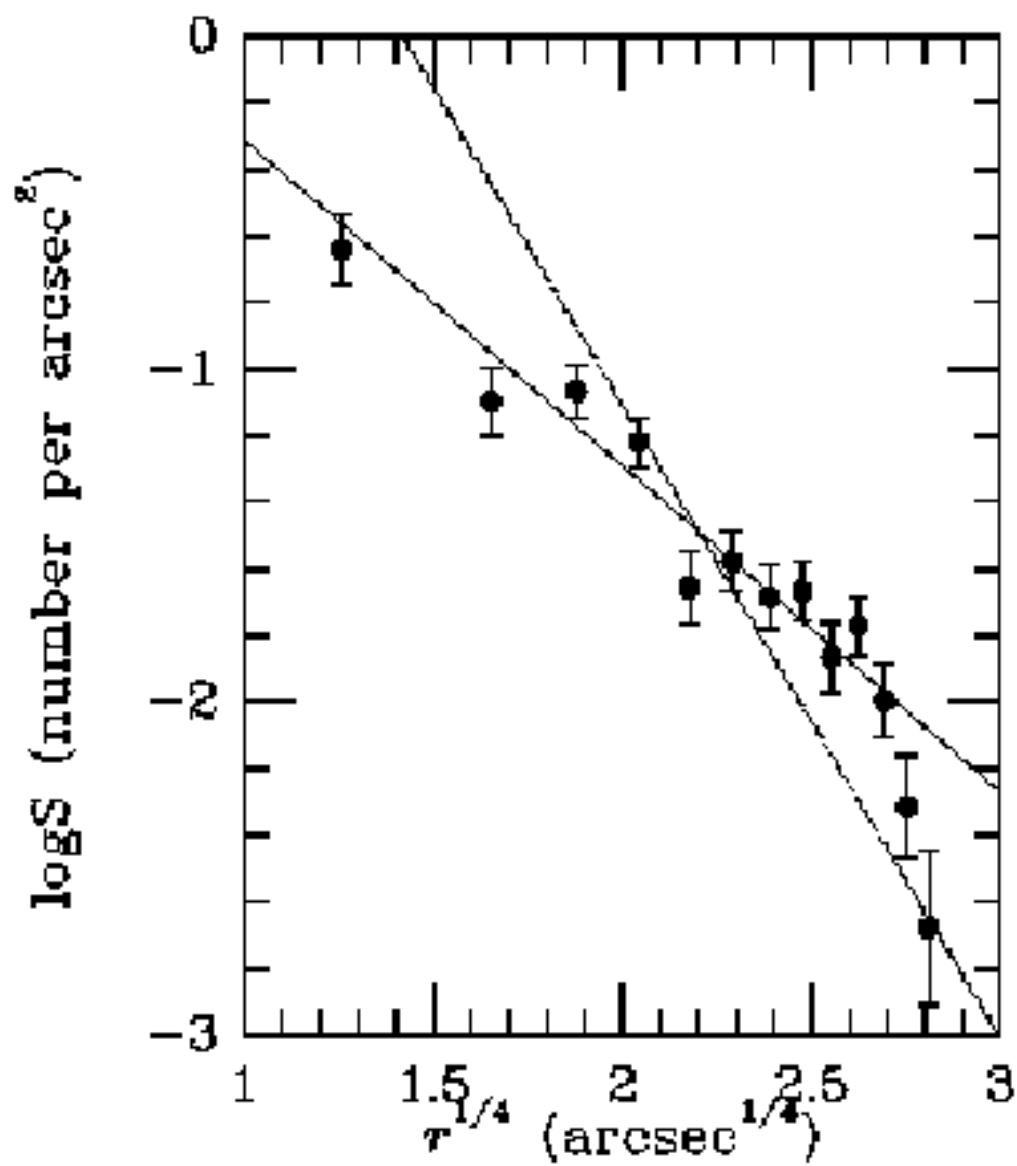
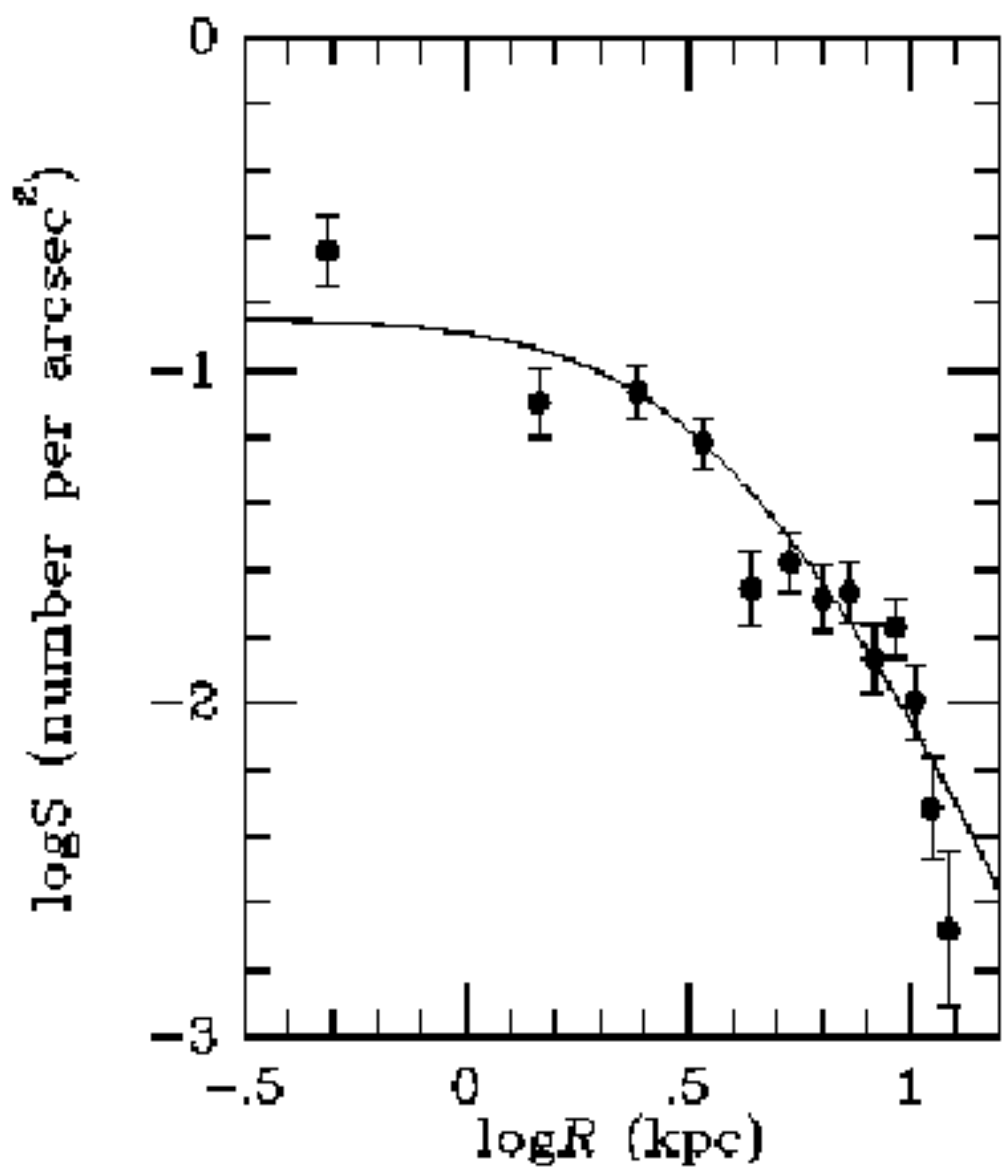
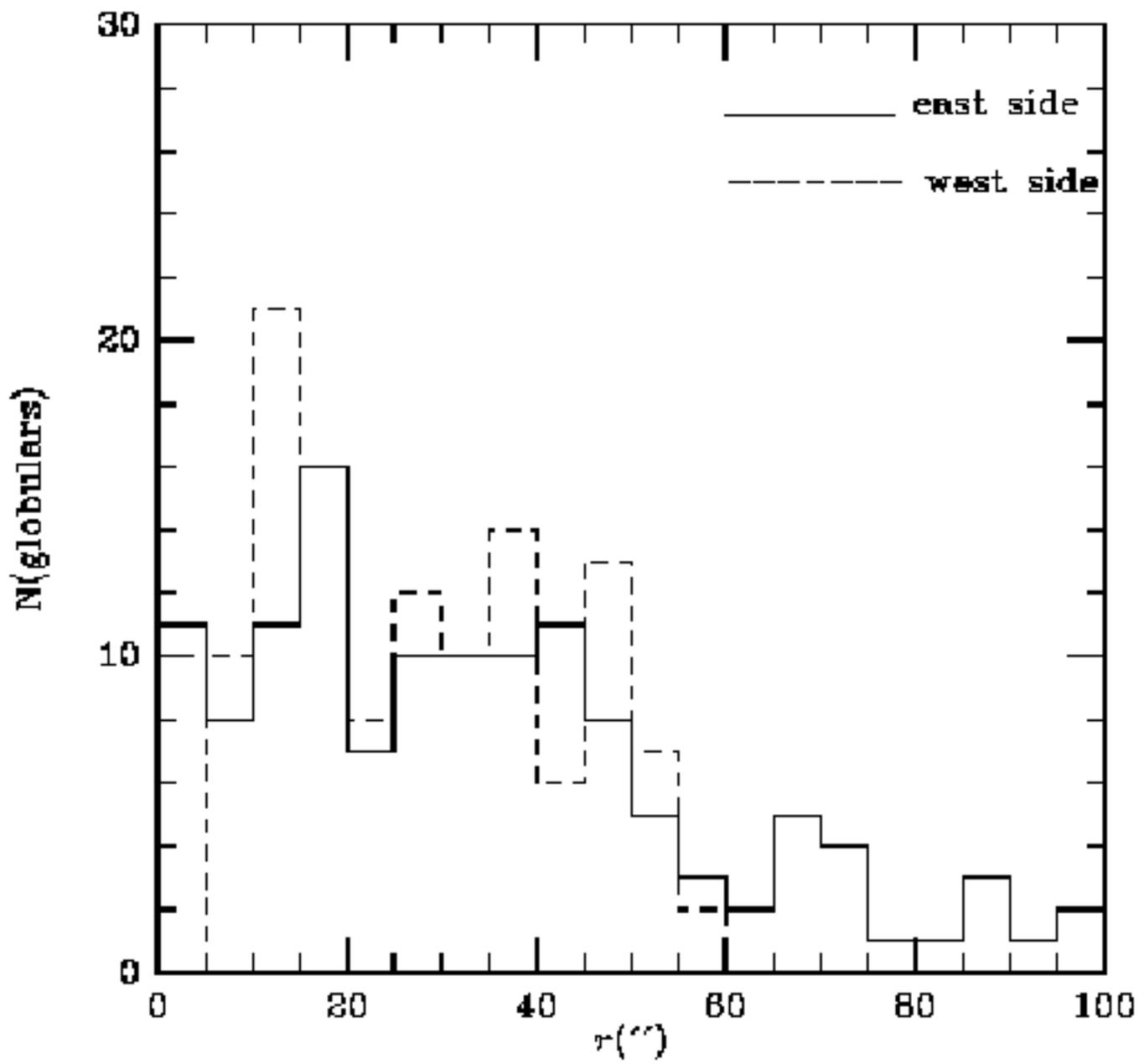
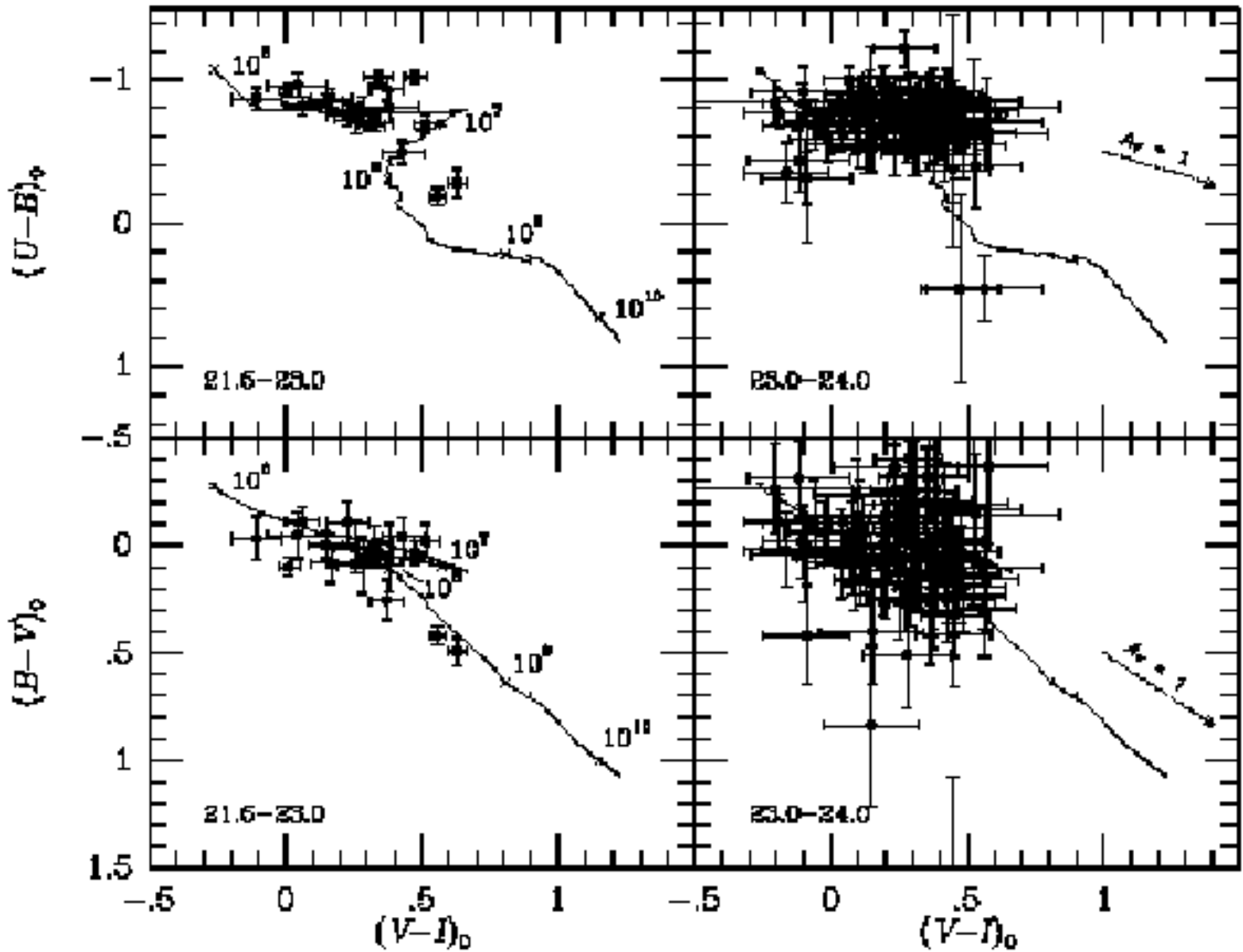
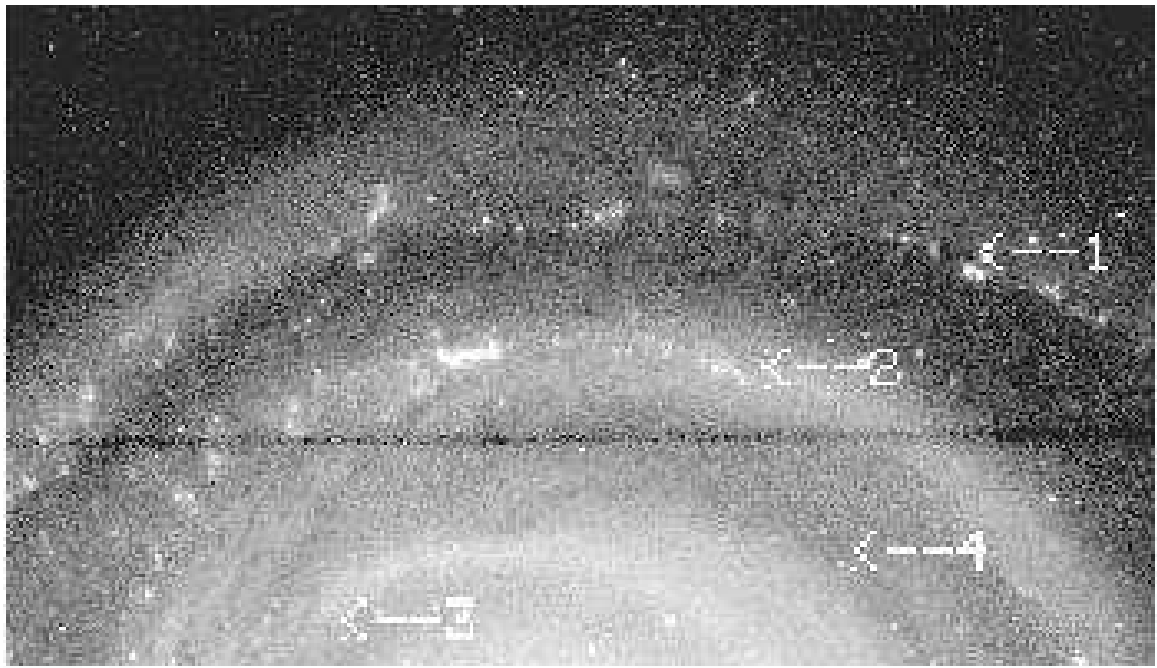
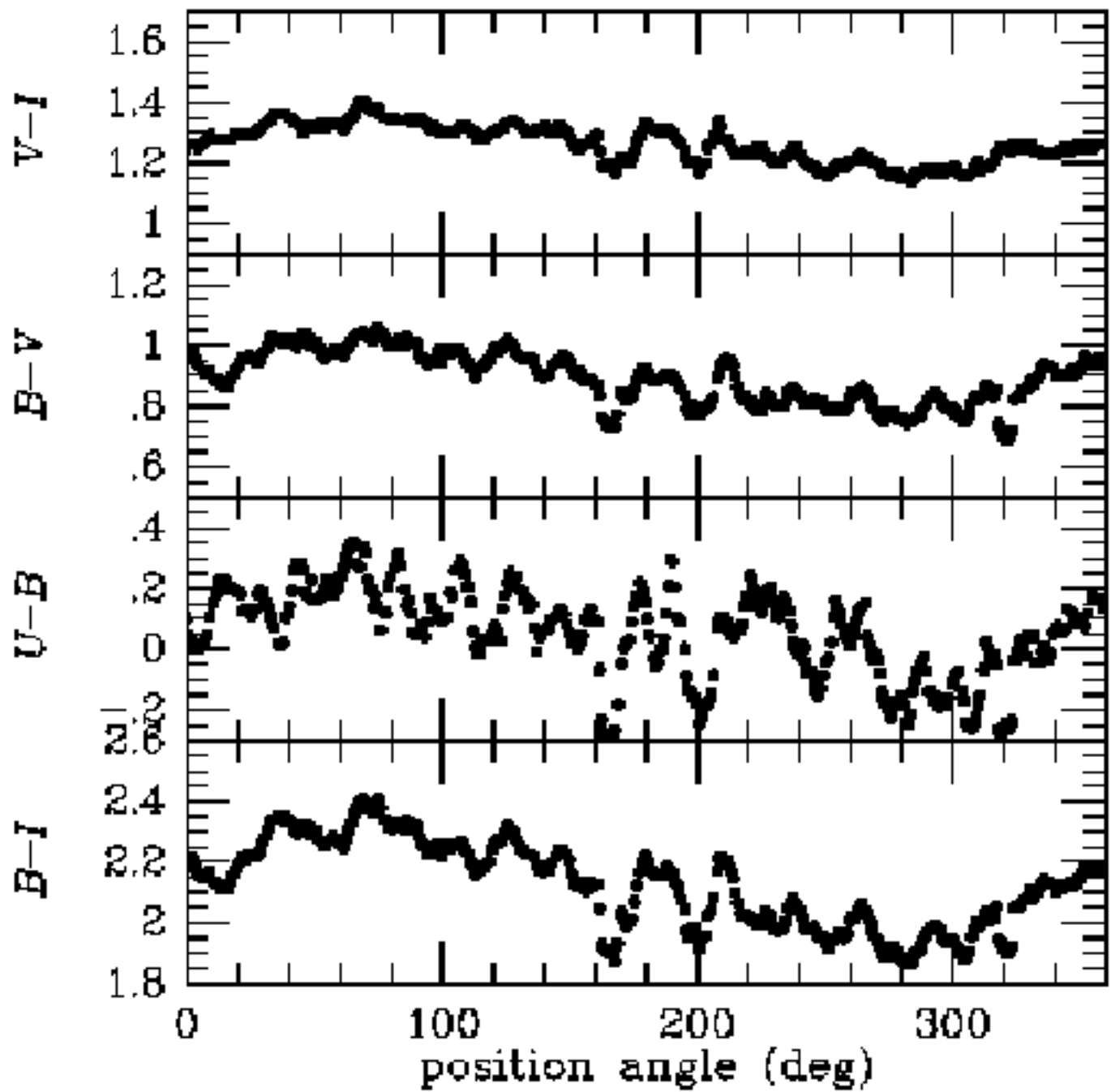
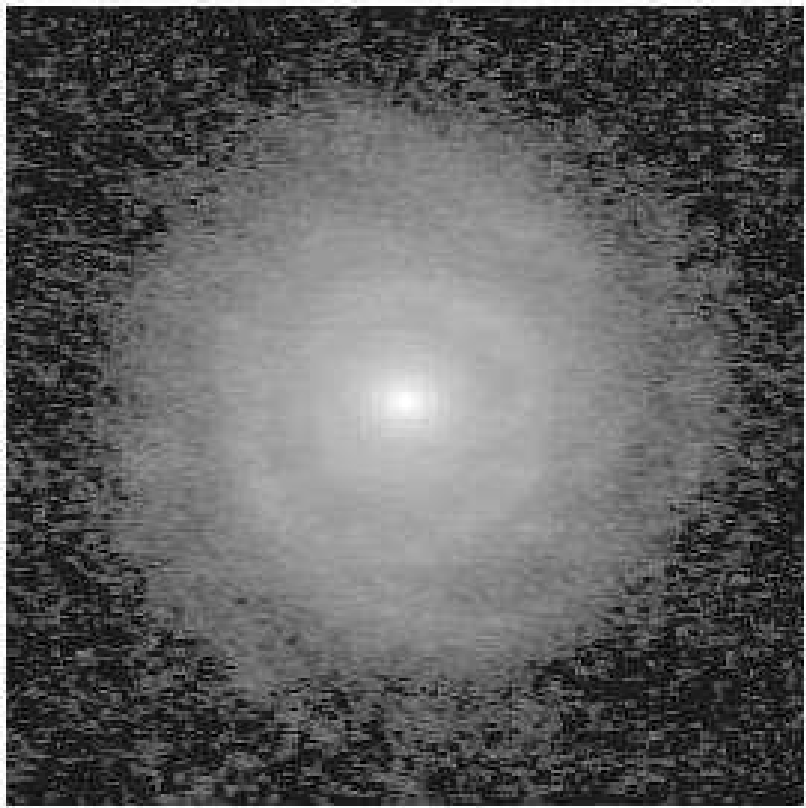
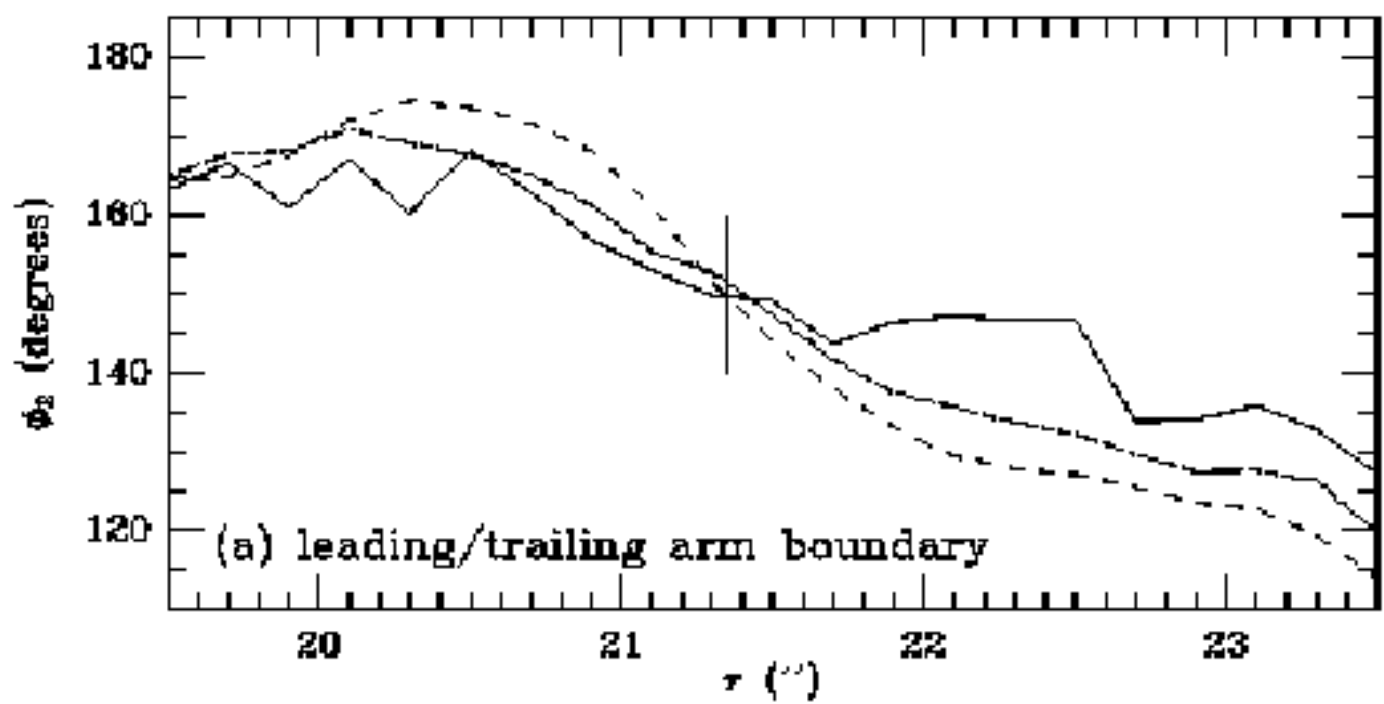
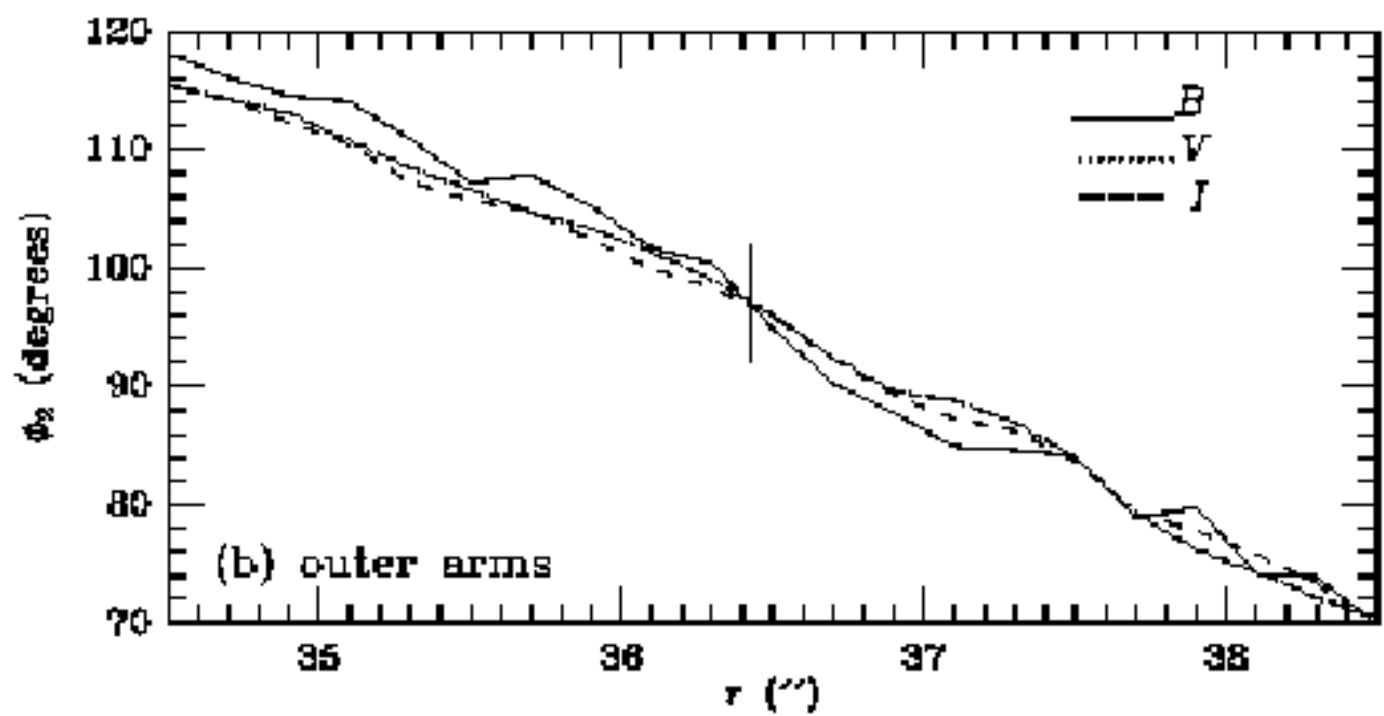
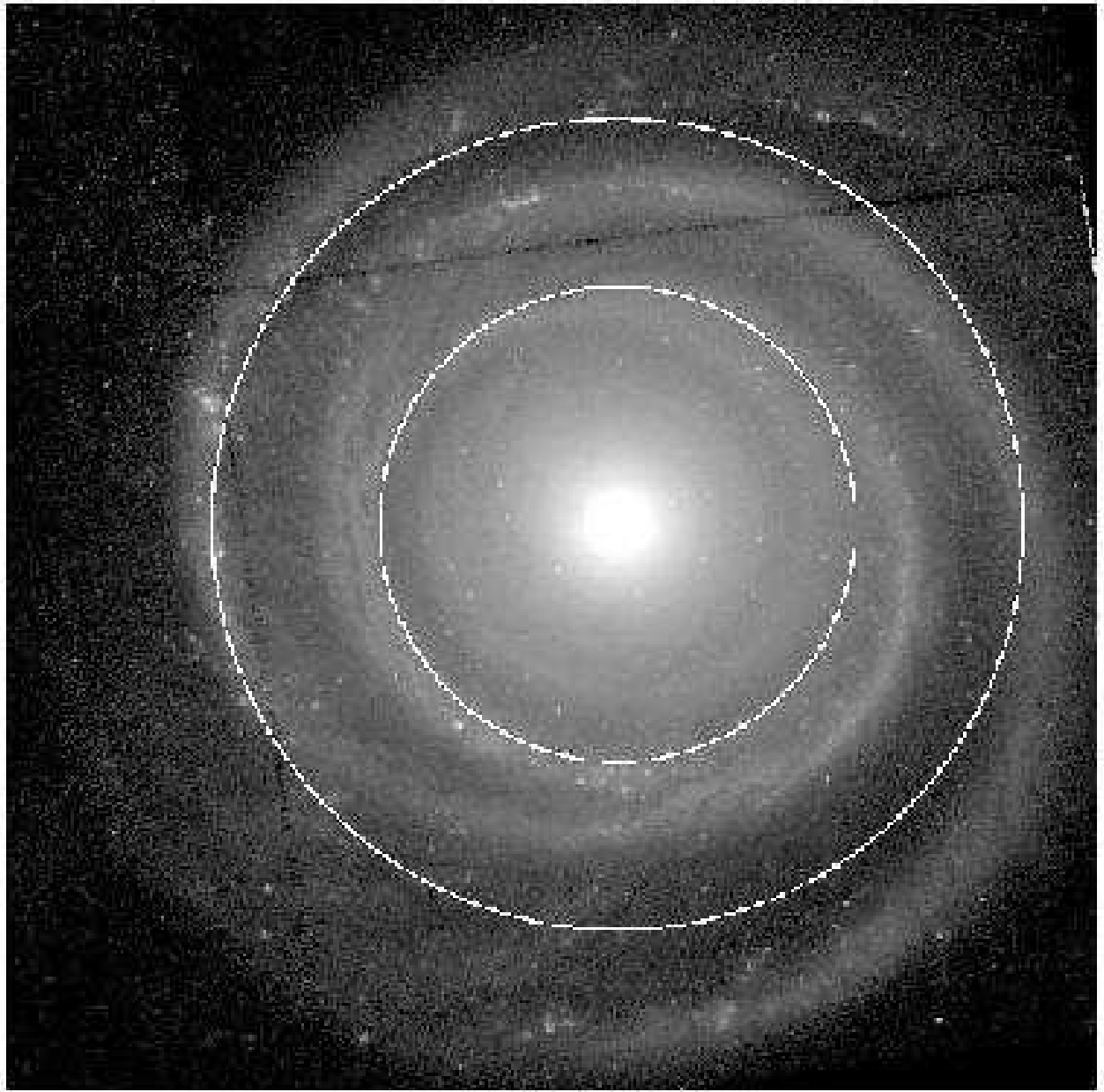
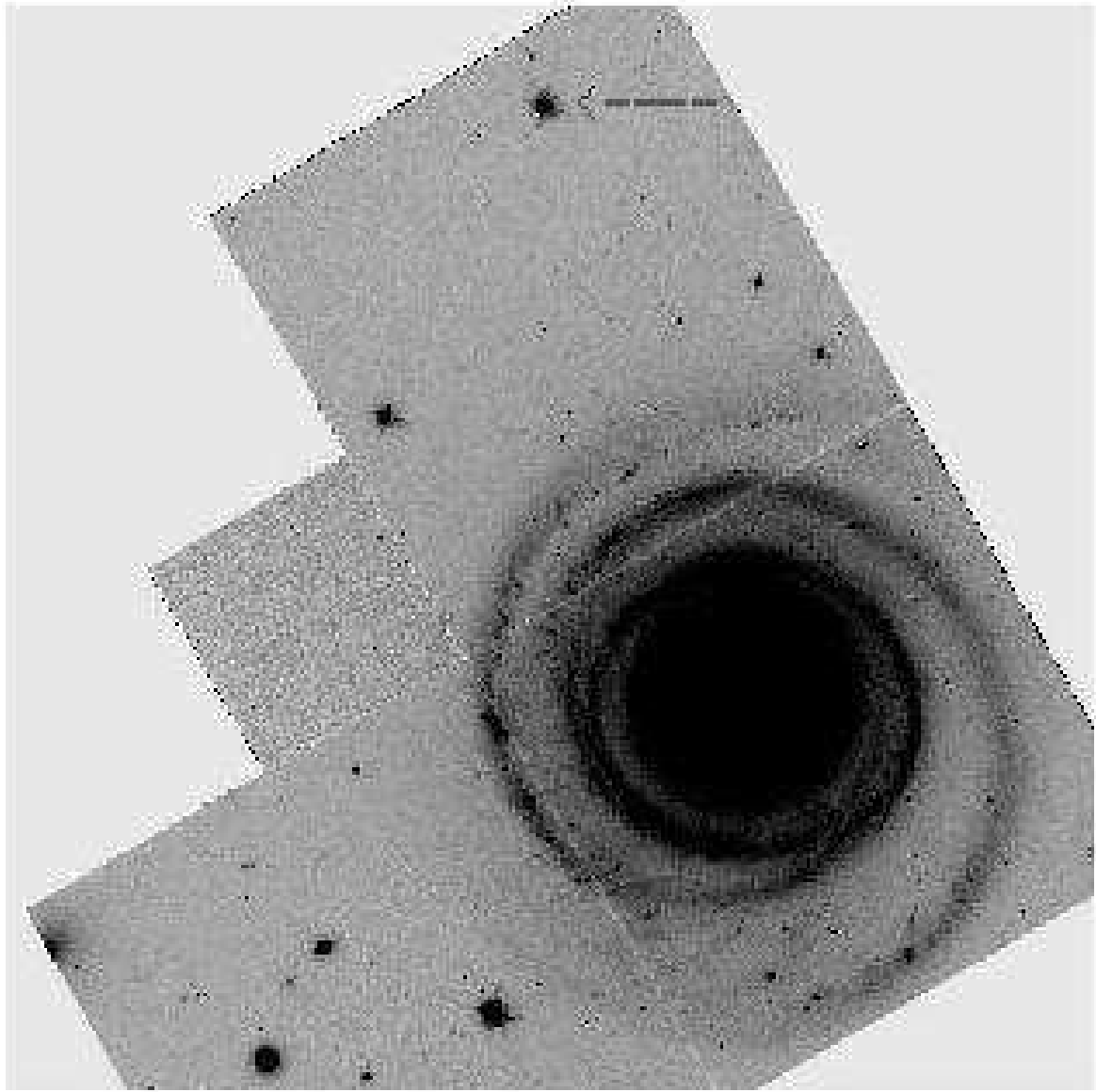
\end{document}